\pdfoutput=1
\documentclass{aa}  

\usepackage{graphicx}
\usepackage{longtable}
\usepackage{pdflscape}
\usepackage[normalem]{ulem}
\usepackage{txfonts}
\usepackage{natbib}
\begin{document}

   \title{Chemical Segregation in Hot Cores With Disk Candidates}

   \subtitle{An investigation with ALMA}

   \author{V. Allen\inst{1,2}
          \and
          F. F. S. van der Tak\inst{1,2}\fnmsep
          \and {\'{A}}. S{\'{a}}nchez-Monge\inst{3}
          \and R. Cesaroni\inst{4}
          \and M. T. Beltr{\'{a}}n\inst{4}
          }

   \institute{Kapteyn Astronomical Institute, University of Groningen,
              the Netherlands\\
              \email{allen@astro.rug.nl, vdtak@sron.nl}
             \and SRON, Groningen, the Netherlands
             \and I. Physikalisches Institut, K{\"{o}}ln, Germany
             \and INAF, Osservatorio Astrofisico di Arcetri, Firenze, Italy
             }

   \date{Received June 2016; accepted May 2017}

 
  \abstract
   {In the study of high-mass star formation, hot cores are empirically defined stages where chemically rich emission is detected toward a massive YSO.  It is unknown whether the physical origin of this emission is a disk, inner envelope, or outflow cavity wall and whether the hot core stage is common to all massive stars.}
   {We investigate the chemical make up of several hot molecular cores to determine physical and chemical structure.  We use high spectral and spatial resolution sub-millimeter observations to determine how this stage fits into the formation sequence of a high mass star.}
   {The sub-millimeter interferometer ALMA (Atacama Large Millimeter Array) was used to observe the G35.20-0.74N and G35.03+0.35 hot cores at 350 GHz in Cycle 0.  We analyzed spectra and maps from four continuum peaks (A, B1, B2 and B3) in G35.20-0.74N, separated by 1000-2000 AU, and one continuum peak in G35.03+0.35. We made all possible line identifications across 8 GHz of spectral windows of molecular emission lines down to a 3$\sigma$ line flux of 0.5~K and determined column densities and temperatures for as many as 35 species assuming local thermodynamic equilibrium (LTE).}
   {In comparing the spectra of the four continuum peaks, we find each has a distinct chemical composition expressed in over 400 different transitions. In G35.20, B1 and B2 contain oxygen- and sulfur-bearing organic and inorganic species but few nitrogen-bearing species whereas A and B3 are strong sources of O-, S-, and N-bearing organic and inorganic species (especially those with the CN-bond).  Column densities of vibrationally excited states are observed to be equal to or greater than the ground state for a number of species.  Deuterated methyl cyanide is clearly detected in A and B3 with D/H ratios of 8 and 13$\%$, respectively, but is much weaker at B1 and undetected at B2.  No deuterated species are detected in G35.03, but similar molecular abundances to G35.20 were found in other species.  We also find co-spatial emission of isocyanic acid (HNCO) and formamide (NH$_2$CHO) in both sources  indicating a strong chemical link between the two species.}
   {The chemical segregation between N-bearing organic species and others in G35.20 suggests the presence of multiple protostars, surrounded by a disk or torus.}


   \maketitle
%

\section{Introduction}

  \begin{table*}[!ht]
  \centering
  \caption{Source continuum characteristics}
  \label{statsTable}
  \begin{tabular}{cccccccc}

  \hline\hline
  Continuum Peak & Right Ascension & Declination & Size ($''$)$^a$ & $S_{\nu}$ (Jy)$^b$ & $T_\mathrm{kin}$ (K)$^c$ & $N$(H$_2$) (cm$^{-2}$)$^d$ & Mass (M$_{\sun}$) $^e$ \\
  \hline
  G35.20 A & 18:58:12.948 & +01:40:37.419 & 0.58 & 0.65 & 285  & $2.4\times10^{25}$ & 13.0 \\
  G35.20 B1 & 18:58:13.030 & +01:40:35.886 & 0.61 & 0.19 & 160 & $6.4\times10^{24}$ & 3.8 \\
  G35.20 B2 & 18:58:13.013 & +01:40:36.649 & 0.65 & 0.12 & 120 & $3.3\times10^{24}$ & 2.2 \\
  G35.20 B3 & 18:58:13.057 & +01:40:35.442 & 0.58 & 0.08 & 300 & $2.5\times10^{24}$ & 1.4 \\
  G35.03 A & 18:54:00.645 & +02:01:19.235 & 0.49 & 0.21 & 275 & $1.1\times10^{25}$ & 4.4 \\
  \hline 
  \end{tabular}
  \tablefoot{$^a$: Deconvolved average diameter of the 50$\%$ contour of the 870~$\mu$m continuum. \\
  $^b$:Integrated flux density within the 10$\sigma$ contour of the 870~$\mu$m continuum. \\
  $^c$:Average kinetic temperature based on CH$_3$CN line ratios as calculated using RADEX. For details, see \S~\ref{radex} \\
  $^d$:Calculated from source size, continuum flux density, and kinetic temperature (\S~\ref{radex}). \\
  $^e$:Sources mass calculated as in \citet{Alvaro2014} using the average kinetic temperatures.}
  \end{table*}

  Studying the formation of high-mass stars ($>$ 8 M$_{\sun}$) is important because they drive the chemical evolution of their host galaxies by injecting energy (through UV radiation, strong stellar winds, and supernovae) and heavy elements into their surroundings \citep{Zinnecker2007}.  In the study of high-mass star formation, several models have been proposed to explain the earliest processes involved. In particular, the work of \citet{McKeeTan} describes a process similar to that of low mass stars including a turbulent accretion disk and bipolar outflows (see also \citet{Tan2014}), the model by Bonnell and Smith (2011) proposes that matter is gathered competitively from low-turbulence surroundings between many low mass protostars funneling more material to the most massive core, and the model by \citet{Keto2007} which uses gravitationally trapped hypercompact HII regions to help a massive protostar to acquire more mass. All of these models predict the existence of disks as a mechanism to allow matter to accrete onto the protostar despite high radiation pressure \citep{Krumholz2009}.  However, until recently only a few candidate disks around B-type protostars were known.  Several disks have been detected through the study of complex organic molecules (COMs), molecular species bearing carbon and at least 6 atoms, allowing for the detection of more disks \citep{{Cesaroni2006},{Kraus2010},{BeltrandeWit}}.
  \par While the earliest stages of high-mass star formation have not yet been clearly determined, it is well known that a chemically rich stage exists, known as a hot molecular core (see \citet{Tan2014} for a review of high mass star formation). In this stage COMs are released from the icy surfaces of dust grains or formed in the hot circumstellar gas \citep{Herbst2009}.  These hot cores are dense (n$_H$ $>$ 10$^7$~cm$^{-3}$), warm (100-500~K), and compact ($<$ 0.05 pc) and are expected to last up to 10$^5$ years.  The signpost of the hot core stage is a rich molecular emission spectrum including many COMs like methanol (CH$_3$OH) and methyl cyanide (CH$_3$CN).  These species may be formed on dust grain surfaces in a cooler place (or time) and released from grain surfaces as forming star heats the grains.  Alternatively, they may form in the hot gas surrounding these massive young objects as the higher temperature allows for endothermic reactions to take place more readily.  In reality, it is likely that both formation paths are necessary to achieve the molecular abundances seen around hot cores.  High spatial and spectral resolution observations can help us to disentangle the different COMs and their spatial distribution during this phase.  Disks candidates have been discovered in a few HMC sources, suggesting a link between disks and HMC chemistry.  Studying the chemistry of such regions can help us to understand the process of high-mass star formation as chemical differences across small physical scales provide clues to the different evolutionary stages involved.
  \par With the advent of ALMA (Atacama Large Millimeter Array), it is now possible to make highly sensitive, high spectral and spatial resolution observations of less abundant molecular species.  The search continues for the precursors of life like the simplest amino acid, glycine (H$_2$NCH$_2$COOH), but complex organic species with up to 12 atoms have already been detected\footnote{https://www.astro.uni-koeln.de/cdms/molecules}.  These include important precursors to amino acids like aminoacetonitrile (H$_2$NCH$_2$CN), detected by \citet{Belloche2008}; the simplest monosaccharide sugar glycolaldehyde (CH$_2$OHCHO), first observed in a hot molecular core outside the galactic center by \citet{BeltranGlyco}; and formamide (NH$_2$CHO) extensively studied by \citet{Ana2015}.  With ALMA we have the ability to detect hot cores and study their properties in detail to determine how the spatial distribution of COMs influences the formation of massive stars.  Despite advances in technology, astronomers have yet to determine whether the emission from the hot core arises from the inner envelope (spherical geometry) or from a circumstellar disk (flat geometry).  It is also possible that these hot cores could be outflow cavity walls as has been recently modeled for low-mass stars by \citet{Droz2015}.
  \par In this paper we study the chemical composition and spatial distribution of species in two high-mass star-forming regions which have been shown to be strong disk-bearing candidates, G35.20-0.74N and G35.03+0.35 (hereafter G35.20 and G35.03 respectively).  We present a line survey of the hot core in G35.03 and in four continuum peaks in the G35.20 hot core containing $\sim$ 18 different molecular species (plus 12 vibrationally excited states and 22 isotopologues) of up to 10 atoms and $>$400 emission lines per source.  We also present our analysis of the chemical segregation within Core B of G35.20 depicting a small-scale ($<$1000~AU) separation of nitrogen chemistry and temperature difference.  A chemical separation on the scale of a few 1000s of AU within a star forming region has been seen before in Orion KL \citep{Caselli1993}, W3(OH) and W3(H$_2$O) \citep{Wyrowski1999}, and AFGL2591 \citep{Izaskun2012}.
  \par The distance to both sources has been estimated from parallax measurements to be 2.2 kpc for G35.20 \citep{Zhang2009} and 2.32 kpc for G35.03 \citep{parallax}. G35.20 has a bolometric luminosity of $3.0\times10^4$ L$_{\sun}$ \citep{Alvaro2014} and has been previously studied in \citet{Alvaro2013} and \citet{Alvaro2014} where they report the detection of a large (r$\sim$2500 AU) Keplerian disk around Core B and a tentative one in Core A.  The bolometric luminosity of G35.03 is $1.2\times10^4$~L$_{\sun}$ and was reported to have a Keplerian disk (r$\sim$1400-2000~AU) around the hot core A in \citet{Beltran2014}.

   \begin{figure}[h]
   \centering
      \includegraphics[width=\hsize]{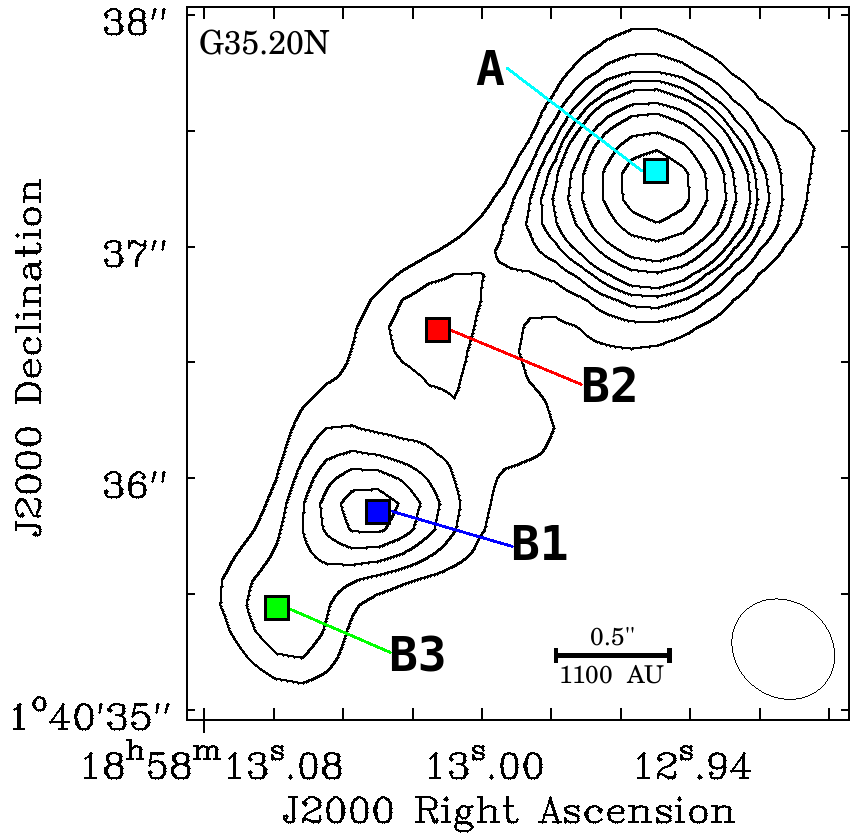}
      \caption{Image of the 870 $\mu$m continuum emission from Cycle 0 ALMA observations of G35.20.  Contour levels are 0.03, 0.042, 0.055, 0.067, 0.08, 0.10, 0.13, 0.18, and 0.23 Jy/beam ($\sigma$ $=$ 1.8 mJy/beam).  The pixel-sized colored squares mark each of the spectral extraction points.}
         \label{ContPts3520}
   \end{figure}

   \begin{figure}[h]
   \centering
      \includegraphics[width=\hsize]{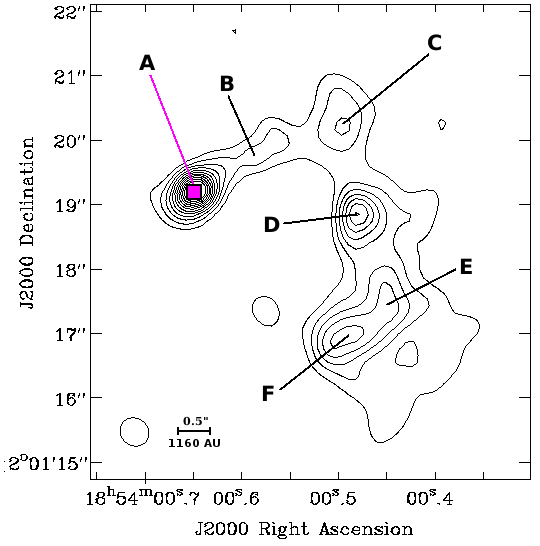}
      \caption{Image of the 870 $\mu$m continuum emission from Cycle 0 ALMA observations of G35.03.  Contour levels are 8.6, 16.8, 24.9, 33, 41.2, 49.3, 57.4, 65.5, 73.6, 81.8, and 89.9 mJy/beam ($\sigma$ $=$ 3.0 mJy/beam).  The pixel-sized colored square marks the spectral extraction point.  The cores identified in \citet{Beltran2014} are labeled A-F.}
         \label{ContPts3503}
   \end{figure}

\begin{table*}[hbt!]
\centering
\caption{Line detections and measurements for H$_2$CS}
\label{H2CS}
{\scriptsize
\begin{tabular}{cccccccccccc}

\hline\hline
 &  & \multicolumn{2}{c}{G35.20 A} & \multicolumn{2}{c}{G35.20 B1} & \multicolumn{2}{c}{G35.20 B2} & \multicolumn{2}{c}{G35.20 B3} & \multicolumn{2}{c}{G35.03 A} \\
Transition & Frequency & FWHM & T$_{peak}$ & FWHM & T$_{peak}$ & FWHM & T$_{peak}$ & FWHM & T$_{peak}$ & FWHM & T$_{peak}$ \\
  & (MHz) & (km/s) &  (K) & (km/s) & (K) & (km/s) & (K) & (km/s) & (K) & (km/s) & (K) \\
\hline
\multicolumn{12}{c}{\bf{H$_2$CS $\nu$=0}} \\
\hline
10$_{1,10}$-9$_{1,9}$ & 338083 & 5.9 $\pm$ 0.1 & 51.7 $\pm$ 0.9 & 2.7 $\pm$ 0.1 & 27.3 $\pm$ 1.0 & 2.4 $\pm$ 0.1 & 28 $\pm$ 1 & 2.54 $\pm$ 0.06 & 44.3 $\pm$ 0.9 & 6.6 $\pm$ 0.1 & 21.0 $\pm$ 0.3 \\
10$_{1,9}$-9$_{1,8}$ & 348532 & 5.8 $\pm$ 0.1 & 59.2 $\pm$ 1.3 & 2.6 $\pm$ 0.1 & 31 $\pm$ 1 & 2.3 $\pm$ 0.1 & 31 $\pm$ 1 & 2.58 $\pm$ 0.04 & 52.1 $\pm$ 0.7  & 6.33 $\pm$ 0.09 & 21.3 $\pm$ 0.3\\
\hline
\multicolumn{12}{c}{\bf{H$_2$C$^{33}$S}} \\
\hline
10$_{1,10}$-9$_{1,9}$ & 335160 & 7.5 $\pm$ 0.2 & 3.91 $\pm$ 0.09 & 1.6 $\pm$ 0.5 & 0.4 $\pm$ 0.1 & 1.5 $\pm$ 0.6 & 0.4 $\pm$ 0.2 & 2.5 $\pm$0.2 & 1.47 $\pm$ 0.08 & \multicolumn{2}{c}{$<$ 3$\sigma$} \\
\hline
\multicolumn{12}{c}{\bf{H$_2$C$^{34}$S}} \\
\hline
10$_{0,10}$-9$_{0,9}$ & 337125 & \multicolumn{2}{c}{blended} & 1.2 $\pm$ 0.7 & 0.7 $\pm$ 0.4 & 1.5 $\pm$ 0.2 & 1.0 $\pm$ 0.1 & 1.9 $\pm$ 0.1 & 2.5 $\pm$ 0.1  & \multicolumn{2}{c}{$<$ 3$\sigma$}\\
10$_{4,6}$-9$_{4,5}$ & 337460 & \multicolumn{2}{c}{blended with CH$_3$OH $\nu$=1} & \multicolumn{2}{c}{in abs. feature} & 1.5 $\pm$ 0.4 & 0.55 $\pm$ 0.09 & \multicolumn{2}{c}{blended with CH$_3$OH $\nu$=1}  & \multicolumn{2}{c}{blended with CH$_3$OH $\nu$=1} \\
10$_{2,9}$-9$_{2,8}$ & 337475 & 6.3 $\pm$ 0.3 & 16.0 $\pm$ 0.3 & 1.66 $\pm$ 0.08 & 3.8 $\pm$ 0.2 & 1.7 $\pm$ 0.1 & 3.8 $\pm$ 0.2 & 3.1 $\pm$ 0.2 & 4.9 $\pm$ 0.2  & \multicolumn{2}{c}{blended with CH$_3$OH $\nu$=1} \\
10$_{3,8}$-9$_{3,7}$ & 337555 & \multicolumn{2}{c}{blended} & 2.0 $\pm$ 0.2 & 0.78 $\pm$ 0.05 & 1.7 $\pm$ 0.5 & 0.6 $\pm$ 0.2 & 2.03 $\pm$ 0.04 & 2.35 $\pm$ 0.03  & \multicolumn{2}{c}{$<$ 3$\sigma$}\\
10$_{3,7}$-9$_{3,6}$ & 337559 & \multicolumn{2}{c}{blended} & 2.16 $\pm$ 0.09 & 1.55 $\pm$ 0.04 & 1.3 $\pm$ 0.1 & 0.54 $\pm$ 0.04 & 2.37 $\pm$ 0.06 & 2.26 $\pm$ 0.03  & \multicolumn{2}{c}{$<$ 3$\sigma$}\\
10$_{2,8}$-9$_{2,7}$ & 337933 & \multicolumn{2}{c}{blended} & 1.2 $\pm$ 0.5 & 0.8 $\pm$ 0.2 & 1.8 $\pm$ 0.5 & 0.7 $\pm$ 0.1 & 2.3 $\pm$ 0.1 & 2.4 $\pm$ 0.1  & \multicolumn{2}{c}{$<$ 3$\sigma$}\\
\hline
\end{tabular}
}
\end{table*}

\section{Observations and Method}

\subsection{Observations}
   G35.20 and G35.03 were observed with ALMA in Cycle 0 between May and June 2012 (2011.0.00275.S). The sources were observed in Band 7 (350 GHz) with the 16 antennas of the array in the extended configuration (baselines in the range 36-400 m) providing sensitivity to structures 0.4'' - 2''.  The digital correlator was configured in four spectral windows (with dual polarization) of 1875 MHz and 3840 channels each, providing a resolution of $\sim$0.4 km/s. The four spectral windows covered the frequency ranges [336 849.57-338 723.83] MHz, [334 965.73-336 839.99] MHz, [348 843.78-350 718.05] MHz, and [346 891.29-348 765.56] MHz.  The rms noise of the continuum maps are 1.8 mJy/beam for G35.20 and 3 mJy/beam for G35.03.  For full details see \citet{Alvaro2014} and \citet{Beltran2014}.

   \begin{figure*}[!thb]
   \centering
      \includegraphics[width=\hsize]{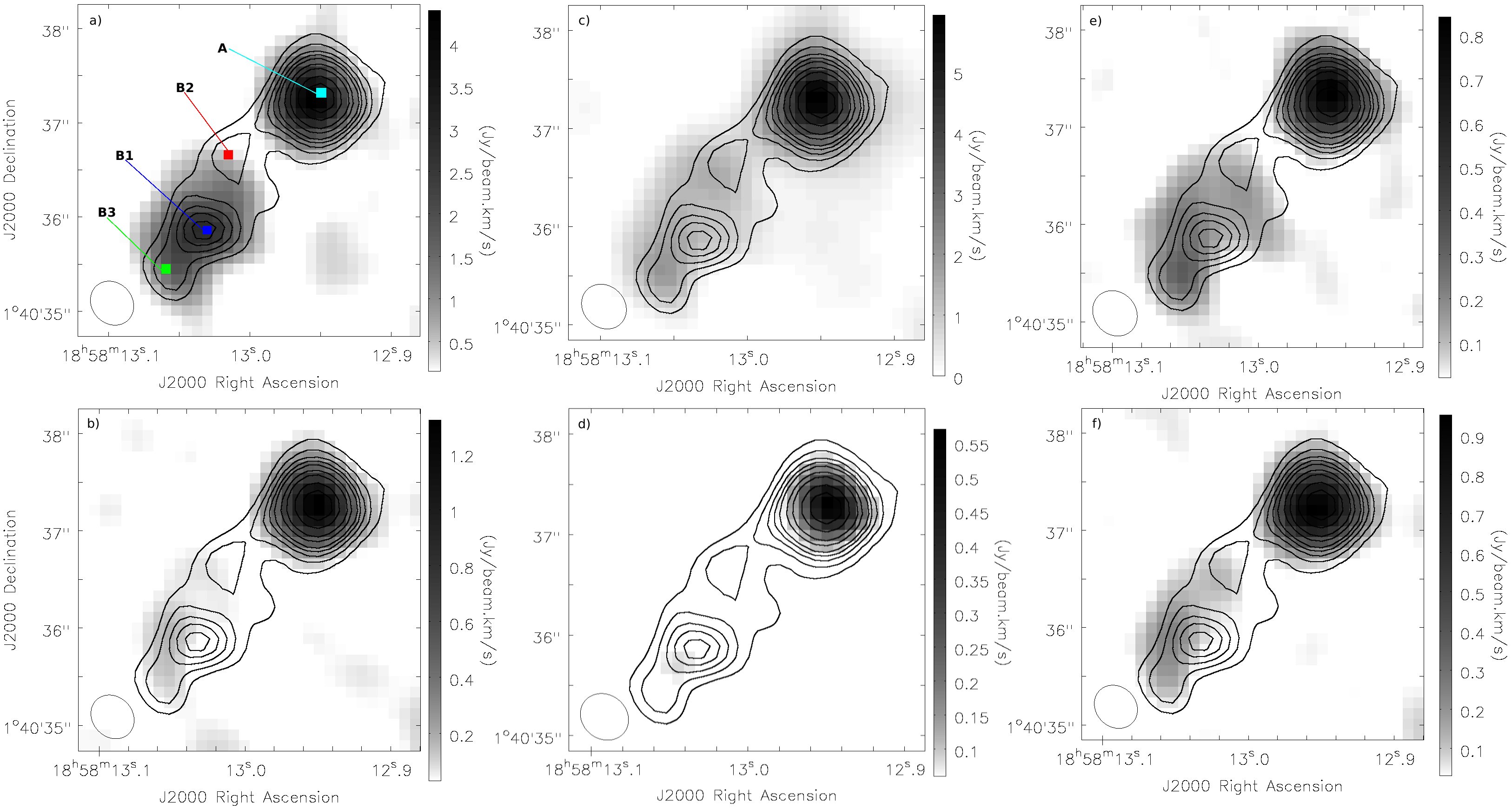}
      \caption{Integrated intensity maps of six species across G35.20, where the contours are the 870~$\mu$m continuum with the same levels as Figure~\ref{ContPts3520}. a) shows the CH$_3$OCHO $\nu$=0 emission at 336.086~GHz integrated from 18.5 to 38~km/s. b) shows the CH$_3$OCHO $\nu$=1 emission at 348.084~GHz integrated from 26 to 38.5~km/s. c) shows the H$_2$CS emission at 338.083~GHz integrated from 24.5 to 38.5~km/s.  d) shows (CH$_2$OH)$_2$ emission at 335.030~GHz integrated from 25-36.5~km/s.  e) shows CH$_3$CHO $\nu$=0 emission at 335.318~GHz integrated from 22.5 to 37~km/s.  f) shows CH$_3$CHO $\nu$=2 emission at 349.752~GHz integrated from 24 to 29~km/s.  It can clearly be seen between panels a) and b) and between e) and f) that vibrationally excited states have a much smaller emitting region.  It is also clear in panel d) that ethylene glycol ((CH$_2$OH)$_2$) is only seen in Core A.}
         \label{maps3520}
   \end{figure*}

\subsection{Line Identification Process}
Spectra were extracted from the central pixel of the continuum peak in core A and the three continuum peaks in core B (B1, B2, B3) in G35.20 and the continuum peak in core A in G35.03 using CASA\footnote{Common Astronomy Software Applications is available from http://casa.nrao.edu/} (see Figures~\ref{ContPts3520} and \ref{ContPts3503} for spectra extraction positions and continuum levels and Table~\ref{statsTable} for the J2000 coordinates and a summary of statistics).  The other peaks (B-F in G35.03 and C-G in G35.20) were not analyzed because they do not show hot core chemistry, i.e. little or no emission from COMs. The three continuum peaks in G35.20 B were chosen to investigate the chemical structure across the disk shown in \citet{Alvaro2014} (who analyzed B as a single core), while the disk in G35.03 A only has a single continuum point associated with the hot core, so analysis for this source was from this peak.  G35.20 A was analyzed as the strongest continuum source in the region with hot core chemistry and was also analyzed at the single continuum peak.  Line parameters (listed in Appendix B) were determined using Gaussian profiles fit to spectral lines from each continuum peak using Cassis\footnote{CASSIS has been developed by IRAP-UPS/CNRS (http://cassis.irap.omp.eu).}, primarily using the CDMS (Cologne Database for Molecular Spectroscopy; \citet{CDMS}) database and JPL (Jet Propulsion Laboratory; \citet{JPL}) database for CH$_2$DOH, C$_2$H$_5$OH, NH$_2$CHO, CH$_3$CHO, and CH$_3$OH ($\nu$=2) transitions. 
\par The process of identifying all species present in these spectra consisted of several parts.  Bright lines (T$_B$ $>$ 5~K) from known species were identified first (ie. those from \citet{Alvaro2014}: CH$_3$OH, CH$_3$OCHO, CH$_3$CN, simple molecules) numbering about 100 lines per source.  The remaining bright lines ($>$ 5~K) were identified by choosing the most likely molecular candidate,  namely the transition with the higher Einstein coefficient, limited to a minimum of about 10$^{-7}$ s$^{-1}$, or with a upper level energy (E$_{up}$) within the expected range, generally less than 500~K, composed of C,H,O, and/or N and within 2 km/s ($\sim$2~MHz) of the transition's rest frequency.  This brings the total to about 200 per source.  Finally, for any remaining unidentified lines $>$ 3$\sigma$ ($\sim$ 0.5~K) a potential species was selected, then the entire spectrum was checked for non-detections of expected transitions of this species.  The total number of identified lines was over 400 for each source (including partially blended and blended transitions where it was evident or implied by the line shape that another transition was present).  It is noted in Appendix~\ref{measurements} if the line identity is uncertain in case of strong blending or multiple probable candidates.
\par The remaining total of unidentified and unclear identity (where there is more than one potential species) lines is about 80 for the peaks in B and G35.03 with an additional 30 in G35.20 A.  These unknown transitions could be either species whose transitions for this frequency regime have not yet been measured/calculated or species where a likely identity was not clear.  The peak intensities of the unknown lines were all less than 5~K.  Line parameters were measured by fitting a Gaussian profile to the emission line with the Cassis "Line Spectrum" tool.  In some cases, partially blended lines were fit together with one or more extra Gaussians for a more accurate measurement, though the errors were then larger.  The full line survey can be found in Appendices A and B, but an example is given in Table~\ref{H2CS}, where the parameters obtained for H$_2$CS (thioformaldehyde) are listed.  The line identities are first presented ordered by frequency, then, to emphasize the chemistry of these objects, the tables of measured line parameters are sorted by molecular species.  
\par To validate the line identifications, fits were made simultaneously to all identified species using the XCLASS software \citet{XCLASS}\footnote{The software can be downloaded from here: https://xclass.astro.uni-koeln.de/}. This program models the data by solving the radiative transfer equation for an isothermal object in one dimension, taking into account source size and dust attenuation.  The residuals between the fitted lines and the observed spectra are between 5 and 25$\%$, validating the XCLASS fits and our line identifications.  The observed spectra and the XCLASS fits can be found in Appendix~\ref{xclassfits} and further information about the XCLASS analysis is detailed in \S~\ref{xclasssection}.

  \begin{table}[!bht]
  \centering
  \caption{Table of source line characteristics.  Column 1 lists the name of the peak.  Column 2 shows the number of molecular species with one or more transition detected.  The total in parentheses indicates the number of XCLASS catalog entries including isotopologues and vibrationally excited transitions separately. Column 3 gives the range of upper level energies observed.  Column 4 is the average line width for each peak.  Column 5 is the average velocity of the lines at each peak. Averages are calculated from all Gaussian line measurements as listed in Appendix B.}
  \label{linestatsTable}
  {\small
  \begin{tabular}{ccccc}

  \hline\hline
  Continuum Peak & Species (total) & E$_{up}$ & <FWHM> & <v$_{LSR}$> \\
   & & (K) & (km/s) & (km/s) \\
  \hline
  G35.20 A & 23 (52) & 17-1143 & 5.2 & 32.2 \\
  G35.20 B1 & 21 (42) & 17-1074 & 2.1 & 29.2 \\
  G35.20 B2 & 21 (41) & 17-973 & 1.9 & 32.3 \\
  G35.20 B3 & 22 (50) & 17-1143 & 2.4 & 28.5 \\
  G35.03 A & 22 (46) & 17-1143 & 4.7 & 45.3 \\
  \hline 
  \end{tabular}
  }
  \end{table}

   \begin{figure*}[!thb]
   \centering
      \includegraphics[width=\hsize]{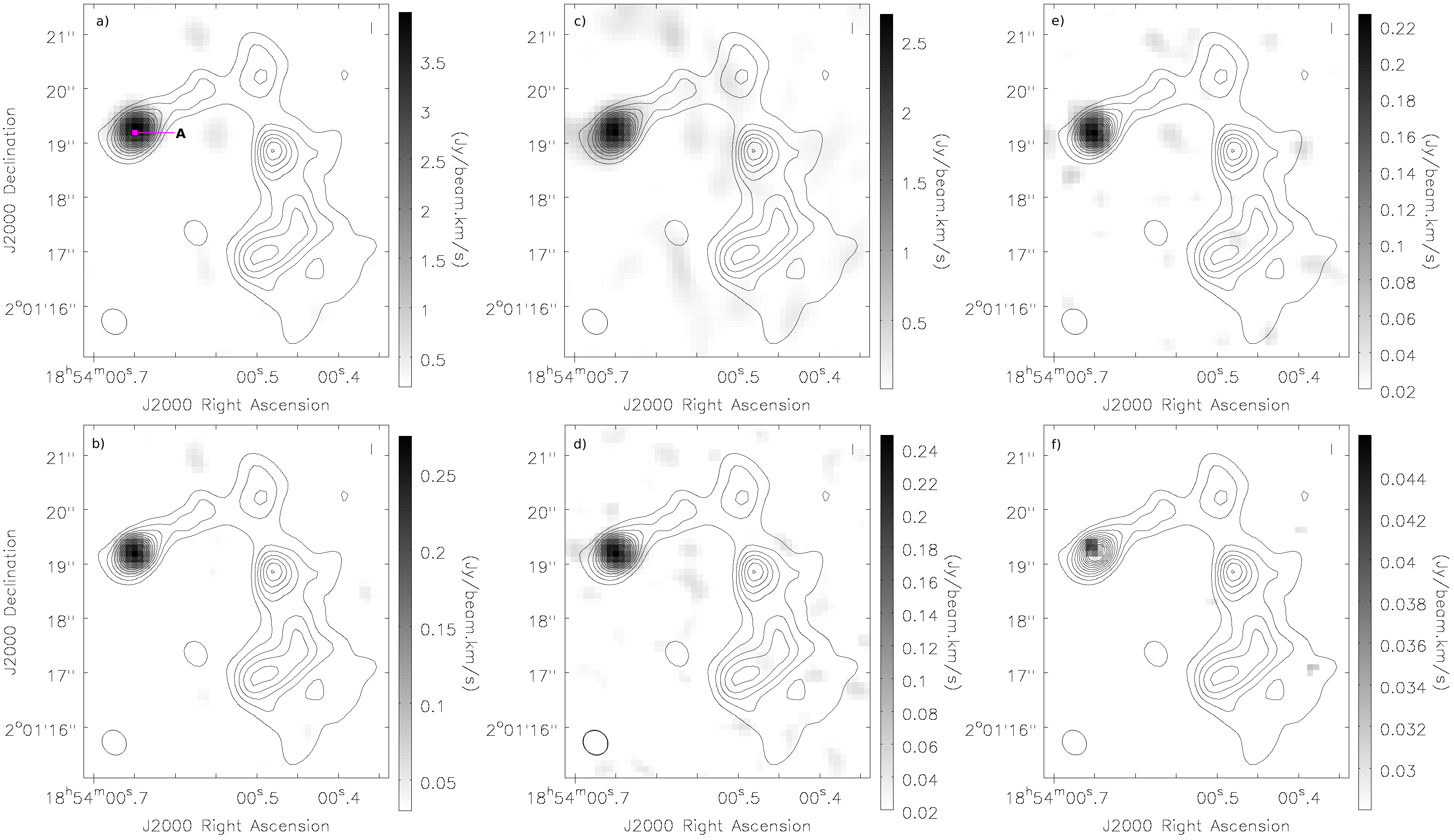}
      \caption{Integrated intensity maps of six species across G35.03, where the contours are the 870~$\mu$m continuum with the same levels as Figure~\ref{ContPts3503}. a) shows the CH$_3$OCHO $\nu$=0 emission at 336.086~GHz integrated from 37 to 57~km/s. b) shows the CH$_3$OCHO $\nu$=1 emission at 348.084~GHz integrated from 42 to 50~km/s. c) shows the H$_2$CS emission at 338.083~GHz integrated from 37 to 52~km/s.  d) shows (CH$_2$OH)$_2$ emission at 335.030~GHz integrated from 38.5 to 48.5~km/s.  e) shows CH$_3$CHO $\nu$=0 emission at 335.318~GHz integrated from 39.5 to 48.5~km/s.  f) shows CH$_3$CHO $\nu$=2 emission at 349.752~GHz integrated from 42 to 47~km/s.  It is clear between panels a) and b) and between e) and f) that vibrationally excited states have a much smaller emitting region.  It is also clear in panel d) that ethylene glycol ((CH$_2$OH)$_2$) is observed in this source.}
         \label{maps3503}
   \end{figure*}

\subsection{Image analysis}
To confirm our identifications of several complex organic species, maps were made of unblended transitions.  Similar spatial distributions and velocity profiles of transitions with similar upper energy levels are consistent with these being the same species.  Figure~\ref{maps3520} shows integrated intensity (moment zero) maps of methyl formate $\nu$=0 and $\nu$=1 transitions, thioformaldehyde, ethylene glycol, acetaldehyde $\nu$=0 and $\nu$=2 transitions in G35.20 and Figure \ref{maps3503} shows the same transitions in G35.03.  During this process, we discovered a difference in spatial extent between N-bearing species and O-bearing species in G35.20 core B.  The N-bearing species peak at the location of continuum peak B3 and are generally not found at the other side of the disk near continuum peak B2.  We will comment on this difference in detail in \S~\ref{ChemsegSect}.  Channel maps were made in CASA for 20 different species for interesting isolated lines with a range of upper energy levels (see Table \ref{linestatsTable}) in order to determine the spatial distribution of various species.  Zeroth (integrated intensity), first (velocity), and second (dispersion) moment maps were also made for these species.  A selection of integrated intensity maps can be found in Figures~\ref{maps3520} and \ref{maps3503}.

\section{Results and Analysis}

\subsection{Line Detections}
A total of 431 different transitions were identified in 52 different catalog entries (18 "regular" $\nu$=0 main isotopes species plus 34 vibrationally excited states and isotopologues).  Table~\ref{linestatsTable} shows the number of species detected per source and Table~\ref{XCLASSsummary} shows the number of unblended and partially blended transitions detected per species in each source.  In addition, a few species were identified from a single transition and are listed in Table~\ref{singles}.
\par The peak with the most transtions is the weakest continuum source: B3.  The strongest continuum source, G35.20 A, suffers greatly from blending (therefore having less unblended transitions) but is also quite chemically diverse (containing 23 identified species versus 22 in B3).  G35.03 A, the second strongest continuum source, contains the third most molecular species, mainly due to deuterated species not being present.  Regarding line flux, B3 generally has the brightest emission of core B except in a few cases where B1 has slightly brighter lines.  Overall B2 has the weakest emission,  but still a diverse range of species.  The lines in G35.03 A are less bright than G35.20 A, and generally brighter than B3.  The line fluxes from continuum peak G35.20 A are higher than any of the peaks in B except in a few cases where B3 has higher line fluxes.

\begin{figure}[!h]
   \centering
      \includegraphics[width=\hsize]{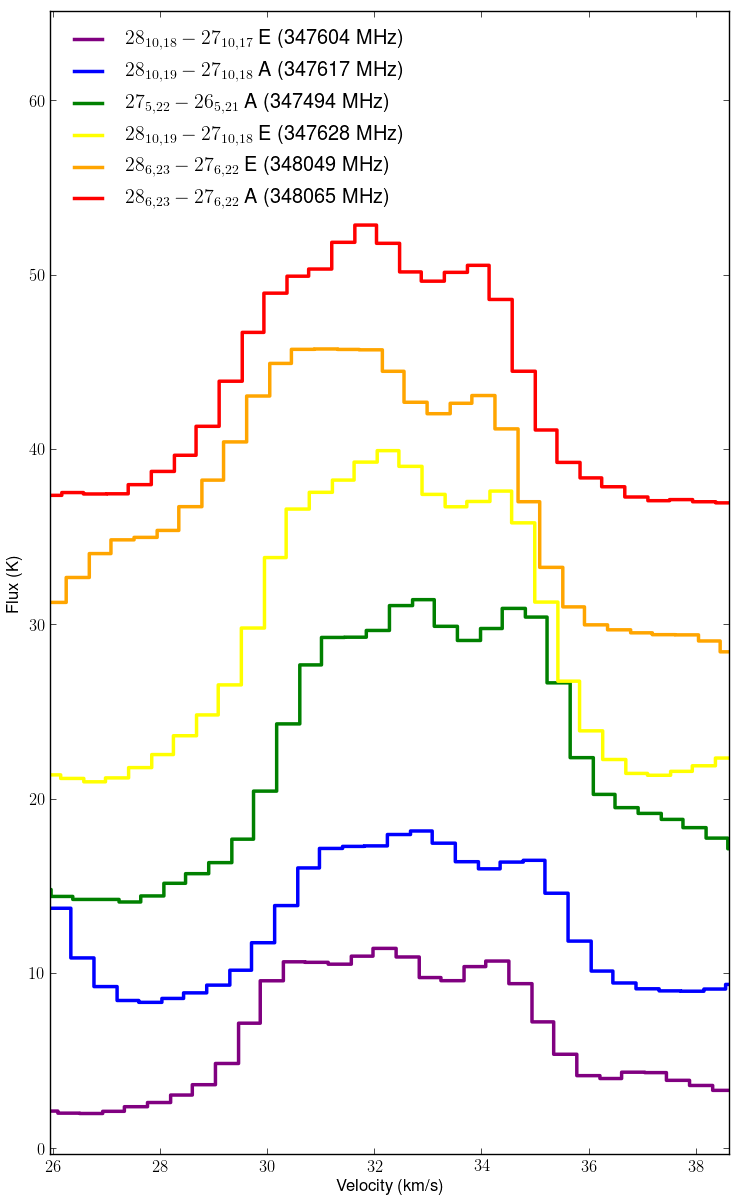}
      \caption{Profiles of methyl formate transitions toward source G35.20 A showing double peaked emission lines.  Features at the edge of the frame are separate lines. If the source is more compact than the beam, this could indicate rotation.}
         \label{ochoDoubles}
\end{figure}

\subsection{Line Profiles}
Most lines are fit by single Gaussians, but some profiles are more complex.  Table~\ref{linestatsTable} shows a summary of line properties at each peak.  The average measured line width for G35.20 A was 5.2~km/s with an average v$_{LSR}$ of 32.2~km/s.  In G35.20 A, 23$\%$ of identified unblended lines are double peaked and nearly all of the rest are broad (FWHM in A is 5-8~km/s compared to 1-3~km/s at the B peaks, see below) suggesting that rotation of an unresolved structure is present (See Figure~\ref{ochoDoubles}).  As the double peaked transitions tend to have higher upper energies (typically $\sim$300~K), we propose that these originate in a warmer region closer to the central source, therefore indicating Keplerian-type rotation. This effect is especially prominent in the methyl formate (CH$_3$OCHO), ethanol (C$_2$H$_5$OH), and deuterated methanol (CH$_2$DOH) lines.  Fits were made to each of the two components for CH$_3$OCHO using Cassis and the peaks were found to be separated by about 2.5~km/s.  Double peaked lines are indicated in the line property tables in Appendix B.  Line blending is prominent for G35.20 A, possibly due to the object being more compact and therefore less resolved than core B.  This could also be a consequence of G35.20 A being more chemically rich or having intrinsically broader line widths.  There are a number of emission lines that are weakly detected in A and undetected at any other continuum peak.  This is possibly because A is the brightest source in both line and continuum emission, so these species may also be present at the continuum peaks in B, but are lost in the noise.  The emission lines from G35.20 A were fit with a single Gaussian for consistency, even where double peaked lines appeared, as the goal was chemical not kinematic analysis.

\par The average line widths for the emission lines from continuum peaks B1, B2 and B3 were 2.1, 1.9, and 2.4~km/s, respectively.  The v$_{LSR}$ of each of the continuum peaks in core B corresponds well with the velocity gradient of the disk observed in \citet{Alvaro2014}.  At B3 in the south-east of the core, the average measured v$_{LSR}$ is 28.5~km/s, at B1, the brightest in continuum in the center of the core, the average v$_{LSR}$ is 29.2~km/s, and at B2 in the northernmost part of core B the average v$_{LSR}$ is 32.3~km/s.  For the continuum peaks B1, B2, and B3, only the emission component was measured and taken into account for LTE modeling.
\par The spectra of sources B1, B2, and B3 show apparent absorption features, which originate in holes in the observations due to emission larger than about 2 $''$ being resolved out.  In the spectra from B1, apparent red-shifted absorption features are seen in every bright line except SO$_2$ and SO.  In CH$_3$CN, the absorption is less pronounced, but the emission lines are asymmetrically blue.  In the spectra of B2, the apparent absorption features are blue-shifted and are obvious in all lines and are especially deep ($\sim$2.5~K) for CH$_3$CN.  In B3, the apparent absorption features can be seen weakly in all species but are strong ($\sim$ 5~K) for CH$_3$OH $\nu$=0 transitions.  

\par
\begin{figure}[h]
   \centering
      \includegraphics[width=\hsize]{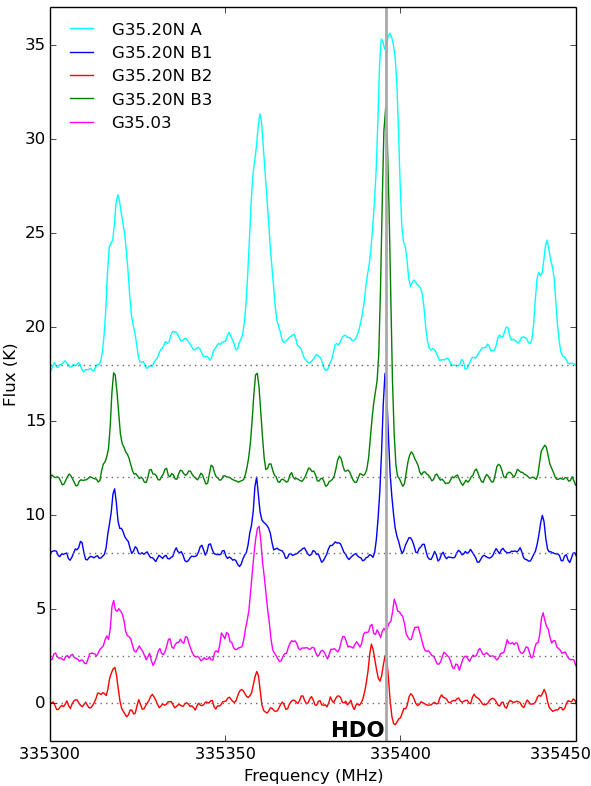}
      \caption{Sample spectrum for the frequency range 335.3-335.45 GHz in the rest frame of each peak to indicate the diversity of these sources. G35.03 and G35.20 A do not appear to have any absorption features (in this range), but it is notable that the lines for these two sources are broader.  The HDO emission line at 335.396 is especially strong in B3, double peaked in G35.20 A, possibly has two velocity components in B2, and is either very weak or offset by several km/s in G35.03.}
         \label{5sources}
  \end{figure}

 G35.03 A generally has weaker lines than the brightest sources in G35.20 (A and B3) and broader lines than those in B1, B2, and B3 with an average FWHM of 4.7~km/s.  The measured average v$_{LSR}$ of the emission lines from this continuum peak was 45.3~km/s.  There are no strong absorption features or double peaked emission lines.  Figure~\ref{5sources} shows the different properties of each source in example spectra.  

\vspace{-0.4cm}

\subsection{Kinetic Gas Temperatures}
\label{radex}
To estimate the kinetic temperature (T$_{kin}$) for each region without assuming LTE, we use RADEX \citep{Radex}, a radiative transfer code that assumes an isothermal and homogeneous medium,  treats optical depth using a local escape probability, and uses collisional rate coefficients from the LAMDA database \citep{{Lamda},{Green1986}}.  We use this software to calculate line intensity ratios across a range of kinetic temperatures and densities and determine whether it is reasonable to assume LTE.
\par We used the CH$_3$CN line ratios for these sources as this species is a known tracer \citep{KSWang2013} of kinetic temperature as a near-symmetric top molecule where transitions with different energy levels have similar critical densities.  We consider as input parameters the ratios of the peaks of unblended CH$_3$CN lines.  The transitions used were 19$_8$-18$_8$, 19$_6$-18$_6$, 19$_5$-18$_5$, 19$_4$-18$_4$, 19$_3$-18$_3$, and 19$_2$-18$_2$ with a column density of 5~$\times$~10$^{15}$~cm$^{-2}$.  The line ratios were modeled for kinetic temperatures between 100 and 500~K and for H$_2$ densities between 10$^6$ and 10$^9$~cm$^{-3}$.  Errors were calculated from the measured error on the Gaussian fit of each spectral line.
\par We find that B2 is the coolest region with an average T$_{kin}$ of 120~K and a range from 90-170~K.  Next hottest is B1 with an average T$_{kin}$ of 160~K and a range from 120-220~K.  G35.20 A is significantly hotter than these with an average T$_{kin}$ of 285~K and a range from 150-450~K.  B3 is consistently the hottest, ranging from 175-490~K with an average T$_{kin}$ of 300~K.  The kinetic temperatures in G35.03 are also very high, ranging from 100-450~K with an average T$_{kin}$ of 275~K.  
\par The varying temperatures for different transition ratios may indicate a temperature gradient within the sampled gas, which requires advanced methods like RATRAN \citep{ratran} or LIME \citep{lime} to model.  The K=6/K=4 ratio consistently traces the lowest temperature.  The K=8/K=3 ratio traces the highest temperature for A, B3, and G35.03, while the highest temperatures for B1 and B2 are traced by the K=6/K=3 and K=6/K=5 ratios, respectively. 
\par These average kinetic temperatures were used in calculating the mass of the core and H$_2$ column density based on the 870~$\mu$m continuum flux as in \citet{Alvaro2014}.  Using a dust opacity of 1.75~cm$^{2}$~g$^{-1}$ and a gas-to-dust ratio of 100, Core A has mass of 13.0~M$_{\sun}$, B1 3.8~M$_{\sun}$, B2 2.2~M$_{\sun}$, B3 1.4~M$_{\sun}$, and G35.03 has a mass of 4.4~M$_{\sun}$.  G35.20 A generally has a lower kinetic temperature than B3, but higher energy transitions are observed and it is also much more massive, with a continuum flux density 10 times higher.

\subsection{Molecular column densities}
\label{xclasssection}
To estimate the column densities of each detected species, we used the XCLASS software. For any given set of parameters (source size, temperature, column density, velocity and linewidth) XCLASS determines the opacity for each spectral channel, for each species, and these opacities are added to produce a spectrum of the opacity changing with frequency. In a last step, the opacity is converted into brightness temperature units to be directly compared with the observed spectrum. The fitting process compares the synthetic spectrum to the observed one, and minimizes the $\chi^2$ by changing the five parameters indicated above. As input parameters, we limited the linewidth and v$_{LSR}$ to $\pm$ 1 km/s from the measured values so the transitions could easily be identified by the fitting algorithm.  Source size, excitation temperature (T$_{ex}$), and column density (N$_{col}$) were allowed to vary widely to begin with, then were better constrained around the lowest $\chi^2$ fits per parameter per species.  For species that were observed to be located only in the regions of the hot cores (mostly complex organic molecules), the source size was varied from 0.1-1.5$''$ to be comparable with the observed emission extent and the T$_{ex}$ was allowed to vary from 50-500~K.  The column density was allowed to vary from 10$^{13}$-10$^{19}$~cm$^{-2}$.  For species that were observed to emit over a more extended region (H$_2$CS and SO$_2$), the source size input range was varied between 1.0-3.5$''$, the T$_{ex}$ input range was 20-200~K, and the column density input range was 10$^{12}$-10$^{16}$~cm$^{-2}$.

   \begin{figure}[ht!]
     \centering
      \includegraphics[width=\hsize]{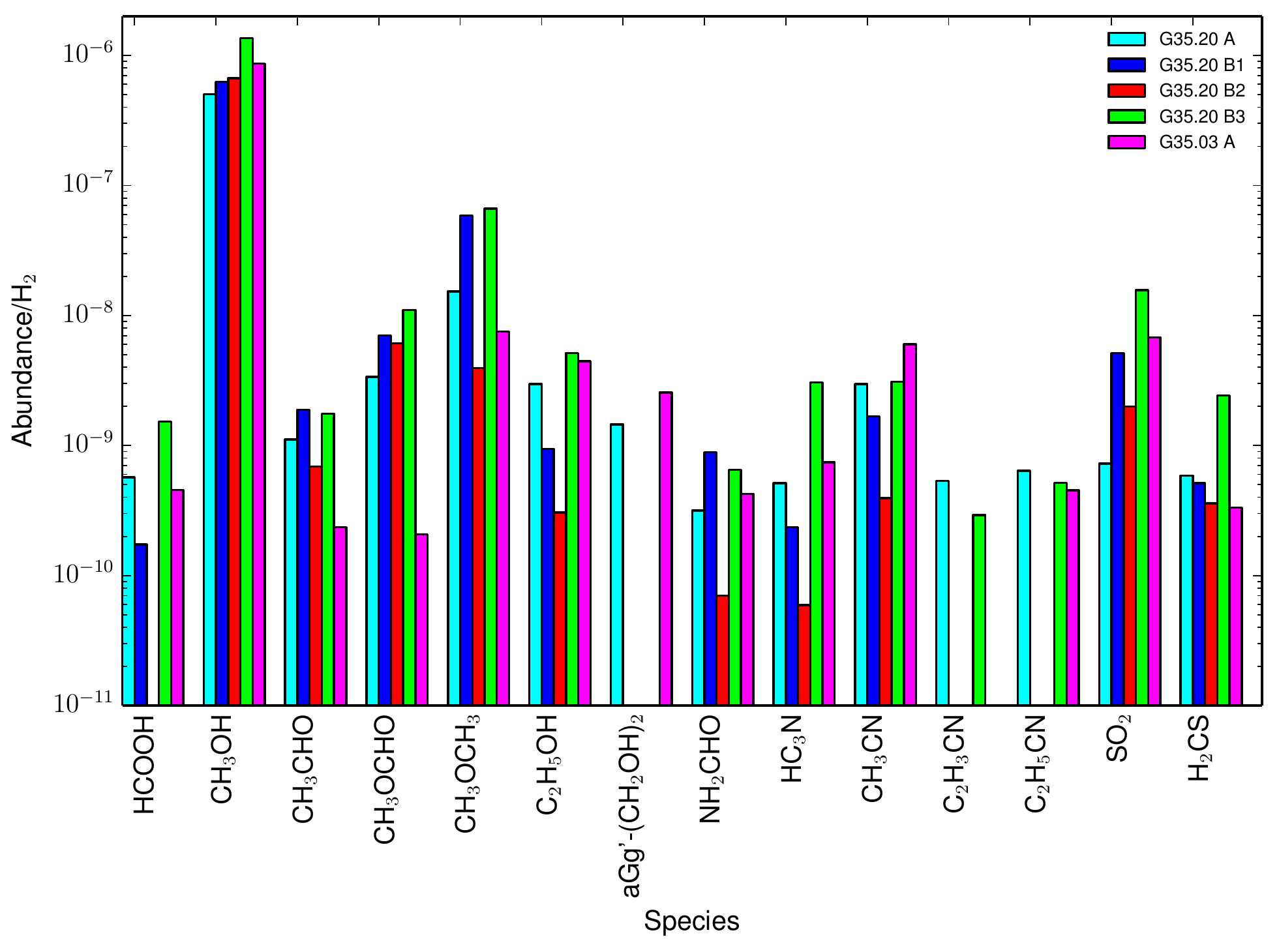}
      \caption{The abundances vs. H$_2$ as determined using the XCLASS software package for each of the cores modeled.  All main isotope species modeled from more than one transition are shown.  The column densities for vibrationally excited states were added to the $\nu$=0 state for CH$_3$OH, CH$_3$CHO, CH$_3$CN, and HC$_3$N to determine abundances. The CN-bearing species in both plots clearly indicate the missing emission in B1 and B2 for C$_2$H$_3$CN and C$_2$H$_5$CN and reduced abundances in B2 for CH$_3$CN and HC$_3$N.  We stress that, as these species do not always trace the same gas, these abundances are lower limits.}
         \label{bargraphs}
   \end{figure}

\par For a few species (SiO, H$^{13}$CO$^+$, C$^{17}$O, H$^{13}$CN, and C$^{34}$S) only one transition was observed, so we kept the source size fixed at the measured extent of the emission at 3$\sigma$ and the excitation temperature fixed at 50~K and 100~K to determine the column densities at these two possible temperatures.  The results for the single line transitions are given in Table~\ref{singles}.  
\par Figure~\ref{bargraphs} presents a summary of the abundances observed per core as modeled using XCLASS.  The excitation temperatures ranged from about 100-300~K generally with a few species outside this range.  The H$_2$ column densities used were based on the 870~$\mu$m continuum emission (values shown in Table~\ref{statsTable}) as determined in \citet{Alvaro2014} section 4.1.  These mass and column density estimates are lower limits as \citet{Alvaro2014} determined that in our observations we recover 30$\%$ of the flux compared to SCUBA 850~$\mu$m observations.  The modeled values for column density and excitation temperature were checked against rotational diagrams from Cassis and found to be in agreement.  The column densities determined using Cassis are lower than those from XCLASS, but this is an effect of a less robust optical depth analysis and assuming the source size fills the beam.
\par Uncertainties on excitation temperatures tend to be 10-20$\%$, but for some species the fit results are upper or lower limits.  For entries that are not upper or lower limits, the range of errors is 1-160~K with an average temperature error of 40~K (or 37$\%$). Source size uncertainties are generally 0.1-0.3$''$ with an average error of 0.2$''$, but range from 0.01-1.0$''$.  Error ranges for column densities were typically less than 1 order of magnitude (with an average error of 0.7 orders of magnitude) with a range between 0.2 and 2.8 orders of magnitude.  For species where only one transition is modeled uncertainties for the column densities of these species are up to two orders of magnitude.  Table~\ref{XCLASSsummary} shows the full list of detected species and isotopologues and the number of transitions detected in each core and indicates which were modeled in XCLASS.  The resulting synthetic spectra are shown together with the observed spectra in Appendix~\ref{xclassfits} and can be seen to be very good fits of the data.  The results of the XCLASS analysis are summarized in Appendix~\ref{xclass}.

\begin{sidewaystable*}[h!]
\centering
\setlength{\tabcolsep}{3pt}
\caption{Results of XCLASS LTE modeling for each of our sources. Four columns are shown under each source name: 1) Number of unblended or partially blended transitions per species detected, 2) modeled source size ($''$), 3) excitation temperature (K), and 4) column density (cm$^{-2}$). The errors on each value are shown in Appendix~\ref{errors}.}
\label{XCLASSsummary}
\begin{tabular}{c|cccc|cccc|cccc|cccc|cccc}
 & \multicolumn{4}{c}{\textbf{G35.20 A}} & \multicolumn{4}{c}{\textbf{G35.20 B1}} & \multicolumn{4}{c}{\textbf{G35.20 B2}} & \multicolumn{4}{c}{\textbf{G35.20 B3}} & \multicolumn{4}{c}{\textbf{G35.03 A}} \\
Species & Lines & Size & T$_{ex}$ & N$_{col}$ & Lines & Size & T$_{ex}$ & N$_{col}$ & Lines & Size & T$_{ex}$ & N$_{col}$ & Lines & Size & T$_{ex}$ & N$_{col}$ & Lines & Size & T$_{ex}$ & N$_{col}$ \\
CH$_3$OH ($\nu$=0) & 29 & 0.6 & 164 & $3.5\times10^{18}$ & 28 & 0.3 & 234 & $6.9\times10^{17}$ & 24 & 0.36 & 136 & $5.3\times10^{17}$ & 33 & 0.35 & 155 & $6.2\times10^{17}$ & 29 & 0.40 & 142 & $1.8\times10^{18}$ \\
CH$_3$OH (v$_{12}$=1) & 21 & 0.5 & 218 & $4.8\times10^{18}$ & 24 & 0.36 & 153 & $1.2\times10^{18}$ & 23 & 0.32 & 145 & $6.2\times10^{17}$ & 24 & 0.36 & 189 & $9.8\times10^{17}$ & 21 & 0.30 & 239 & $2.6\times10^{18}$ \\
CH$_3$OH (v$_{12}$=2) & 14 & 0.4 & 227 & $3.8\times10^{18}$ & 18 & 0.18 & 154 & $2.2\times10^{18}$ & 18 & 0.15 & 152 & $1.0\times10^{18}$ & 19 & 0.20 & 172 & $1.8\times10^{18}$ & 14 & 0.20 & 192 & $5.1\times10^{18}$ \\
$^{13}$CH$_3$OH & 5 & 1.0 & 121 & $1.6\times10^{17}$ & 6 & 0.3 & 113 & $3.9\times10^{16}$ & 6 & 1.1 & 97 & $8.0\times10^{15}$ & 7 & 1.20 & 89 & $1.6\times10^{16}$ & 5 & 0.6 & 120 & $5.0\times10^{16}$ \\
CH$_3$ $^{18}$OH & 4 & 1.1 & 121 & $4.4\times10^{16}$ & 6 & 0.3 & 82 & $1.4\times10^{16}$ & 4 & 0.9 & 75 & $1.4\times10^{15}$ & 5 & 1.20 & 111 & $5.8\times10^{15}$ & 4 & 1.2 & 130 & $1.0\times10^{16}$ \\
H$_2$C$^{18}$O & 3 & 0.6 & 188 & $8.5\times10^{14}$ & 4 & 0.3 & 41 & $5.7\times10^{14}$ & 4 & 0.32 & 29 & $1.5\times10^{15}$ & 4 & 1.09 & 125 & $4.2\times10^{14}$ & 1 & 0.50 & 62 & $2.1\times10^{14}$ \\
CH$_3$OCHO ($\nu$=0) & 28 & 0.98 & 103 & $8.1\times10^{16}$ & 38 & 0.5 & 285 & $4.5\times10^{16}$ & 33 & 0.5 & 64 & $2.0\times10^{16}$ & 31 & 0.59 & 67 & $2.7\times10^{16}$ & 28 & 0.2 & 100 & $2.3\times10^{15}$ \\
CH$_3$OCH$_3$ & 10 & 0.34 & 229 & $3.7\times10^{17}$ & 11 & 0.22 & 156 & $3.8\times10^{17}$ & 8 & 0.4 & 67 & $1.3\times10^{16}$ & 19 & 0.49 & 97 & $1.7\times10^{17}$ & 10 & 0.7 & 150 & $8.3\times10^{16}$ \\
CH$_3$CHO ($\nu$=0) & 14 & 0.4 & 234 & $1.7\times10^{16}$ & 19 & 0.33 & 206 & $7.0\times10^{16}$ & 18 & 0.26 & 88 & $2.3\times10^{15}$ & 20 & 1.16 & 170 & $2.1\times10^{15}$ & 14 & 0.5 & 120 & $1.3\times10^{15}$ \\
CH$_3$CHO ($\nu_{15}$=1) & 7 & 1.1 & 278 & $5.5\times10^{15}$ & 7 & 0.2 & 300.0 & $2.3\times10^{14}$ & 4 & $\star$ &  &  & 13 & 0.56 & 224 & $1.4\times10^{15}$ & 7 & 0.5 & 220 & $1.3\times10^{15}$ \\
CH$_3$CHO ($\nu_{15}$=2) & 4 & 0.7 & 295 & $4.3\times10^{15}$ & 1 & $\star$ &  &  & 0 & $\star$ &  &  & 3 & 0.12 & 102 & $8.1\times10^{14}$ & 0 & $\star$ &  &  \\
HCOOH & 5 & 0.6 & 103 & $1.4\times10^{16}$ & 4 & 0.3 & 156 & $1.1\times10^{15}$ & 2 & $\star$ &  &  & 7 & 0.4 & 173 & $3.8\times10^{15}$ & 5 & 0.7 & 100 & $5.0\times10^{15}$ \\
C$_2$H$_5$OH & 30 & 0.8 & 281 & $7.1\times10^{16}$ & 33 & 0.7 & 78 & $6.0\times10^{15}$ & 33 & 0.12 & 297 & $1.0\times10^{15}$ & 32 & 0.43 & 178 & $1.3\times10^{16}$ & 30 & 0.4 & 150 & $4.9\times10^{16}$ \\
(CH$_2$OH)$_2$ & 23 & 0.6 & 172 & $3.5\times10^{16}$ & 0 & $\star$ &  &  & 0 & $\star$ &  &  & 0 & $\star$ &  &  & 34 & 0.70 & 100 & $2.8\times10^{16}$ \\
NH$_2$CHO & 3 & 0.7 & 98 & $7.6\times10^{15}$ & 3 & 0.3 & 50 & $5.7\times10^{15}$ & 2 & 0.6 & 45 & $2.3\times10^{14}$ & 4 & 0.3 & 53 & $1.6\times10^{15}$ & 3 & 0.8 & 43 & $4.7\times10^{15}$ \\
NH$_2$$^{13}$CHO & 2 & $\dagger$ &  & $1.4\times10^{14}$ & 2 & $\dagger$ &  & $1.0\times10^{14}$ & 2 & $\dagger$ &  & $5.1\times10^{12}$ & 2 & $\dagger$ &  & $5.1\times10^{13}$ & 2 & $\dagger$ &  & $8.5\times10^{13}$ \\
CH$_3$CN ($\nu$=0) & 10 & 0.45 & 208 & $3.9\times10^{16}$ & 9 & 0.32 & 129 & $7.2\times10^{15}$ & 9 & 0.62 & 132 & $1.3\times10^{15}$ & 10 & 0.35 & 208 & $6.5\times10^{15}$ & 10 & 0.30 & 186 & $3.0\times10^{16}$ \\
CH$_3$CN ($\nu_8$=1) & 11 & 0.6 & 359 & $3.2\times10^{16}$ & 5 & 0.29 & 283 & $3.5\times10^{15}$ & 4 & $\star$ &  &  & 13 & 0.8 & 213 & $1.3\times10^{15}$ & 11 & 0.30 & 216 & $3.6\times10^{16}$ \\
CH$_3$ $^{13}$CN & 2 & 1.0 & 263 & $5.0\times10^{14}$ & 4 & 0.7 & 262 & $2.0\times10^{14}$ & 3 & 0.44 & 55 & $1.4\times10^{14}$ & 9 & 0.7 & 118 & $5.4\times10^{14}$ & 2 & 0.35 & 70 & $1.4\times10^{15}$ \\
CH$_2$DCN & 4 & 0.5 & 82.3 & $4.1\times10^{15}$ & 7 & 0.1 & 180 & $3.0\times10^{13}$ & 1 & $\star$ &  &  & 6 & 0.8 & 90 & $1.1\times10^{15}$ & 1 & $\star$ &  &  \\
C$_2$H$_5$CN & 13 & 0.9 & 125 & $1.5\times10^{16}$ & 8 & $\star$ &  &  & 3 & $\star$ &  &  & 18 & 0.8 & 160 & $4.5\times10^{14}$ & 13 & 0.5 & 78 & $5.0\times10^{15}$ \\
CH$_3$$^{13}$CH$_2$CN & 2 & $\dagger$ &  & $6.9\times10^{14}$ & 0 & $\star$ &  &  & 0 & $\star$ &  &  & 0 & $\star$ &  &  & 0 & $\star$ &  &  \\
$^{13}$CH$_3$CH$_2$CN & 1 & $\dagger$ &  & $6.9\times10^{14}$ & 0 & $\star$ &  &  & 0 & $\star$ &  &  & 0 & $\star$ &  &  & 0 & $\star$ &  &  \\
CH$_3$CHDCN & 13 & 0.6 & 97 & $2.9\times10^{15}$ & 0 & $\star$ &  &  & 0 & $\star$ &  &  & 0 & $\star$ &  &  & 0 & $\star$ &  &  \\
C$_2$H$_3$CN & 6 & 0.6 & 77 & $1.3\times10^{16}$ & 0 & $\star$ &  &  & 0 & $\star$ &  &  & 11 & 0.8 & 207 & $7.3\times10^{14}$ & 0 & $\star$ &  &  \\
HCCCN ($\nu=0$) & 1 & 1.2 & 176 & $1.7\times10^{15}$ & 1 & 0.9 & 119 & $9.2\times10^{14}$ & 1 & 0.8 & 122 & $2.0\times10^{14}$ & 1 & 1.2 & 251 & $1.8\times10^{15}$ & 1 & 0.6 & 140 & $2.3\times10^{15}$ \\
HC$^{13}$CCN ($\nu=0$) & 1 & $\dagger$ &  & $3.4\times10^{13}$ & 0 & $\star$ &  &  & 1 & $\star$ &  &  & 1 & $\dagger$ &  & $3.8\times10^{13}$ & 1 & $\star$ &  &  \\
HCC$^{13}$CN ($\nu=0$) & 0 & $\dagger$ &  & $3.4\times10^{13}$ & 0 & $\star$ &  &  & 0 & $\star$ &  &  & 1 & $\dagger$ &  & $3.8\times10^{13}$ & 0 & $\star$ &  &  \\
HCCCN ($\nu_6$=1) & 1 & 0.6 & 200 & $3.1\times10^{15}$ & 0 & $\star$ &  &  & 0 & $\star$ &  &  & 2 & 0.4 & 256 & $2.4\times10^{15}$ & 0 & $\star$ &  & \\
HCC$^{13}$CN ($\nu_6=1$) & 1 & $\dagger$ &  & $6.2\times10^{13}$ & 0 & $\star$ &  &  & 0 & $\star$ &  &  & 1 & $\dagger$ &  & $5.4\times10^{13}$ & 0 & $\star$ &  &  \\
HCCCN ($\nu_7$=1) & 2 & 0.37 & 251 & $5.3\times10^{15}$ & 3 & 0.8 & 219 & $5.8\times10^{14}$ & 2 & $\star$ &  &  & 3 & 0.98 & 288 & $2.1\times10^{15}$ & 2 & 0.8 & 194 & $5.9\times10^{15}$ \\
H$^{13}$CCCN ($\nu_7=1$) & 2 & $\dagger$ &  & $1.0\times10^{15}$ & 0 & $\star$ &  &  & 0 & $\star$ &  &  & 2 & $\dagger$ &  & $4.6\times10^{13}$ & 0 & $\star$ &  &  \\
HC$^{13}$CCN ($\nu_7=1$) & 1 & $\dagger$ &  & $1.0\times10^{15}$ & 0 & $\star$ &  &  & 0 & $\star$ &  &  & 1 & $\dagger$ &  & $4.6\times10^{13}$ & 0 & $\star$ &  &  \\
HCC$^{13}$CN ($\nu_7=1$) & 1 & $\dagger$ &  & $1.0\times10^{15}$ & 1 & $\star$ &  &  & 0 & $\star$ &  &  & 2 & $\dagger$ &  & $4.6\times10^{13}$ & 0 & $\star$ &  &  \\
HCCCN ($\nu_7=2$) & 1 & 0.20 & 350 & $2.2\times10^{15}$ & 0 & $\star$ &  &  & 0 & $\star$ &  &  & 4 & 0.35 & 473 & $1.3\times10^{15}$ & 0 & $\star$ &  &  \\
H$_2$CS & 2 & 1.3 & 165 & $1.4\times10^{16}$ & 2 & 2.0 & 230 & $3.3\times10^{15}$ & 2 & 1.0 & 50 & $1.2\times10^{15}$ & 2 & 0.9 & 87 & $6.1\times10^{15}$ & 2 & 1.7 & 96 & $3.7\times10^{15}$ \\
H$_2$C$^{34}$S & 1 & $\dagger$ &  & $5.3\times10^{14}$ & 5 & $\dagger$ &  & $1.9\times10^{14}$ & 6 & $\dagger$ &  & $8.9\times10^{13}$ & 5 & $\dagger$ &  & $6.1\times10^{14}$ & 0 & $\star$ &  & \\
SO$_2$ & 5 & 1.4 & 260 & $1.7\times10^{16}$ & 5 & 1.44 & 279 & $3.3\times10^{16}$ & 5 & 2.42 & 114 & $6.6\times10^{15}$ & 5 & 1.3 & 288 & $3.9\times10^{16}$ & 5 & 1.2 & 281 & $7.5\times10^{16}$ \\
$^{33}$SO$_2$ & 1 & $\dagger$ & & $1.7\times10^{14}$ & 1 & $\dagger$ &  & $2.5\times10^{14}$ & 1 & $\dagger$ &  & $1.4\times10^{14}$ & 2 & $\dagger$ &  & $1.9\times10^{15}$ & 1 & $\dagger$ &  & $3.4\times10^{15}$ \\
$^{34}$SO$_2$ & 1 & $\dagger$ & & $1.2\times10^{14}$ & 1 & $\dagger$ &  & $1.2\times10^{14}$ & 1 & $\dagger$ &  & $1.6\times10^{14}$ & 1 & $\dagger$ &  & $1.3\times10^{15}$ & 1 & $\star$ &  &  \\
\hline
CH$_2$DOH $\ddagger$& 18 &  & 150 & $5.0\times10^{16}$ & 35 &  & 140.0 & $9.0\times10^{15}$ & 31 &  & 240.0 & $6.0\times10^{15}$ & 32 &  & 160.0 & $1.5\times10^{16}$ & 1 &  &  & \\
CH$_3$OCHO ($\nu$=1) $\ddagger$ & 8 &  & 120 & $1.9\times10^{17}$ & 6 &  & 140.0 & $9.0\times10^{15}$ & 4 &  &  &  & 8 &  &  &  & 4 &  &  & \\
\end{tabular}
\tablefoot{Star ($\star$) symbols indicate that a species was not modeled in XCLASS for this peak.  $\dagger$ indicates that this species was coupled to the main isotopologue for fitting and the isotope ratio was calculated keeping the source size and excitation temperature the same as the main isotope.  The column density indicated in these cases reflects the best fit isotope ratio.  To improve the fits for various HC$_3$N states, the $^{12}$C/$^{13}$C isotope ratio was fixed at 50.  $\ddagger$ these species were analyzed using Cassis because they were not yet incorporated in the XCLASS database.
}
\end{sidewaystable*}

\clearpage

   \begin{figure}[ht!]
   \resizebox{\hsize}{!}
            {\includegraphics{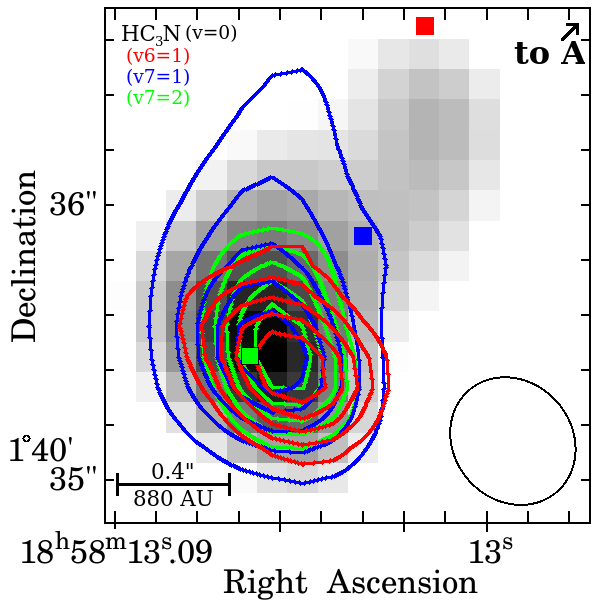}}
      \caption{Integrated intensity of HC$_{3}$N J = 37-36 emission is shown: ($\nu$=0) (greyscale), $\nu_7$=1 (blue contours), $\nu_6$=1 (red contours), and $\nu_7$=2 (green contours) The green contours are 0.05, 0.069, 0.088, 0.106, and 0.125 Jy/beam km/s.  Blue contours are 0.2, 0.422, 0.644, 0.866, and 1.088 Jy/beam km/s. Red contours are 0.043, 0.067, 0.092, 0.117, and 0.141 Jy/beam km/s. Sources B1, B2, and B3 are marked with colored boxes as in Figure~\ref{ContPts3520}.}
         \label{HC3Nvibe}
   \end{figure}

\par In the following subsections, we outline any special considerations used in modeling specific molecules. Section~\ref{COMs} outlines the treatment of most complex organic molecules and their isotopologues and excited states. Section~\ref{hc3n} details the special treatment of the observed HC$_3$N emission.  Section~\ref{Sbearing} explains the fitting methods for SO$_2$ and H$_2$CS.  Section~\ref{simples} shows how simple molecules with only one transition are modeled.  Section~\ref{unmodeled} summarizes how the few species not included in the XCLASS database are handled.

\subsubsection{Complex organic molecules}
\label{COMs}
We modeled ten different species containing six or more atoms: methanol (CH$_3$OH), ethanol (C$_2$H$_5$OH), methyl formate (CH$_3$OCHO), acetaldehyde (CH$_3$CHO), dimethyl ether (CH$_3$OCH$_3$), formamide (NH$_2$CHO), ethylene glycol ((CH$_2$OH)$_2$), methyl cyanide (CH$_3$CN), vinyl cyanide (C$_2$H$_3$CN), and ethyl cyanide (C$_2$H$_5$CN).  
\par Several species with $^{13}$C isotopologues were detected, along with many cases of deuteration.  The $^{18}$O isotopologue for methanol and formaldehyde were detected in all cores, but in no other species.  This is due, in part, to limited laboratory data where the properties of these transitions have not yet been measured/calculated.  
\par High energy transitions in our sources are observed to emit from a much smaller area than lower energy transitions (see figures~\ref{maps3520} and \ref{maps3503}) and many are not observed in B1 or B2.  Because these vibrationally excited states emit from a smaller region, we assume that this emission originates from a denser and possibly hotter region and therefore, the continuum derived H$_2$ column density is a lower limit.  For this reason the column densities for these species cannot be easily converted to abundances, and cannot be precisely compared to their $\nu$=0 states.  Nevertheless, noting their derived excitation temperatures and densities is useful in comparing the properties of the different regions of gas.

\vspace{-0.1cm}
\setlength{\tabcolsep}{2pt}
\begin{table}[htb]
\centering
\caption{Columns 2 and 3 list vibrational temperatures for HC$_3$N with corresponding column densities.  Fluxes for $^{13}$C isotopologues were multiplied by 50 to be comparable to galactic isotope ratios. Columns 4 and 5 correspond to the kinetic temperatures (from RADEX) and the average excitation temperatures (from all XCLASS modeled HC$_3$N vibrational states) and column 6 is the total column density from the XCLASS fits.}
\label{Tvib}
\begin{tabular}{cccccc}

\hline\hline
Source & T$_{vib}$ & N$_{col}$ (vib) & T$_{kin}$ & T$_{ex}$ & N$_{col}$ (XCLASS) \\
 & (K) & (cm$^{-2}$) & (K) & (K) & (cm$^{-2}$) \\
 \hline
G35.20 A & 210$\pm$80 & $4^{+11}_{-3}\times10^{15}$ & 285$^{+165}_{-135}$ & 280 & $1.2\times10^{16}$ \\
G35.20 B1 & 160$\pm$20 & $6.2^{+340}_{-6.0}\times10^{14}$ & 160$^{+60}_{-40}$ & 210 & $1.6\times10^{15}$\\
G35.20 B2 & 120$\pm$60 & $5^{+27}_{-4}\times10^{14}$ & 120$^{+50}_{-30}$ & 130 & $2\times10^{14}$\\
G35.20 B3 & 160$\pm$20 & $2.4^{+1.4}_{-0.9}\times10^{15}$ & 300$^{+190}_{-125}$ & 310 &$8.4\times10^{15}$ \\
G35.03 A & n/a & n/a & 275$\pm$175 & 170 & $8.1\times10^{15}$ \\
\hline
\end{tabular}
\end{table}

\vspace{-0.4cm}

\begin{table*}[!bht]
\centering
\caption{Table of column densities (cm$^{-2}$) determined using XCLASS for species with single transition detections.  Each peak was modeled with the excitation temperatures fixed at 50~K and 100~K.  The source sizes are the measured diameter of the 3$\sigma$ emission in arcseconds ($''$).}
\label{singles}
{\tiny
\begin{tabular}{cccccccccccccccc}

\hline\hline
 & \multicolumn{3}{c}{G35.20 A} & \multicolumn{3}{c}{G35.20 B1} & \multicolumn{3}{c}{G35.20 B2} & \multicolumn{3}{c}{G35.20 B3} & \multicolumn{3}{c}{G35.03 A} \\
Species & Size & 50~K & 100~K & Size & 50~K & 100~K & Size & 50~K & 100~K & Size & 50~K & 100~K & Size & 50~K & 100~K \\
\hline
HN$^{13}$C & 1.5 & $3.1\times10^{13}$ & $4.1\times10^{13}$ & 0.8 & $1.7\times10^{13}$ & $1.6\times10^{13}$ & 1.0 & $1.6\times10^{13}$ & $1.6\times10^{13}$ & 1.5 & $1.1\times10^{14}$ & $1.2\times10^{14}$ & 1.2 & $2.5\times10^{12}$ & $5.8\times10^{12}$ \\
SO & 1.5 & $3.6\times10^{13}$ & $3.6\times10^{13}$ & 1.4 & $3.7\times10^{17}$ & $3.6\times10^{17}$ & 1.2 & $3.7\times10^{16}$ & $1.7\times10^{16}$ & 1.2 & $3.7\times10^{16}$ & $1.1\times10^{17}$ & 1.3 & $2.4\times10^{17}$ & $1.1\times10^{17}$ \\
H$^{13}$CO+ & 1.5 & $1.6\times10^{13}$ & $5.3\times10^{12}$ & 0.8 & $8.3\times10^{12}$ & $8.2\times10^{12}$ & 1.0 & $3.1\times10^{13}$ & $3.1\times10^{13}$ & 1.5 & $6.5\times10^{12}$ & $2.8\times10^{13}$ & 1.5 & $1.7\times10^{13}$ & $2.1\times10^{13}$ \\
SiO & 1.4 & $7.1\times10^{13}$ & $7.2\times10^{13}$ & 1.1 & $9.0\times10^{13}$ & $8.7\times10^{13}$ & 1.1 & $6.2\times10^{12}$ & $6.2\times10^{12}$ & 0.8 & $1.2\times10^{11}$ & $2.8\times10^{11}$ & 1.1 & $5.7\times10^{13}$ & $1.1\times10^{14}$ \\
C$^{34}$S & 1.7 & $2.8\times10^{15}$ & $1.2\times10^{15}$ & 1.2 & $3.4\times10^{14}$ & $2.9\times10^{14}$ & 1.0 & $4.6\times10^{14}$ & $3.5\times10^{14}$ & 1.1 & $5.3\times10^{14}$ & $6.8\times10^{14}$ & 1.5 & $8.1\times10^{14}$ & $7.1\times10^{14}$ \\
HNCO & 1.3 & $2.3\times10^{16}$ & $7.5\times10^{16}$ & 1.0 & $1.2\times10^{16}$ & $6.9\times10^{15}$ & 1.0 & $2.2\times10^{15}$ & $9.4\times10^{14}$ & 1.1 & $2.3\times10^{16}$ & $8.4\times10^{15}$ & 1.1 & $1.8\times10^{16}$ & $7.7\times10^{15}$ \\
HDCO & 1.5 & $1.6\times10^{15}$ & $2.6\times10^{13}$ & 1.2 & $4.5\times10^{14}$ & $1.0\times10^{15}$ & 1.0 & $1.7\times10^{14}$ & $2.6\times10^{14}$ & 1.5 & $1.6\times10^{15}$ & $1.5\times10^{15}$ &  & N/A & \\
HDO & 1.3 & $1.8\times10^{18}$ & $4.8\times10^{17}$ & 0.9 & $5.6\times10^{17}$ & $3.3\times10^{16}$ & 0.7 & $1.8\times10^{17}$ & $7.8\times10^{15}$ & 1.0 & $9.2\times10^{17}$ & $7.6\times10^{16}$ &  & N/A & \\
\hline 
\end{tabular}
}
\end{table*}

\subsubsection{Cyanoacetylene (HC$_3$N) and Vibrational Temperature}
\label{hc3n}
Between two and ten different states were detected in each source for this species, but with only a few transitions, so the isotopologues were coupled to the main isotopologue for each vibrational state and fixed at a $^{12}$C/$^{13}$C isotope ratio of 50.  The $\nu$=0 state was modeled for all regions and the isotopologue HC$^{13}$CCN $\nu$=0 was coupled with HC$_{3}$N $\nu$=0 to improve the uncertainty (from fitting one transition to fitting two).  The fit for HCC$^{13}$CN $\nu$=0 was also coupled with HC$_{3}$N $\nu$=0 for B3, as this is the only location where this species was detected.  
\par Each of the vibrational states ($\nu_6$=1, $\nu_7$=1, $\nu_7$=2) were modeled separately due to their differing spatial extent (Figure~\ref{HC3Nvibe}) and the source size was observed to be more compact with higher excitation.  No vibrationally excited states were modeled for B2, as they were not detected in the observations, and only the $\nu_7$=1 state was modeled for B1 and G35.03.  HC$_3$N $\nu_6$=1 was modeled for A and B3 and was coupled with HCC$^{13}$CN $\nu_6$=1 with the $^{12}$C/$^{13}$C isotope ratio fixed at 50.  HC$_3$N $\nu_7$=1 was also modeled coupled with the three different $^{13}$C isotopologues of HC$_3$N $\nu_7$=1 for A and B3 with the isotope ratio fixed at 50.  HC$_3$N $\nu_7$=2 was only modeled for A and B3 where the emission becomes very compact.
\par
\begin{figure}[htb!]
   \resizebox{\hsize}{!}
            {\includegraphics{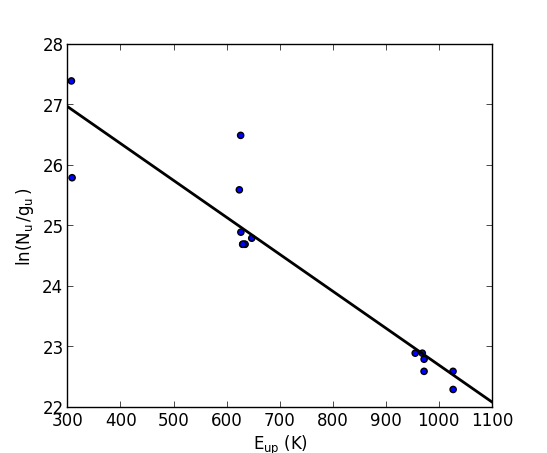}}
      \caption{The vibrational diagram for all of the HC$_3$N transitions from G35.20 B3 including ground and vibrationally excited states with J=37-36 and J=38-37. Fluxes for $^{13}$C isotopologues were multiplied by 50 to be comparable to galactic isotope ratios.  The vibrational temperature calculated for peak B3 is 160~$\pm$~20~K}
         \label{HC3NTvib}
   \end{figure}

 We determined vibrational temperatures from all of the observed HC$_3$N lines for each peak and found them to be in agreement with our RADEX and XCLASS results (see Table~\ref{Tvib} and Figure~\ref{HC3NTvib}).  The temperatures ranged from 120-210~K which indicates that our assumption of LTE is reasonable, even where species are vibrationally excited.  The vibrational temperature for peak B3 is smaller than the kinetic temperature, but consistent within errors (see Table~\ref{Tvib}.
\par The ratio of intensities of HC$_3$N $\nu_7$ and $\nu_0$ transitions indicates the proportion of vibrationally excited to ground-state molecules in the region \citep{Wyrowskihc3n1999}.  For G35.20 B1 and B2 this ratio is $\sim$0.15, and for A and B3 it is $\sim$0.3.  Our sources are all similar to the other hot cores studied in \citet{Wyrowskihc3n1999} with G35.20 A being similar to SgrB2N, B1 and B2 similar to Orion KL and W3(H$_2$O), and B3 similar to G29.96-0.02.  Vibrational temperature analysis for G35.03 A could not be completed as the only unblended HC$_3$N lines detected were from the vibrational state $\nu_7$=1.

\subsubsection{Sulfur bearing molecules}
\label{Sbearing}
Sulfur bearing molecules SO$_2$ and H$_2$CS were modeled with their detected isotopologues coupled to the main isotopologue varying the isotope ratio.  Sulfur isotope ratios in the ISM have been shown to be 15-35 for $^{32}$S/$^{34}$S and 4-9 for $^{34}$S/$^{33}$S \citep{Chin1996}.  Solar isotope ratios are 22.6 for $^{32}$S/$^{34}$S and 5.5 for $^{34}$S/$^{33}$S \citep{Anders1989}. Our best fit isotope ratio for $^{32}$SO$_2$/$^{33}$SO$_2$ was between 16 and 100.  The ratio of $^{32}$SO$_2$/$^{33}$SO$_2$ in space has been reported for Orion KL in \citet{Gizela2013}, with varying ratios for different parts of the region ranging from 5.8-125 reporting a ratio of 25 in the Orion Hot Core.  The best fit isotope ratio for our observations of $^{32}$SO$_2$/$^{34}$SO$_2$ was around 33.  The main isotopologue fit of H$_2$CS was made based on 3 transitions and was modeled with H$_2$C$^{34}$S coupled (only the abundance ratio was varied).  The best fit isotopic ratio for H$_2$C$^{32}$S/H$_2$C$^{34}$S was 11 where the ratio reported for SgrB2 by \citet{Belloche2013} was 22.

\subsubsection{Simple molecules}
\label{simples}
For the following simple species (those with less than six atoms), only a single transition was observed, so in order to estimate their column densities, the source size and excitation temperatures were fixed.  The temperatures were modeled at 50~K and 100~K for all but C$^{17}$O, which was modeled at 20~K and the source size was fixed at the measured extent of the 3$\sigma$ emission.  Several species were previously demonstrated to have quite extended emission (H$^{13}$CO$^+$, C$^{17}$O, SiO) in \citet{{Alvaro2014},{Beltran2014}}.  A summary of the results for these species is in Table~\ref{singles}.
\begin{itemize}
  \item Formyl Cation (H$^{13}$CO$^+$ 4-3) - only the emission component of this species was modeled.  Extended emission shown in \citet{{Alvaro2014},{Beltran2014}}.
  \item Carbon Monoxide (C$^{17}$O 3-2) - at the location of our pixel sources there was a lot of uncertainty in identifying of this line due to severe line blending at this frequency.  For G35.20 A this could not be modeled due to line confusion.  Extended emission indicates that this species is seen in the surrounding cloud, so a larger source size and a lower temperature were used. 
  \item Heavy (Deuterated) Water (HDO 3$_{3,1}$-4$_{2,2}$) - This transition, along with all other deuterated species, was not clearly detected in G35.03, so was not modeled there.  For the other peaks, the emission was fairly extended and the best fit source sizes were between 0.6$''$ (at B2) and 1.5$''$ (at B3).
\end{itemize}

    \begin{figure}[h]
   \centering
      \includegraphics[width=\hsize]{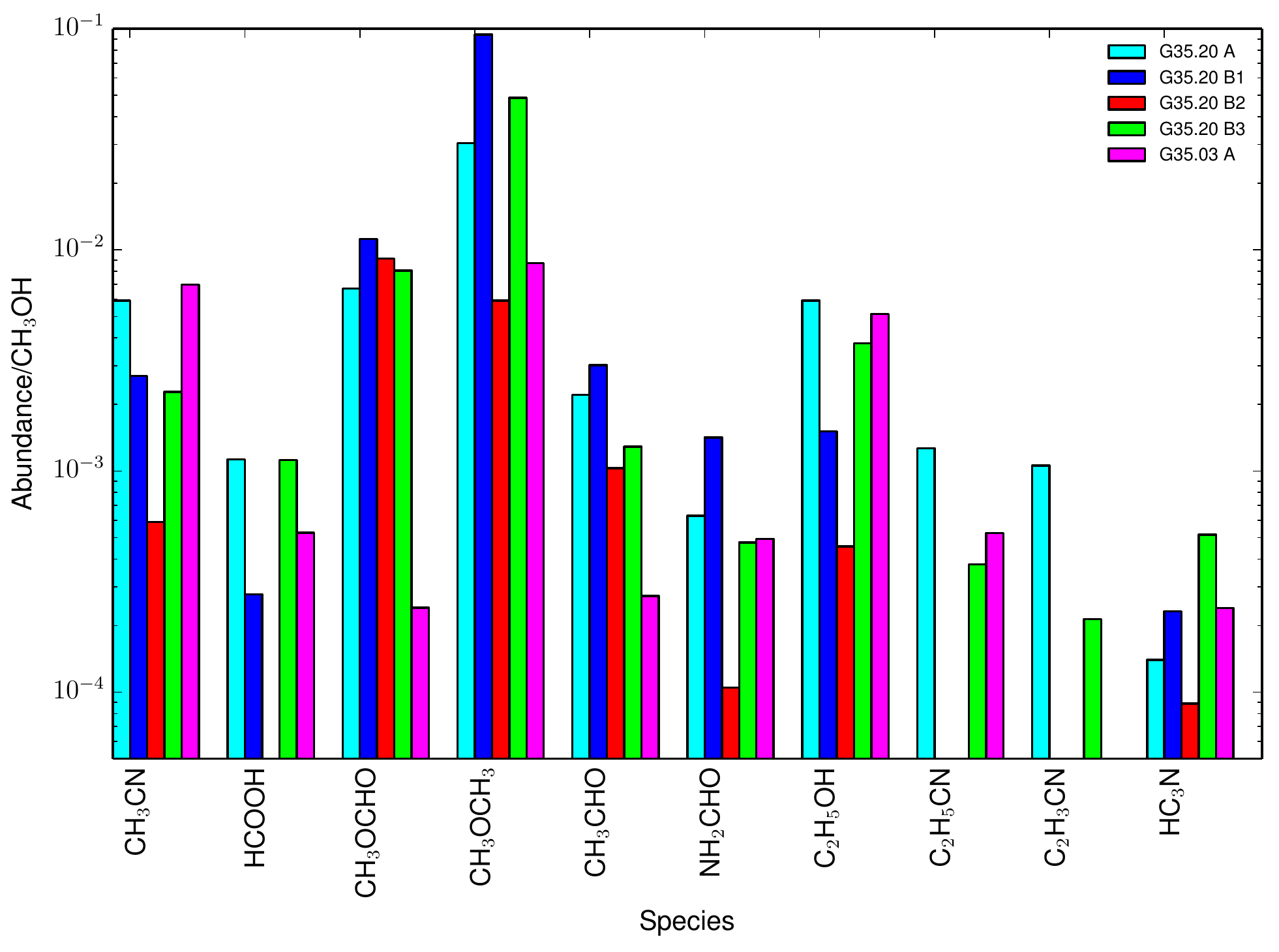}
      \caption{Molecular abundances vs. CH$_3$OH. The column densities for vibrationally excited states were added to the $\nu$=0 state for CH$_3$OH, CH$_3$CHO, CH$_3$CN, and HC$_3$N to determine abundances.}
         \label{vsch3oh}
   \end{figure}

\subsubsection{Species Analyzed with Cassis}
\label{unmodeled}
Some species could not be modeled with XCLASS as they were not yet included in its database.  These species were measured and analyzed with Cassis.
\begin{itemize}
  \item Deuterated Methanol (CH$_2$DOH) - rotational diagrams were created using Cassis for all peaks in G35.20.  The rotational temperatures ranged from 140-240~K and column densities were 0.6-5.0 x 10$^{16}$ cm$^{-2}$.  The CH$_3$OH $\nu$=0 rotational diagrams were made using Cassis to compare to these values to determine deuteration fraction.
  \item Vibrationally excited Methyl Formate (CH$_3$OCHO $\nu$=1) - rotational diagrams were made from all CH$_3$OCHO transitions and the high energy $\nu$=1 transitions continued the trend of the rotational diagrams well.  Therefore the reported rotational temperatures and column densities are those of all transitions for that peak.
  \item Doubly Deuterated Formaldehyde (D$_2$CO) - this species was not modeled because only a single partially blended transition was detected.
\end{itemize}

\begin{figure*}[!hbt]
   \resizebox{\hsize}{!}
            {\includegraphics[width=10.5cm]{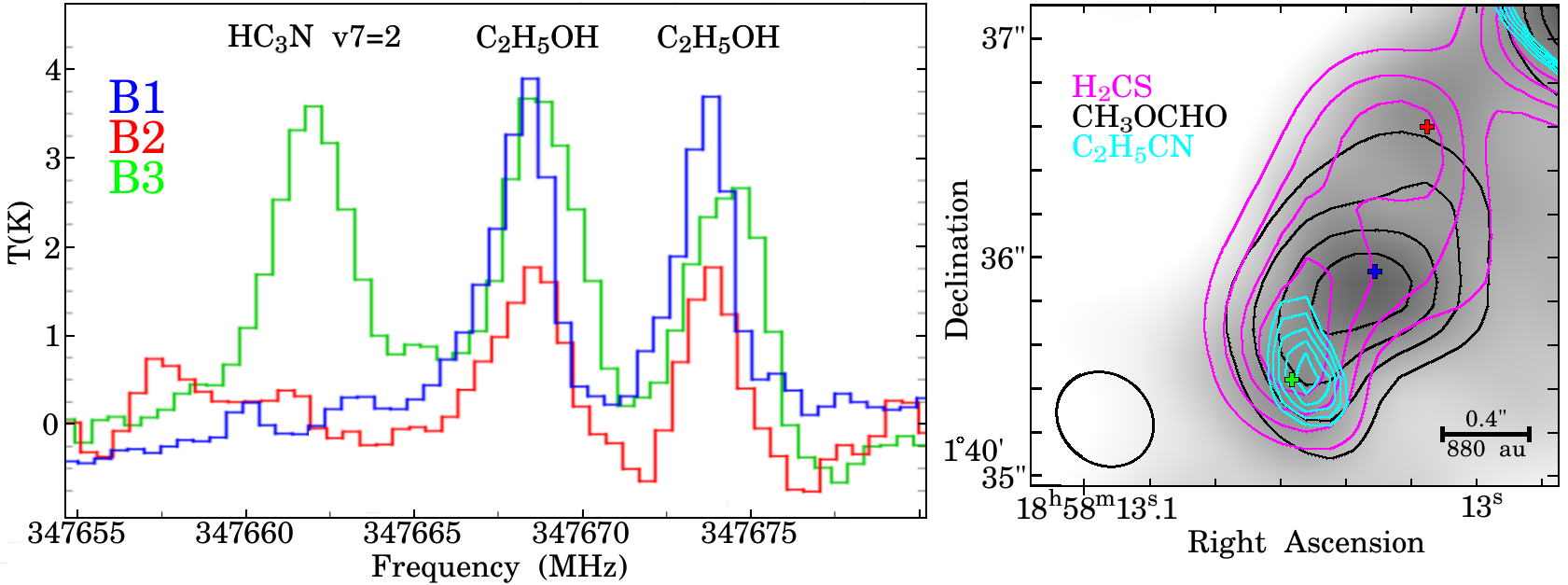}}
      \caption{G35.20 Core B shows clear evidence for small-scale chemical segregation.  On the left are spectra extracted from each continuum peak in core B (corresponding to the red, blue, and green crosses in the map to the right). It can clearly be seen in the spectra that the N-bearing species (HC$_3$N) is only strong in B3, where the O-bearing species (C$_2$H$_5$OH) is strong in all 3 regions. 
On the right, the integrated intensity contours of H$_2$CS 10$_{1,9}$-9$_{1,8}$ (0.55, 0.94, 1.34, and 1.73 Jy/beam km/s), CH$_3$OCHO 27$_{9,18}$-26$_{9,17}$ (0.70, 1.28, 1.85, and 2.43 Jy/beam km/s), and C$_2$H$_5$CN 40$_{1,39}$-39$_{1,38}$ (0.085, 0.100, 0.115, 0.130, and 0.145 Jy/beam km/s) are shown overlaid on the continuum (greyscale) for Core B of G35.20. While the O- and S-bearing organics are distributed across core B, the N-bearing species is only found toward the southwestern part.
}
         \label{ChemSeg}
\end{figure*} 

\section{Discussion}
\subsection{Overall chemical composition}
It is interesting to note that, despite originating from different clouds, G35.03 and G35.20 have similar (within an order of magnitude) abundances of all modeled species except deuterated isotopologues (see \S~\ref{DHdiscuss}).  We find peak B3 shows the highest abundances within G35.20 B versus H$_2$ of all species  except for NH$_2$CHO and CH$_3$CHO of which peak B1 has the highest abundance and H$_2$CO where peak B2 has the highest abundance.  
\par It is possible that comparing the column densities of various complex organic molecules to that of H$_2$ is a less effective method of comparing abundances between these sources.  The value for H$_2$ column density derived from the continuum (and therefore the dust) does not necessarily reflect the density of the warm dense gas where COMs are formed.  Given the uncertainty of the H$_2$ column densities, we also estimated the abundances of some molecules relative to CH$_3$OH, whose emission is less resolved than the continuum emission. Figure~\ref{vsch3oh} shows the relative abundance for several species. This figure confirms the main result of Figure~\ref{bargraphs} namely that abundances in B3 are higher than the other continuum peaks in G35.20 B except in NH$_2$CHO and CH$_3$CHO. We thus conclude that B3 appears to be the most chemically rich of the three sources in G35.20 B. The ratio versus methanol for our sources are less than any of the different types of objects reviewed in \citet{Herbst2009}.  Comparing the ratio of CH$_3$CN to CH$_3$OH in our sources to those in \citet{Oberg2013}, we see that to reach a similar ratio in NGC~7538~IRS9, the gas would be over 7000~AU from the center.
\par In \citet{Oberg2014}, it is suggested that the ratio of abundances of CH$_3$CHO + CH$_3$OCHO (X-CHO) and CH$_3$OCH$_3$ + C$_2$H$_5$OH (X-CH$_3$) is related to the temperature and type of source.  Laboratory experiments have shown that higher abundance of CHO-bearing molecules indicates the importance of cold ice COM chemistry. If X-CHO/X-CH$_3$ is near 1, then the source is a cooler, lower-mass source and if a ratio much less than 1 corresponds to a hotter and more massive source. In line with their observation, we also see a higher ratio of X-CHO/X-CH$_3$ at the coolest peak, G35.20 B2, where the ratio is 1.6, and low ratios at the hottest peaks, G35.20 A, B1, B3, and G35.03 A (0.25, 0.15, 0.18, and 0.04 respectively).  From figure 1 in \citet{Oberg2014}, peak G35.20 B2 could be a massive hot core, but it could also be low-mass, whereas peaks G35.20 A, B1, B3, and G35.03 A definitely fall into the massive hot core regime where warm ice chemistry becomes more important.
\par It is interesting that our XCLASS model fits show higher or nearly equal column densities for several vibrationally excited states versus their ground states.  The XCLASS analysis is satisfactory as long as the energy of the lines span a relatively small range, but a single temperature model is inadequate to fit lines with very different excitation energy. Because of the presence of temperature gradients in these sources, the ground state lines and vibrationally excited lines can be fitted with significantly different temperatures since they trace gas originating from smaller areas with equal or higher column densities.

\subsection{Chemical segregation in G35.20}
\label{ChemsegSect}
\citet{Alvaro2013} show evidence for a Keplerian disk in core B of G35.20.  When analyzing the chemical structure of this core at continuum peaks B1, B2, and B3, we see a striking chemical difference within this disk, which argues against a simple axisymmetric disk scenario.  Our data show clear evidence for chemical segregation of the G35.20 core on a scale of 100s of AU.  Nitrogen-bearing species, especially those containing the cyanide (CN) group (HC$_3$N, C$_2$H$_5$CN, etc.), are only observed in A and the southern part of B (peak B3) except for CH$_3$CN $\nu$=0, which appears in all four locations though the abundance compared to CH$_3$OH at B3 is 4 times that at B2; HN$^{13}$C where the abundance versus CH$_3$OH at B3 is 6 times more than at B2; and HC$_3$N $\nu$=0, where the abundance compared to CH$_3$OH at B3 is 7.5 times that at B2 (see Figure~\ref{vsch3oh}).  The linear scale for this separation of chemistry is less than 1000 AU, which is the smallest observed chemical separation in a star-forming region to date.  Figures~\ref{HC3Nvibe} and \ref{ChemSeg} show that cyanides (HC$_3$N, C$_2$H$_5$CN, etc.) are only observed toward A and the southern part of B, with higher abundances at peak B3 and low abundances or missing emission towards B1 and B2.
\par There are three plausible scenarios to explain this chemical differentiation.  First, Core B could be a disk in the process of fragmenting on scales that are not well resolved in this dataset, where each of the fragments are developing their own chemistry.  Second, there could also be two or three distinct sources within core B, each uniquely influencing the chemistry of their surroundings which could be due to evolutionary age and/or physical conditions.  If the higher kinetic temperature of this region is driving the nitrogen enrichment, \citet{Crockett2015} showed that cyanides can also be made more easily in hot gas phase than other COMs.  If the age is a factor, then an age difference between sources would affect the chemistry of the surroundings.  With enhanced abundances of almost all species, it is possible that B3 contains the hottest source in a multi core system sharing a circumcluster disk with sources at B1 and B2.
\par Third, G35.20B could be a disrupted disk, where it is possible that there are chemical changes within the rotation period of the disk, which is 9700-11100 years (based on the observed radial velocity 3.5-4~km/s and minimum linear diameter of 2500 AU and assuming an edge-on circular orbit).  This is quite a short period of time chemically, though warm-up chemical models like those seen in \citet{Crockett2015} show a sharp increase in abundance from 10$^{-10}$ to 10$^{-8}$ over about 5000 years for CH$_3$CN.  Though N-bearing species are limited to the east side of the disk, N- and O- bearing species formamide (NH$_2$CHO) and isocyanic acid (HNCO) have a more extended range but show significantly reduced emission at B2 as seen in Figure~\ref{hncoForma}. These chemical relationships will be further investigated in a following paper using chemical models.

\subsection{HNCO and Formamide co-spatial emission}
\begin{figure}[tbh]
\resizebox{\hsize}{!}
        {\includegraphics[width=10.5cm]{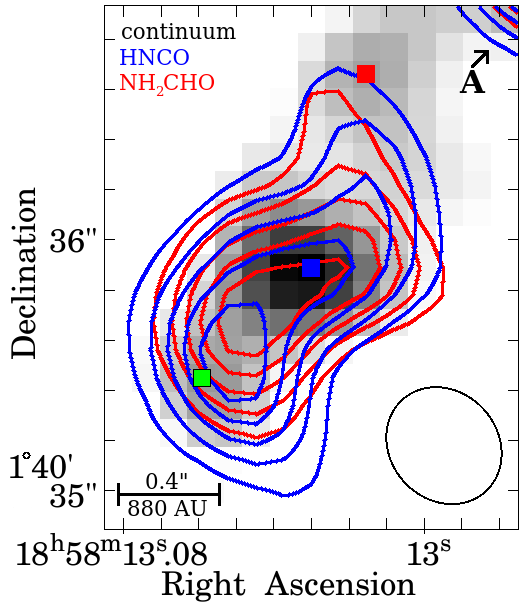}}
  \caption{Formamide 16$_{2,15}$-15$_{2,14}$ (red contours) and HNCO 16$_{1,16}$-15$_{1,15}$ (blue contours) emission is shown overlaid on the dust continuum (greyscale) for core B.  These N- and O- bearing species are present in B3 and B1, but B2 is just outside the outermost contour (indicating 1 $\sigma$).  Red contours are 0.20, 0.33, 0.47, 0.61, and 0.74 Jy/beam km/s and blue contours are 0.40, 0.75, 1.10, 1.45, and 1.80 Jy/beam km/s.  B1, B2, and B3 are marked with colored boxes as in Figure~\ref{ContPts3520}.}
      \label{hncoForma}
\end{figure}

It has previously been proposed \citep{{Bisschop2007},{Mendoza2014},{Ana2015}} based on single dish observations that formamide (NH$_2$CHO) forms through the hydrogenation of HNCO because there appears to be a constant abundance ratio across a large range of source luminosities and masses.  Figure~\ref{hncoForma} shows that these two species have almost the same spatial extent in G35.20 B and their emission peaks are only 0.15$''$ (or $\sim$300 AU) apart in G35.20 B.  The separation is less than 0.11$''$ (240 AU) in G35.20 A.  The velocity intervals spanned by the line peak velocities in each pixel differ by only 0.5~-~1~km/s.  Our modeled abundance values show N(HNCO)/N(NH$_2$CHO) is between 2 and 8 for HNCO at 50~K and between 1 and 10 for HNCO modeled at 100~K.  
\par In G35.03, the HNCO and formamide emissions have a separation of less than 0.11$''$ (255 AU), with the velocity peak differences between 0.5 and 1.0~km/s.  The striking physical connection between these two species makes a strong case for the formation of formamide predominantly through the hydrogenation of HNCO.  \citet{Coutens2016} has also recently observed co-spatial emission in HNCO and Formamide in the low mass protobinary system IRAS16293.

\subsection{Deuteration}
\label{DHdiscuss}
We detect seven deuterated species in G35.20, four of which only with one or two observed transitions.  We determined the deuterium fractionation of the other three, CH$_2$DCN, CH$_2$DOH, and CH$_3$CHDCN, using rotation diagrams in Cassis for consistency (because CH$_2$DOH was not in the XCLASS database).  From these rotation diagrams, we calculated the D/H values based on the best-fit column densities obtained using the opacity function in Cassis.  Relatively little has previously been written about the D/H ratio in methyl cyanide (CH$_3$CN).  In its place of first discovery, Orion KL, the D/H ratio is 0.4-0.9$\%$ \citep{Gerin1992}.  In a recent paper by \citet{Belloche2016}, CH$_2$DCN was detected in Sgr B2 with a D/H of 0.4$\%$.  A D/H for methyl cyanide of 1.3\% was also reported in \citet{Taquet2014} in low-mass protostar IRAS 16293-2422. Our values for G35.20 are significantly higher, and the varying deuterium fractionation across core B is quite pronounced for this species.  The D/H range in methyl cyanide for each continuum peak is 1-11$\%$ at A, 0.3-6$\%$ at B1, and 7-21$\%$ at B3.  Only one unblended transition of CH$_2$DCN was detected at continuum peak B2, so the D/H could not be determined. The D/H percentages for methyl cyanide determined using the XCLASS fits were 10$\%$ at A, 0.4$\%$ at B1, and 15$\%$ at B3 which fall within the ranges determined using Cassis.  We are therefore justified in using Cassis to determine D/H for methanol.
\par Deuteration in methanol has been more widely studied.  In low-mass star forming regions the CH$_2$DOH/CH$_3$OH abundance fraction has been observed to be about 37$\%$ \citep{Parise2002} in IRAS 16293, and in prestellar core L1544 it was close to 10$\%$ \citep{L1544}.  For G35.20, the D/H ratio was 3-9$\%$ at peaks B1 and B2, 4-12$\%$ at B3, and 7-17$\%$ at A.  These values are very similar across core B, though slightly enhanced at B3.  It is possible that because the methanol emission is more extended, it is more homogeneous.  The extra few percent at B3 could be linked to the high temperature and the possibility that this region has heated up recently allowing any deuterium enhancement on the grain surfaces to be released in the gas phase.
\par
\begin{figure}[tbh]
 \resizebox{\hsize}{!}
         {\includegraphics[width=10.5cm]{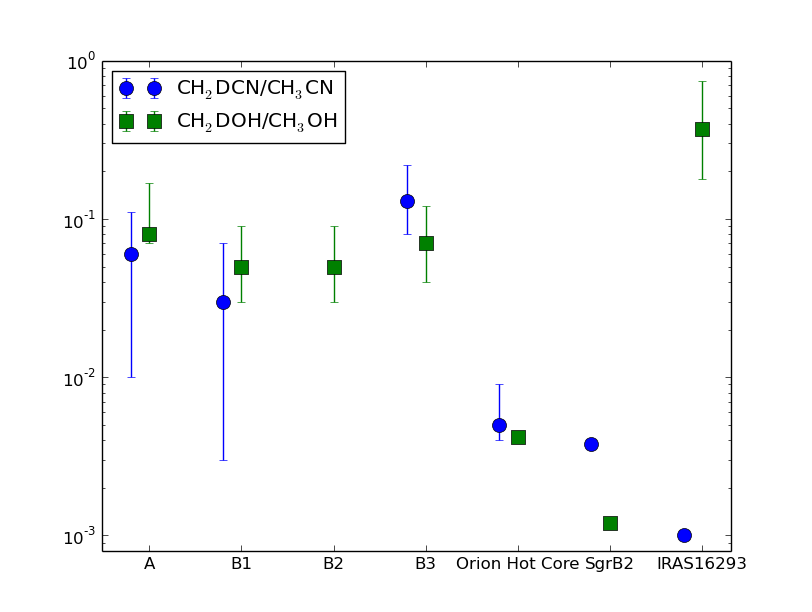}}
   \caption{The D/H fractions for the four continuum peaks in G35.20 compared to the Orion hot core (HC), Sgr B2, and IRAS 16293-2422.  The deuterium fractionation in G35.20 is higher than that of other high-mass star forming regions in Orion and the galactic center, but lower (for methanol) than in the low mass star forming region IRAS 16293.  The methyl cyanide D/H value for IRAS 16293 is from unpublished analysis reported in \citet{Taquet2014}.}
       \label{deuteriumFig}
\end{figure}

 Deuterated ethyl cyanide was detected at A with five unblended transitions, eight partially blended transitions, and two identifiable blended transitions.  The errors are larger for this species due to the line blending, but the D/H value for ethyl cyanide using Cassis was found to be between 3 and 26$\%$ with a best-fit value of 12$\%$ and the D/H from the XCLASS fit is 19$\%$.  A summary of these results is shown in Table~\ref{deuteriumPercent} and Figure~\ref{deuteriumFig}.
\par In contrast, there is almost no sign of deuteration in G35.03.  The presence of CH$_2$DOH is shown through a single line with a brightness temperature of less than 1~K, and HDO is not clearly present as it is either blended with other transitions or offset from v$_{LSR}$ by more than 3 km s$^{-1}$.  HDCO may be present, but is blended with other lines.  Our RADEX analysis indicates that the kinetic temperature of the gas around peaks B3 and G35.03 is over 300~K, so the deuterium fraction is unlikely to be tied to the kinetic temperature in these hot cores.  From our results there is no clear trend with either mass or temperature and deuterium fraction. 
\par A high fraction of deuterium can indicate that an object is very young ($<$ 10$^5$ years) as deuterated species are formed in cold environments where CO has been depleted onto dust grains \citep{Millar1989}.  Once CO returns to the gas-phase, deuterated species are destroyed, so a high deuterium fraction indicates that CO has only sublimated recently.  We conclude that B3 is a much younger region than the hot core in G35.03, and in the case of multiple sources within the disk of core B, sublimation of CO from ice grains has happened at different times or rates across the core.

\begin{table}[h]
\centering
\caption{Deuterium fractionation percentages ($\%$) at continuum peaks in G35.20 as calculated using Cassis.  Deuterated ethyl cyanide is only detected at peak A and determining deuterium fractionation for methyl cyanide was not possible for B2.}
\label{deuteriumPercent}
\begin{tabular}{cccc}

\hline\hline
Source & $\frac{\text{CH2DCN}}{\text{CH3CN}}$ & $\frac{\text{CH2DOH}}{\text{CH3OH}}$ & $\frac{\text{CH3CHDCN}}{\text{CH3CH2CN}}$ \\
\hline
A & 6$\pm{5}$ & 4$^{+4}_{-2}$ & 13$^{+13}_{-10}$\\
B1 & 3$^{+3}_{-2.7}$ & 4$^{+3}_{-2}$ & x\\
B2 & x & 5$^{+3}_{-2}$ & x\\
B3 & 12$^{+9}_{-5}$ & 6$^{+4}_{-3}$ & x\\
\hline
\end{tabular}
\end{table}

\subsection{Comparison to other hot cores}
The hot core and compact ridge in Orion KL (separation $\sim$5000 AU) show a chemical difference between N-bearing and O-bearing species.  In \citet{Caselli1993}, the authors use a time-dependent model to explain the chemistry of both regions. In this model, shells at different distances were collapsing toward the nearby object IRc2, but when accretion stopped the regions heated up and the grain mantles sublimated showing different chemistry.  The model does not perfectly replicate the Orion KL region, but is still a reasonable explanation.  In G35.20, there is no clear nearby accreting (or formerly accreting) object that could have caused this same scenario.  
\par The chemical differentiation between W3(OH) and W3(H$_2$O) \citep{Wyrowski1999} shows that the latter is a strong N-bearing source with various complex organics, but the former only contains a handful of O-bearing species (both contain CH$_3$CN).  In \citet{Qin2016}, they conclude that this region is undergoing global collapse, but W3(OH) contains an expanding HII region, whereas W3(H$_2$O) contains a young stellar object that is accreting material but also has an outflow.  This is similar to G35.20, but on a larger scale (the separation between these two sources is $\sim$7000 AU).
\par \citet{Izaskun2012} observed that AFGL2591 has a hole in the methanol emission (diameter $\sim$3000~AU) that is explained using concentric shells where methanol is mainly in a cooler outer shell and S and N bearing chemistry is driven by molecular UV photo-dissociation and high-temperature gas-phase chemistry within the inner shell where the extinction is lower.  This differs from G35.20 because the hot N-bearing regions are toward the outer edges of the emission with O- and S-bearing species found between.
\par Of the three regions where chemical differentiation has been observed, G35.20 core B is most similar to W3(OH) and W3(H$_2$O).  Chemical differences would  reasonably be seen if Core B contains multiple objects at different evolutionary stages.

\section{Conclusions}
This work describes the chemical composition of G35.20-0.74N and G35.03+0.35 while providing a template for future chemical study of hot cores in this wavelength regime.  Chemical segregation in high-mass star forming regions is observed on a small scale ($<$ 1000 AU) showing that the high spatial resolution capabilities of ALMA are needed to determine whether such segregation is common.  Further observations are needed to determine whether Core B in G35.20-0.74N contains a single or multiple sources.  While the CH$_3$CN emission points to Keplerian rotation \citep{Alvaro2013}, the continuum implies several protostars within and the chemical variation across the proposed disk indicates a complicated source unlike simpler low mass disks. Both of the regions studied showed co-spatial emission from HNCO and NH$_2$CHO indicating a chemical link.  Various deuterated species were detected at G35.20 peak B3 indicating a very young region.  In contrast, G35.03 A shows no obvious deuteration.

\par Higher spatial resolution ALMA observations of this object are planned.  This will allow us to better resolve the emission from core A and better determine the nature of the velocity gradient there.  In addition it may allow us to better determine the origin of the chemical segregation in Core B.
\par The XCLASS software package has a routine that does LTE analysis of each point in a map to demonstrate temperature and density differences pixel by pixel.  Follow up work will be done with this LTE analysis and non-LTE map analysis with RADEX.
\par Time-dependent chemical modeling will help to determine if age is a significant factor in the presence of chemical segregation in star forming regions.  A physical chemical model can also help understand the nature of hot cores.

\begin{acknowledgements}
      We would also like to show our gratitude to the late Malcolm Walmsley, who acted as referee for this paper. His careful reading and numerous insightful comments have shaped this work into its complete form.\\
     This paper makes use of the following ALMA data: ADS/JAO.ALMA 2011.0.00275.S. ALMA is a partnership of ESO (representing its member states), NSF (USA) and NINS (Japan), together with NRC (Canada) and NSC and ASIAA (Taiwan), in cooperation with the Republic of Chile. The Joint  ALMA Observatory is operated by ESO, AUI/NRAO and NAOJ.\\
     This paper made use of information from the Red MSX Source survey database at http://rms.leeds.ac.uk/cgi-bin/public/RMS$\_$DATABASE.cgi which was constructed with support from the Science and Technology Facilities Council of the UK.\\
     \'A.\ S.-M.\ is supported by Deutsche Forschungsgemeinschaft through grant SFB 956 (subproject A6).
\end{acknowledgements}

\bibliography{bib} 

\clearpage

\onecolumn
\begin{appendix} 
\section{Properties of detected lines}
\label{ids}
\begin{longtable}{@{\extracolsep{\fill}}lccc@{}}
\caption{Detected lines organized by frequency with species, upper energy level (K) and Einstein coefficients (s$^{-1}$) indicated.  Most species are detected in all sources.  Ethylene Glycol (aGg’(CH$_2$OH)$_2$) was only detected in G35.20 A and G35.03 A and deuterated ethyl cyanide (CH$_3$CHDCN) was only detected in G35.20 A. For specific transitions see Appendix B.}\\
\hline\hline
\bf{Species} & \bf{Frequency (MHz)} & \bf{E$_{up}$ (K)} & \bf{A$_{ij}$ (s$^{-1}$)} \\
\hline
\endfirsthead
\caption{Detected lines continued.}\\
\hline\hline
\bf{Species} & \bf{Frequency (MHz)} & \bf{E$_{up}$ (K)} & \bf{A$_{ij}$ (s$^{-1}$)} \\
\hline
\endhead
\hline
\endfoot
CH$_3$OH $\nu$=0 & 334970 & 166.0 & 3.42E-8 \\
CH$_3$CHO $\nu$=0 & 334980 & 359.9 & 1.28E-3 \\
CH$_3$OCHO $\nu$=1 & 335016 & 443.5 & 5.33E-4 \\
aGg’(CH$_2$OH)$_2$ & 335030 & 279.4 & 9.34E-4 \\
HC$^{13}$CCN $\nu$=0 & 335092 & 305.6 & 3.01E-3 \\
HDCO & 335097 & 56.2 & 1.04E-3 \\
CCCS & 335109 & 474.6 & 2.98E-3 \\
HCC$^{13}$CN $\nu$=0 & 335124 & 305.6 & 3.01E-3 \\
\hline
\end{longtable}

\clearpage

\begin{landscape}
\section{Line properties per core (organized by species)}
\label{measurements}

\begin{table*}[]
\setlength\LTleft{0pt}
\setlength\LTright{0pt}
\centering
\caption{Single line detections}
\begin{tabular}{ccccccccccccc}

\hline\hline
 & & & \multicolumn{2}{c}{G35.20 A} & \multicolumn{2}{c}{G35.20 B1} & \multicolumn{2}{c}{G35.20 B2} & \multicolumn{2}{c}{G35.20 B3} & \multicolumn{2}{c}{G35.03 A} \\
 Species & Transition & Frequency & FWHM & T$_{peak}$ & FWHM & T$_{peak}$ & FWHM & T$_{peak}$ & FWHM & T$_{peak}$ & FWHM & T$_{peak}$ \\
 & & (MHz) & (km/s) & (K) & (km/s) & (K) & (km/s) & (K) & (km/s) & (K) & (km/s) & (K) \\
\hline
\multicolumn{13}{c}{\bf{2 atom}} \\
\hline
SiO & 8 - 7 & 347330 & 6.18 $\pm$ 0.10 & 10.4 $\pm$ 0.1 & 6.0 $\pm$ 0.4 & 7.9 $\pm$ 0.2 & 1.7 $\pm$ 0.5 & 2.0 $\pm$ 0.4 & 3.1 $\pm$ 0.6 & 2.8 $\pm$ 0.3 & \multicolumn{2}{l}{blended} \\
SO & 10(11)-10(10) & 336553 & 7.1 $\pm$ 0.8 & 4.0 $\pm$ 0.3 & 7.3 $\pm$ 0.3 & 5.1 $\pm$ 0.2 & 3.1 $\pm$ 0.6 & 1.7 $\pm$ 0.3 & 3.28 $\pm$ 0.09 & 6.4 $\pm$ 0.1 & 9.2 $\pm$ 0.3 & 5.7 $\pm$ 0.2 \\
C$^{34}$S & 7 - 6 & 337400 & 4.7 $\pm$ 0.4 & 51.7 $\pm$ 8.4 & 2.7 $\pm$ 0.2 & 24.1 $\pm$ 0.7 & 2.4 $\pm$ 0.1 & 24.5 $\pm$ 0.8 & 2.7 $\pm$ 0.1 & 36.1 $\pm$ 0.6 & 7.6 $\pm$ 0.2 & 23.3 $\pm$ 0.5 \\
C$^{17}$O & 3 - 2 & 337060 & \multicolumn{2}{l}{blended} & 1.9 $\pm$ 0.9 & 2.5 $\pm$ 0.6 & 2.3 $\pm$ 0.3 & 5.4 $\pm$ 0.3 & 2.83 $\pm$ 0.06 & 5.48 $\pm$ 0.08 & 4.6 $\pm$ 0.2 & 9.7 $\pm$ 0.5 \\
\hline
\end{tabular}
\tablefoot{$\star$ double peaked}
\end{table*}

\begin{table*}[]
\centering
\caption{Formaldehyde}
\begin{tabular}{cccccccccccc}

\hline\hline
 & & \multicolumn{2}{c}{G35.20 A} & \multicolumn{2}{c}{G35.20 B1} & \multicolumn{2}{c}{G35.20 B2} & \multicolumn{2}{c}{G35.20 B3} & \multicolumn{2}{c}{G35.03 A} \\
Transition & Frequency & FWHM & T$_{peak}$ & FWHM & T$_{peak}$ & FWHM & T$_{peak}$ & FWHM & T$_{peak}$ & FWHM & T$_{peak}$ \\
 & (MHz) & (km/s) & (K) & (km/s) & (K) & (km/s) & (K) & (km/s) & (K) & (km/s) & (K) \\
\hline
\multicolumn{12}{c}{\bf{H$_2$C$^{18}$O}} \\
\hline
5$_{1, 5}$- 4$_{1, 4}$ o & 335815 & 4.9 $\pm$ 0.2 & 3.7 $\pm$ 0.1 & 2.3 $\pm$ 0.1 & 5.7 $\pm$ 0.2 & 1.6 $\pm$ 0.2 & 3.0 $\pm$ 0.2 & 2.23 $\pm$ 0.03 & 16.4 $\pm$ 0.1 & 3.4 $\pm$ 0.1 & 2.65 $\pm$ 0.08 \\
5$_{4, 2}$- 4$_{4, 1}$ p & 346984 & 4.4 $\pm$ 0.2 & 8.6 $\pm$ 0.2 & 2.2 $\pm$ 0.2 & 2.1 $\pm$ 0.2 & 1.7 $\pm$ 0.2 & 2.0 $\pm$ 0.2 & 2.4 $\pm$ 0.2 & 5.8 $\pm$ 0.1 & \multicolumn{2}{l}{blended} \\
5$_{3, 3}$- 4$_{3, 2}$ o & 347133 & 6.4 $\pm$ 0.1 & 6.93 $\pm$ 0.09 & 2.2 $\pm$ 2.0 & 2.1 $\pm$ 2.0 & 2.1 $\pm$ 0.4 & 1.9 $\pm$ 0.2 & 2.46 $\pm$ 0.04 & 9.7 $\pm$ 0.1 & \multicolumn{2}{l}{blended} \\
5$_{3, 2}$- 4$_{3, 1}$ o & 347144 & \multicolumn{2}{l}{blended} & 1.5 $\pm$ 0.2 & 3.3 $\pm$ 0.2 & 3.1 $\pm$ 0.5 & 1.1 $\pm$ 0.1 & 1.78 $\pm$ 0.06 & 6.8 $\pm$ 0.2 & \multicolumn{2}{l}{blended} \\
\hline
\end{tabular}
\tablefoot{$\diamond$ partially blended with other lines, $\star$ double peaked}
\end{table*}

\clearpage

\begin{table*}[]
\centering
\caption{Formic Acid}
\begin{tabular}{cccccccccccc}

\hline\hline
 & & \multicolumn{2}{c}{G35.20 A} & \multicolumn{2}{c}{G35.20 B1} & \multicolumn{2}{c}{G35.20 B2} & \multicolumn{2}{c}{G35.20 B3} & \multicolumn{2}{c}{G35.03 A} \\
Transition & Frequency & FWHM & T$_{peak}$ & FWHM & T$_{peak}$ & FWHM & T$_{peak}$ & FWHM & T$_{peak}$ & FWHM & T$_{peak}$ \\
 & (MHz) & (km/s) & (K) & (km/s) & (K) & (km/s) & (K) & (km/s) & (K) & (km/s) & (K) \\
 \hline
\multicolumn{12}{c}{\bf{t-HCOOH}} \\
\hline
15$_{9,6}$-14$_{9,5}$ & 337429 & 5.4 $\pm$ 0.4 & 7.2 $\pm$ 0.2 & \multicolumn{2}{l}{blended} & \multicolumn{2}{l}{$\textless$ 3$\sigma$} & 1.7 $\pm$ 0.2 & 1.0 $\pm$ 0.1 & 3.9 $\pm$ 0.2 & 5.6 $\pm$ 0.1 \\
15$_{10,6}$-14$_{10,5}$ & 337436 & 4.4 $\pm$ 0.4 & 6.1 $\pm$ 0.2 & \multicolumn{2}{l}{blended} & \multicolumn{2}{l}{$\textless$ 3$\sigma$} & 1.9 $\pm$ 0.4 & 0.8 $\pm$ 0.1 & 5.5 $\pm$ 0.4 & 4.96 $\pm$ 0.09 \\
15$_{8,7}$-14$_{8,6}$ & 337444 & \multicolumn{2}{l}{blended} & \multicolumn{2}{l}{blended} & \multicolumn{2}{l}{$\textless$ 3$\sigma$} & \multicolumn{2}{l}{blended} & 5.3 $\pm$ 0.1 & 7.38 $\pm$ 0.09 \\
15$_{7,8}$-14$_{7,7}$ & 337492 & \multicolumn{2}{l}{blended} & \multicolumn{2}{l}{blended} & \multicolumn{2}{l}{blended} & 2.1 $\pm$ 0.8 & 7.2 $\pm$ 0.7 & \multicolumn{2}{l}{$\textless$ 3$\sigma$} \\
\hline
\end{tabular}
\tablefoot{$\diamond$ partially blended with other lines}
\end{table*}

\begin{longtable}{p{0.2\textwidth}ccccccccccc}
\caption{Methanol}\\
\hline\hline
 & & \multicolumn{2}{c}{G35.20 A} & \multicolumn{2}{c}{G35.20 B1} & \multicolumn{2}{c}{G35.20 B2} & \multicolumn{2}{c}{G35.20 B3} & \multicolumn{2}{c}{G35.03 A} \\
Transition & Frequency & FWHM & T$_{peak}$ & FWHM & T$_{peak}$ & FWHM & T$_{peak}$ & FWHM & T$_{peak}$ & FWHM & T$_{peak}$ \\
 & (MHz) & (km/s) & (K) & (km/s) & (K) & (km/s) & (K) & (km/s) & (K) & (km/s) & (K) \\
\hline
\endfirsthead
\caption{Methanol continued.}\\
\hline\hline
 & & \multicolumn{2}{c}{G35.20 A} & \multicolumn{2}{c}{G35.20 B1} & \multicolumn{2}{c}{G35.20 B2} & \multicolumn{2}{c}{G35.20 B3} & \multicolumn{2}{c}{G35.03 A} \\
Transition & Frequency & FWHM & T$_{peak}$ & FWHM & T$_{peak}$ & FWHM & T$_{peak}$ & FWHM & T$_{peak}$ & FWHM & T$_{peak}$ \\
 & (MHz) & (km/s) & (K) & (km/s) & (K) & (km/s) & (K) & (km/s) & (K) & (km/s) & (K) \\
\hline
\endhead
\hline
\endfoot
\multicolumn{12}{l}{{$\diamond$ partially blended with other lines, $\star$ double peaked.}} \\
\endlastfoot

\multicolumn{12}{c}{\bf{CH$_3$OH $\nu$=0}} \\
\hline
11$_{0, 11}$- 10$_{2, 8}$ & 334970 & \multicolumn{2}{l}{blended} & \multicolumn{2}{l}{blended} & \multicolumn{2}{l}{blended} & 1.43 $\pm$ 0.06 & 1.22 $\pm$ 0.04 & \multicolumn{2}{l}{out of window} \\
2$_{2,1}$-3$_{1,2}$- - & 335134 & 5.28 $\pm$ 0.09 & 70.1 $\pm$ 0.7 & 2.89 $\pm$ 0.10 & 37.27 $\pm$ 1.14 & 2.4 $\pm$ 0.1 & 28 $\pm$ 1 & 2.40 $\pm$ 0.04 & 48.7 $\pm$ 0.7 & 5.2 $\pm$ 0.2 & 26.0 $\pm$ 0.7 \\
16$_{1,15}$-16$_{-1,16}$ & 335222 & 6.0 $\pm$ 0.3 & 1.83 $\pm$ 0.06 & \multicolumn{2}{l}{blended} & \multicolumn{2}{l}{blended} & 1.4 $\pm$ 0.2 & 0.9 $\pm$ 0.1 & 2.3 $\pm$ 0.1 & 1.27 $\pm$ 0.05 \\
7$_{3,4}$-8$_{0,8}$ & 335363 & \multicolumn{2}{l}{blended} & \multicolumn{2}{l}{blended} & \multicolumn{2}{l}{blended} & $\diamond$1.2 $\pm$ 0.2 & 0.8 $\pm$ 0.1 & \multicolumn{2}{l}{$\textless$ 3$\sigma$} \\
7$_{1,7}$-6$_{1,6}$++ & 335582 & 6.2 $\pm$ 0.1 & 77 $\pm$ 2 & 3.47 $\pm$ 0.11 & 48.98 $\pm$ 1.19 & 3.3 $\pm$ 0.2 & 45 $\pm$ 2 & 2.89 $\pm$ 0.06 & 55 $\pm$ 1 & 6.1 $\pm$ 0.1 & 34.5 $\pm$ 0.4 \\
25$_{8,18}$-26$_{7,19}$- - & 335702 & 5.8 $\pm$ 0.1 & 7.3 $\pm$ 0.1 & 1.3 $\pm$ 0.4 & 0.5 $\pm$ 0.1 & \multicolumn{2}{l}{blended} & 3.1 $\pm$ 0.2 & 1.10 $\pm$ 0.06 & \multicolumn{2}{l}{$\textless$ 3$\sigma$} \\
14$_{7,8}$-15$_{6,9}$++ & 336438 & 5.42 $\pm$ 0.08 & 47.8 $\pm$ 0.6 & 2.52 $\pm$ 0.05 & 19.26 $\pm$ 0.51 & 2.2 $\pm$ 0.1 & 11.1 $\pm$ 0.5 & 2.34 $\pm$ 0.01 & 27.8 $\pm$ 0.4 & 4.23 $\pm$ 0.09 & 17.3 $\pm$ 0.4 \\
\hline
\end{longtable}

\clearpage

\begin{table*}[]
\setlength\LTleft{0pt}
\setlength\LTright{0pt}
\centering
\setlength{\tabcolsep}{5pt}
\caption{Acetaldehyde}
\begin{tabular}{cccccccccccc}

\hline\hline
 & & \multicolumn{2}{c}{G35.20 A} & \multicolumn{2}{c}{G35.20 B1} & \multicolumn{2}{c}{G35.20 B2} & \multicolumn{2}{c}{G35.20 B3} & \multicolumn{2}{c}{G35.03 A} \\
Transition & Frequency & FWHM & T$_{peak}$ & FWHM & T$_{peak}$ & FWHM & T$_{peak}$ & FWHM & T$_{peak}$ & FWHM & T$_{peak}$ \\
 & (MHz) & (km/s) & (K) & (km/s) & (K) & (km/s) & (K) & (km/s) & (K) & (km/s) & (K) \\
 \hline
\multicolumn{12}{c}{\bf{CH$_3$CHO $\nu$=0}} \\
\hline
18$_{1,4}$-17$_{1,4}$ & 334980 & $\star$4.8 $\pm$ 0.3 & 2.07 $\pm$ 0.07 & 0.8 $\pm$ 0.1 & 0.55 $\pm$ 0.09 & 0.9 $\pm$ 0.2 & 0.5 $\pm$ 0.1 & 1.4 $\pm$ 0.4 & 1.0 $\pm$ 0.1 & \multicolumn{2}{l}{$\textless$ 3$\sigma$} \\
18$_{0,18}$-17$_{0,17}$E & 335318 & 5.1 $\pm$ 0.1 & 8.9 $\pm$ 0.2 & 2.0 $\pm$ 0.2 & 3.3 $\pm$ 0.2 & 2.3 $\pm$ 0.1 & 2.0 $\pm$ 0.1 & 1.7 $\pm$ 0.1 & 3.7 $\pm$ 0.2 & 4.2 $\pm$ 0.2 & 2.42 $\pm$ 0.10 \\
18$_{0,18}$-17$_{0,17}$A & 335358 & 5.38 $\pm$ 0.09 & 12.8 $\pm$ 0.2 & 2.2 $\pm$ 0.5 & 3.8 $\pm$ 0.3 & 1.7 $\pm$ 0.2 & 1.7 $\pm$ 0.2 & 2.5 $\pm$ 0.1 & 5.6 $\pm$ 0.1 & \multicolumn{2}{l}{$\textless$ 3$\sigma$} \\
18$_{7,11}$-17$_{7,10}$E & 346934 & \multicolumn{2}{l}{blended} & \multicolumn{2}{l}{blended} & 1.4 $\pm$ 0.3 & 1.55 $\pm$ 0.10 & 2.5 $\pm$ 0.1 & 2.8 $\pm$ 0.2 & 3.1 $\pm$ 1.4 & 5.8 $\pm$ 0.9 \\
\hline
\end{tabular}
\tablefoot{$\diamond$ partially blended with other lines, $\star$ double peaked}
\end{table*}

\clearpage

\begin{longtable}{p{0.2\textwidth}ccccccccccc}
\caption{Methyl Formate}\\
\hline\hline
 & & \multicolumn{2}{c}{G35.20 A} & \multicolumn{2}{c}{G35.20 B1} & \multicolumn{2}{c}{G35.20 B2} & \multicolumn{2}{c}{G35.20 B3} & \multicolumn{2}{c}{G35.03 A} \\
Transition & Frequency & FWHM & T$_{peak}$ & FWHM & T$_{peak}$ & FWHM & T$_{peak}$ & FWHM & T$_{peak}$ & FWHM & T$_{peak}$ \\
 & (MHz) & (km/s) & (K) & (km/s) & (K) & (km/s) & (K) & (km/s) & (K) & (km/s) & (K) \\
\hline
\endfirsthead
\caption{Methyl Formate continued.}\\
\hline\hline
 & & \multicolumn{2}{c}{G35.20 A} & \multicolumn{2}{c}{G35.20 B1} & \multicolumn{2}{c}{G35.20 B2} & \multicolumn{2}{c}{G35.20 B3} & \multicolumn{2}{c}{G35.03 A} \\
Transition & Frequency & FWHM & T$_{peak}$ & FWHM & T$_{peak}$ & FWHM & T$_{peak}$ & FWHM & T$_{peak}$ & FWHM & T$_{peak}$ \\
 & (MHz) & (km/s) & (K) & (km/s) & (K) & (km/s) & (K) & (km/s) & (K) & (km/s) & (K) \\
\hline
\endhead
\hline
\endfoot
\multicolumn{12}{l}{{$\diamond$ partially blended with other lines, $\star$ double peaked.}} \\
\endlastfoot

\multicolumn{12}{c}{\bf{CH$_3$OCHO $\nu$=0}} \\
\hline
15$_{6,10}$-14$_{5,10}$E & 335145 & \multicolumn{2}{l}{blended} & $\diamond$2.0 $\pm$ 0.2 & 0.62 $\pm$ 0.02 & \multicolumn{2}{l}{$\textless$ 3$\sigma$} & \multicolumn{2}{l}{$\textless$ 3$\sigma$} & \multicolumn{2}{l}{blended} \\
28$_{4,24}$-27$_{5,23}$E & 335158 & \multicolumn{2}{l}{blended} & $\diamond$2.2 $\pm$ 0.5 & 0.7 $\pm$ 0.1 & \multicolumn{2}{l}{$\textless$ 3$\sigma$} & \multicolumn{2}{l}{blended} & \multicolumn{2}{l}{blended} \\
28$_{4,24}$-27$_{5,23}$A & 335183 & 6.0 $\pm$ 0.1 & 3.26 $\pm$ 0.05 & 2.3 $\pm$ 0.5 & 0.77 $\pm$ 0.09 & \multicolumn{2}{l}{$\textless$ 3$\sigma$} & \multicolumn{2}{l}{$\textless$ 3$\sigma$} & 4.4 $\pm$ 0.2 & 2.42 $\pm$ 0.09 \\
13$_{2,11}$-12$_{1,12}$A & 335336 & \multicolumn{2}{l}{blended} & \multicolumn{2}{l}{$\textless$ 3$\sigma$} & \multicolumn{2}{l}{$\textless$ 3$\sigma$} & \multicolumn{2}{l}{$\textless$ 3$\sigma$} & \multicolumn{2}{l}{$\textless$ 3$\sigma$} \\
15$_{6,9}$-14$_{5,10}$A & 335403 & \multicolumn{2}{l}{blended} & 1.90 $\pm$ 0.11 & 0.85 $\pm$ 0.04 & 1.1 $\pm$ 0.3 & 0.5 $\pm$ 0.1 & \multicolumn{2}{l}{blended} & \multicolumn{2}{l}{$\textless$ 3$\sigma$} \\
15$_{6,9}$-14$_{5,10}$E & 335454 & 3.1 $\pm$ 1.1 & 1.3 $\pm$ 0.6 & \multicolumn{2}{l}{$\textless$ 3$\sigma$} & \multicolumn{2}{l}{$\textless$ 3$\sigma$} & \multicolumn{2}{l}{$\textless$ 3$\sigma$} & \multicolumn{2}{l}{blended} \\
26$_{5,22}$-25$_{4,21}$A & 335828 & 5.6 $\pm$ 0.8 & 2.6 $\pm$ 0.2 & 1.9 $\pm$ 0.2 & 0.87 $\pm$ 0.06 & \multicolumn{2}{l}{$\textless$ 3$\sigma$} & 1.8 $\pm$ 0.2 & 0.82 $\pm$ 0.08 & 2.5 $\pm$ 0.1 & 1.27 $\pm$ 0.06 \\
\hline
\end{longtable}

\clearpage

\begin{table*}[]
\centering
\caption{Dimethyl ether}
\begin{tabular}{cccccccccccc}

\hline\hline
 & & \multicolumn{2}{c}{G35.20 A} & \multicolumn{2}{c}{G35.20 B1} & \multicolumn{2}{c}{G35.20 B2} & \multicolumn{2}{c}{G35.20 B3} & \multicolumn{2}{c}{G35.03 A} \\
Transition & Frequency & FWHM & T$_{peak}$ & FWHM & T$_{peak}$ & FWHM & T$_{peak}$ & FWHM & T$_{peak}$ & FWHM & T$_{peak}$ \\
 & (MHz) & (km/s) & (K) & (km/s) & (K) & (km/s) & (K) & (km/s) & (K) & (km/s) & (K) \\
 \hline
\multicolumn{12}{c}{\bf{CH$_3$OCH$_3$ $\nu$=0}} \\
\hline
21$_{2,19}$-20$_{3,18}$AA & 337421 & $\star$5.4 $\pm$ 0.2 & 13.2 $\pm$ 0.4 & 2.05 $\pm$ 0.05 & 10.8 $\pm$ 0.2 & 1.7 $\pm$ 0.2 & 3.2 $\pm$ 0.3 & 2.53 $\pm$ 0.04 & 22.5 $\pm$ 0.3 & 4.7 $\pm$ 0.2 & 6.1 $\pm$ 0.2 \\
7$_{4, 4}$- 6$_{3, 3}$ EA & 337712 & \multicolumn{2}{l}{blended} & $\diamond$1.5 $\pm$ 1.2 & 2.2 $\pm$ 1.7 & \multicolumn{2}{l}{$\textless$ 3$\sigma$} & $\diamond$2.40 $\pm$ 0.04 & 5.35 $\pm$ 0.07& \multicolumn{2}{l}{$\textless$ 3$\sigma$} \\
7$_{4, 4}$- 6$_{3, 3}$ EE & 337722 & $\star$6.8 $\pm$ 3.4 & 10.5 $\pm$ 1.9 & 1.89 $\pm$ 0.06 & 7.9 $\pm$ 0.2 & 1.6 $\pm$ 0.2 & 3.3 $\pm$ 0.3 & 2.25 $\pm$ 0.05 & 19.8 $\pm$ 0.4 & 4.1 $\pm$ 0.1 & 4.57 $\pm$ 0.08 \\
7$_{4, 3}$- 6$_{3, 3}$ EA & 337731 & $\star$6.7 $\pm$ 1.7 & 16.4 $\pm$ 6.8 & 2.45 $\pm$ 0.05 & 11.3 $\pm$ 0.2 & 2.1 $\pm$ 0.2 & 3.3 $\pm$ 0.3 & 3.00 $\pm$ 0.08 & 24.8 $\pm$ 0.6 & 5.5 $\pm$ 0.2 & 6.5 $\pm$ 0.1 \\
\hline
\end{tabular}
\tablefoot{$\diamond$ partially blended with other lines, $\star$ double peaked}
\end{table*}

\begin{table*}[]
\centering
\caption{Ethanol}
\begin{tabular}{cccccccccccc}

\hline\hline
 & & \multicolumn{2}{c}{G35.20 A} & \multicolumn{2}{c}{G35.20 B1} & \multicolumn{2}{c}{G35.20 B2} & \multicolumn{2}{c}{G35.20 B3} & \multicolumn{2}{c}{G35.03 A} \\
Transition & Frequency & FWHM & T$_{peak}$ & FWHM & T$_{peak}$ & FWHM & T$_{peak}$ & FWHM & T$_{peak}$ & FWHM & T$_{peak}$ \\
 & (MHz) & (km/s) & (K) & (km/s) & (K) & (km/s) & (K) & (km/s) & (K) & (km/s) & (K) \\
 \hline
\multicolumn{12}{c}{\bf{t-C$_2$H$_5$OH}} \\
\hline
24$_{1,23}$ - 23$_{2,21}$ & 335192 & $\diamond$4 $\pm$ 2 & 1.2 $\pm$ 0.5 & \multicolumn{2}{l}{$\textless$ 3$\sigma$} & \multicolumn{2}{l}{$\textless$ 3$\sigma$} & \multicolumn{2}{l}{$\textless$ 3$\sigma$} & 2.5 $\pm$ 0.2 & 1.00 $\pm$ 0.09 \\
23$_{7,16}$-23$_{6,17}$ & 335441 & $\star$5.1 $\pm$ 0.2 & 6.4 $\pm$ 0.2 & 1.6 $\pm$ 0.2 & 1.98 $\pm$ 0.08 & 1.9 $\pm$ 0.5 & 0.6 $\pm$ 0.2 & 2.3 $\pm$ 0.1 & 1.78 $\pm$ 0.08 & 2.8 $\pm$ 0.2 & 1.8 $\pm$ 0.1 \\
23$_{7,17}$-23$_{6,18}$ & 335631 & $\star$5.5 $\pm$ 0.2 & 7.0 $\pm$ 0.2 & 1.9 $\pm$ 0.2 & 2.2 $\pm$ 0.1 & 1.3 $\pm$ 0.7 & 1.1 $\pm$ 0.2 & 2.2 $\pm$ 0.2 & 1.8 $\pm$ 0.1 & 3.6 $\pm$ 0.2 & 2.16 $\pm$ 0.10 \\
13$_{3,11}$-12$_{2,10}$ & 335950 & $\star$5.0 $\pm$ 0.2 & 10.3 $\pm$ 0.4 & 2.1 $\pm$ 0.2 & 3.9 $\pm$ 0.2 & 1.9 $\pm$ 0.2 & 1.8 $\pm$ 0.2 & 1.9 $\pm$ 0.3 & 4.2 $\pm$ 0.2 & 3.1 $\pm$ 0.2 & 2.79 $\pm$ 0.09 \\
\hline
\end{tabular}
\tablefoot{$\diamond$ partially blended with other lines, $\star$ double peaked}
\end{table*}

\begin{table*}[]
\centering
\caption{Ethylene Glycol}
\begin{tabular}{cccccccccccc}

\hline\hline
 & & \multicolumn{2}{c}{G35.20 A} & \multicolumn{2}{c}{G35.20 B1} & \multicolumn{2}{c}{G35.20 B2} & \multicolumn{2}{c}{G35.20 B3} & \multicolumn{2}{c}{G35.03 A} \\
Transition & Frequency & FWHM & T$_{peak}$ & FWHM & T$_{peak}$ & FWHM & T$_{peak}$ & FWHM & T$_{peak}$ & FWHM & T$_{peak}$ \\
 & (MHz) & (km/s) & (K) & (km/s) & (K) & (km/s) & (K) & (km/s) & (K) & (km/s) & (K) \\
 \hline
\multicolumn{12}{c}{\bf{aGg’ (CH$_2$OH)$_2$}} \\
\hline
32$_{6,27}$ $\nu$= 1 - 31$_{6,26}$ $\nu$= 0 & 335030 & 5.7 $\pm$ 0.2 & 4.6 $\pm$ 0.1 & \multicolumn{2}{l}{$\textless$ 3$\sigma$} & \multicolumn{2}{l}{$\textless$ 3$\sigma$} & \multicolumn{2}{l}{$\textless$ 3$\sigma$} & 3.9 $\pm$ 0.1 & 2.72 $\pm$ 0.09 \\
33$_{ 9,25}$ $\nu$= 0 - 32$_{ 9,24}$ $\nu$= 1 & 335180 & 6.1 $\pm$ 0.2 & 3.23 $\pm$ 0.07 & \multicolumn{2}{l}{$\textless$ 3$\sigma$} & \multicolumn{2}{l}{$\textless$ 3$\sigma$} & \multicolumn{2}{l}{$\textless$ 3$\sigma$} & 3.6 $\pm$ 0.2 & 2.15 $\pm$ 0.10 \\
32$_{11,21}$ $\nu$= 1 - 31$_{11,20}$ $\nu$= 0 & 335357 & \multicolumn{2}{l}{blended} & \multicolumn{2}{l}{$\textless$ 3$\sigma$} & \multicolumn{2}{l}{$\textless$ 3$\sigma$} & \multicolumn{2}{l}{$\textless$ 3$\sigma$} & 3.78 $\pm$ 0.06 & 6.48 $\pm$ 0.09 \\
33$_{ 9,24}$ $\nu$= 0 - 32$_{ 9,23}$ $\nu$= 1 & 335397 & \multicolumn{2}{l}{blended} & \multicolumn{2}{l}{$\textless$ 3$\sigma$} & \multicolumn{2}{l}{$\textless$ 3$\sigma$} & \multicolumn{2}{l}{$\textless$ 3$\sigma$} & 4.7 $\pm$ 0.3 & 2.38 $\pm$ 0.09 \\
\hline
\end{tabular}
\tablefoot{$\diamond$ partially blended with other lines}
\end{table*}

\begin{table*}[]
\centering
\caption{Formamide}
\begin{tabular}{cccccccccccc}

\hline\hline
 & & \multicolumn{2}{c}{G35.20 A} & \multicolumn{2}{c}{G35.20 B1} & \multicolumn{2}{c}{G35.20 B2} & \multicolumn{2}{c}{G35.20 B3} & \multicolumn{2}{c}{G35.03 A} \\
Transition & Frequency & FWHM & T$_{peak}$ & FWHM & T$_{peak}$ & FWHM & T$_{peak}$ & FWHM & T$_{peak}$ & FWHM & T$_{peak}$ \\
 & (MHz) & (km/s) & (K) & (km/s) & (K) & (km/s) & (K) & (km/s) & (K) & (km/s) & (K) \\
 \hline
\multicolumn{12}{c}{\bf{NH$_2$CHO $\nu$=0}} \\
\hline
16$_{2,15}$-15$_{2,14}$ & 336136 & 5.8 $\pm$ 0.2 & 30.8 $\pm$ 0.8 & 5.0 $\pm$ 0.1 & 8.0 $\pm$ 0.1 & 2.8 $\pm$ 0.4 & 2.3 $\pm$ 0.1 & 3.17 $\pm$ 0.09 & 8.2 $\pm$ 0.1 & 4.63 $\pm$ 0.09 & 15.4 $\pm$ 0.3 \\
16$_{1,16}$-15$_{0,15}$ & 336161 & 8.5 $\pm$ 0.2 & 9.4 $\pm$ 0.2 & 3.2 $\pm$ 0.6 & 1.54 $\pm$ 0.07 & \multicolumn{2}{l}{$\textless$ 3$\sigma$} & 2.7 $\pm$ 0.2 & 5.6 $\pm$ 0.3 & 4.3 $\pm$ 0.2 & 4.24 $\pm$ 0.10 \\
28$_{2,26}$-28$_{1,27}$ & 347729 & \multicolumn{2}{l}{blended} & \multicolumn{2}{l}{$\textless$ 3$\sigma$} & \multicolumn{2}{l}{$\textless$ 3$\sigma$} & 4.1 $\pm$ 0.5 & 0.85 $\pm$ 0.05 & \multicolumn{2}{l}{$\textless$ 3$\sigma$} \\
 & 349053 & \multicolumn{2}{l}{blended} & \multicolumn{2}{l}{$\textless$ 3$\sigma$} & \multicolumn{2}{l}{$\textless$ 3$\sigma$} & 2.0 $\pm$ 0.2 & 1.66 $\pm$ 0.09 & \multicolumn{2}{l}{$\textless$ 3$\sigma$} \\
\hline
\end{tabular}
\tablefoot{$\diamond$ partially blended with other lines}
\end{table*}

\clearpage

\begin{longtable}{p{0.2\textwidth}ccccccccccc}
\caption{Methyl Cyanide}\\

\hline\hline
 & & \multicolumn{2}{c}{G35.20 A} & \multicolumn{2}{c}{G35.20 B1} & \multicolumn{2}{c}{G35.20 B2} & \multicolumn{2}{c}{G35.20 B3} & \multicolumn{2}{c}{G35.03 A} \\
Transition & Frequency & FWHM & T$_{peak}$ & FWHM & T$_{peak}$ & FWHM & T$_{peak}$ & FWHM & T$_{peak}$ & FWHM & T$_{peak}$ \\
 & (MHz) & (km/s) & (K) & (km/s) & (K) & (km/s) & (K) & (km/s) & (K) & (km/s) & (K) \\
\hline
\endfirsthead
\caption{Methyl Cyanide continued.}\\
\hline\hline
 & & \multicolumn{2}{c}{G35.20 A} & \multicolumn{2}{c}{G35.20 B1} & \multicolumn{2}{c}{G35.20 B2} & \multicolumn{2}{c}{G35.20 B3} & \multicolumn{2}{c}{G35.03 A} \\
Transition & Frequency & FWHM & T$_{peak}$ & FWHM & T$_{peak}$ & FWHM & T$_{peak}$ & FWHM & T$_{peak}$ & FWHM & T$_{peak}$ \\
 & (MHz) & (km/s) & (K) & (km/s) & (K) & (km/s) & (K) & (km/s) & (K) & (km/s) & (K) \\
\hline
\endhead
\hline
\endfoot
\multicolumn{12}{l}{{$\diamond$ partially blended with other lines, $\star$ double peaked.}} \\
\endlastfoot

\multicolumn{12}{c}{\bf{CH$_3$CN $\nu$=0}} \\
\hline
19$_{9}$ - 18$_{9}$ & 348911 & $\star$8.0 $\pm$ 0.2 & 27.3 $\pm$ 0.4 & \multicolumn{2}{l}{blended} & \multicolumn{2}{l}{blended} & 2.0 $\pm$ 0.2 & 8 $\pm$ 1 & \multicolumn{2}{l}{$\textless$ 3$\sigma$} \\
19$_{8}$ - 18$_{8}$ & 349025 & $\star$7.8 $\pm$ 0.2 & 19.1 $\pm$ 0.5 & 1.5 $\pm$ 0.2 & 3.0 $\pm$ 0.2 & 1.6 $\pm$ 0.3 & 1.2 $\pm$ 0.2 & 2.22 $\pm$ 0.06 & 12.2 $\pm$ 0.4 & 7.6 $\pm$ 0.1 & 8.2 $\pm$ 0.1 \\
19$_{7}$ - 18$_{7}$ & 349125 & $\star$6.3 $\pm$ 0.2 & 26.6 $\pm$ 0.8 & 2.1 $\pm$ 0.1 & 6.3 $\pm$ 0.3 & 1.6 $\pm$ 0.2 & 2.6 $\pm$ 0.3 & 2.58 $\pm$ 0.08 & 22.0 $\pm$ 0.5 & 5.97 $\pm$ 0.09 & 10.9 $\pm$ 0.2 \\
19$_{6}$ - 18$_{6}$ & 349212 & $\star$6.6 $\pm$ 0.3 & 47.0 $\pm$ 1.0 & 3.2 $\pm$ 0.3 & 18 $\pm$ 1 & 1.8 $\pm$ 0.2 & 8.4 $\pm$ 1.0 & 2.82 $\pm$ 0.06 & 39.9 $\pm$ 0.8 & 4.4 $\pm$ 0.2 & 13.6 $\pm$ 0.5 \\
19$_{5}$ - 18$_{5}$ & 349286 & $\star$7.0 $\pm$ 0.3 & 48.9 $\pm$ 1.8 & 3.2 $\pm$ 0.2 & 18.4 $\pm$ 1.0 & 1.9 $\pm$ 0.2 & 8.9 $\pm$ 0.6 & 2.70 $\pm$ 0.03 & 41.7 $\pm$ 0.3 & 5.6 $\pm$ 0.5 & 19.4 $\pm$ 0.3 \\
19$_{4}$ - 18$_{4}$ & 349346 & $\star$6.3 $\pm$ 0.2 & 54.8 $\pm$ 1.6 & 3.2 $\pm$ 0.2 & 24 $\pm$ 1 & 2.3 $\pm$ 0.2 & 14.7 $\pm$ 1.2 & 2.89 $\pm$ 0.03 & 47.7 $\pm$ 0.5 & 7.1 $\pm$ 0.2 & 21.9 $\pm$ 0.6 \\
19$_{3}$ - 18$_{3}$ & 349393 & 6.2 $\pm$ 0.2 & 62.3 $\pm$ 1.8 & 3.3 $\pm$ 0.1 & 26.6 $\pm$ 1.0 & 3.35 $\pm$ 0.08 & 25.9 $\pm$ 0.5 & 3.14 $\pm$ 0.06 & 38.8 $\pm$ 0.5 & 6.2 $\pm$ 0.3 & 25.4 $\pm$ 0.9 \\
\hline
\end{longtable}
 
\clearpage

\begin{table*}[]
\centering
\caption{Vinyl Cyanide}
\begin{tabular}{cccccccccccc}

\hline\hline
 & & \multicolumn{2}{c}{G35.20 A} & \multicolumn{2}{c}{G35.20 B1} & \multicolumn{2}{c}{G35.20 B2} & \multicolumn{2}{c}{G35.20 B3} & \multicolumn{2}{c}{G35.03 A} \\
Transition & Frequency & FWHM & T$_{peak}$ & FWHM & T$_{peak}$ & FWHM & T$_{peak}$ & FWHM & T$_{peak}$ & FWHM & T$_{peak}$ \\
 & (MHz) & (km/s) & (K) & (km/s) & (K) & (km/s) & (K) & (km/s) & (K) & (km/s) & (K) \\
 \hline
\multicolumn{12}{c}{\bf{C$_2$H$_3$CN}} \\
\hline
36$_{2,35}$-35$_{2,34}$ & 337040 & $\diamond$4.0 $\pm$ 0.3 & 3.4 $\pm$ 0.2 & \multicolumn{2}{l}{blended} & \multicolumn{2}{l}{$\textless$ 3$\sigma$} & 2.38 $\pm$ 0.09 & 1.96 $\pm$ 0.06 & 8.1 $\pm$ 3.1 & 1.25 $\pm$ 0.10 \\
35$_{ 3,32}$-34$_{ 3,31}$ & 337051 & \multicolumn{2}{l}{blended} & \multicolumn{2}{l}{blended} & \multicolumn{2}{l}{$\textless$ 3$\sigma$} & 2.0 $\pm$ 0.9 & 0.9 $\pm$ 0.3 & 2.5 $\pm$ 0.2 & 2.8 $\pm$ 0.1 \\
37$_{ 1,37}$-36$_{ 1,36}$ & 338214 & $\diamond$7.2 $\pm$ 0.3 & 5.4 $\pm$ 0.1 & \multicolumn{2}{l}{$\textless$ 3$\sigma$} & \multicolumn{2}{l}{$\textless$ 3$\sigma$} & 2.8 $\pm$ 0.1 & 2.49 $\pm$ 0.09 & \multicolumn{2}{l}{$\textless$ 3$\sigma$} \\
35$_{ 2,33}$-34$_{ 2,32}$ & 338278 & 4.31 $\pm$ 0.09 & 3.90 $\pm$ 0.03 & \multicolumn{2}{l}{$\textless$ 3$\sigma$} & \multicolumn{2}{l}{$\textless$ 3$\sigma$} & 2.32 $\pm$ 0.08 & 2.09 $\pm$ 0.06 & 5.7 $\pm$ 0.8 & 1.12 $\pm$ 0.06 \\
\hline
\end{tabular}
\tablefoot{$\diamond$ partially blended with other lines}
\end{table*}

\clearpage

\begin{longtable}{p{0.2\textwidth}ccccccccccc}
\caption{Ethyl Cyanide}\\
\hline\hline
 & & \multicolumn{2}{c}{G35.20 A} & \multicolumn{2}{c}{G35.20 B1} & \multicolumn{2}{c}{G35.20 B2} & \multicolumn{2}{c}{G35.20 B3} & \multicolumn{2}{c}{G35.03 A} \\
Transition & Frequency & FWHM & T$_{peak}$ & FWHM & T$_{peak}$ & FWHM & T$_{peak}$ & FWHM & T$_{peak}$ & FWHM & T$_{peak}$ \\
 & (MHz) & (km/s) & (K) & (km/s) & (K) & (km/s) & (K) & (km/s) & (K) & (km/s) & (K) \\
\hline
\endfirsthead
\caption{Ethyl Cyanide continued.}\\
\hline\hline
 & & \multicolumn{2}{c}{G35.20 A} & \multicolumn{2}{c}{G35.20 B1} & \multicolumn{2}{c}{G35.20 B2} & \multicolumn{2}{c}{G35.20 B3} & \multicolumn{2}{c}{G35.03 A} \\
Transition & Frequency & FWHM & T$_{peak}$ & FWHM & T$_{peak}$ & FWHM & T$_{peak}$ & FWHM & T$_{peak}$ & FWHM & T$_{peak}$ \\
 & (MHz) & (km/s) & (K) & (km/s) & (K) & (km/s) & (K) & (km/s) & (K) & (km/s) & (K) \\
\hline
\endhead
\hline
\endfoot
\multicolumn{12}{l}{{$\diamond$ partially blended with other lines.}} \\
\endlastfoot

\multicolumn{12}{c}{\bf{C$_2$H$_5$CN $\nu$=0}} \\
\hline
55$_{8,48}$-55$_{7,49}$ & 335274 & \multicolumn{2}{l}{$\textless$ 3$\sigma$} & \multicolumn{2}{l}{$\textless$ 3$\sigma$} & \multicolumn{2}{l}{$\textless$ 3$\sigma$} & 1.65 $\pm$ 0.22 & 0.72 $\pm$ 0.08 & \multicolumn{2}{l}{$\textless$ 3$\sigma$} \\
9$_{ 6, 4}$- 8$_{ 5, 3}$ & 335321 & \multicolumn{2}{l}{blended} & \multicolumn{2}{l}{blended} & \multicolumn{2}{l}{$\textless$ 3$\sigma$} & \multicolumn{2}{l}{blended} & \multicolumn{2}{l}{$\textless$ 3$\sigma$} \\
52$_{8,44}$-52$_{7,45}$ & 335895 & 2.89 $\pm$ 0.29 & 1.14 $\pm$ 0.04 & \multicolumn{2}{l}{$\textless$ 3$\sigma$} & \multicolumn{2}{l}{$\textless$ 3$\sigma$} & \multicolumn{2}{l}{$\textless$ 3$\sigma$} & \multicolumn{2}{l}{$\textless$ 3$\sigma$} \\
20$_{ 4,17}$-19$_{ 3,16}$ & 336614 & \multicolumn{2}{l}{blended} & \multicolumn{2}{l}{$\textless$ 3$\sigma$} & \multicolumn{2}{l}{$\textless$ 3$\sigma$} & 1.59 $\pm$ 0.18 & 0.50 $\pm$ 0.05 & \multicolumn{2}{l}{$\textless$ 3$\sigma$} \\
38$_{ 3,36}$-37$_{ 3,35}$ & 337347 & \multicolumn{2}{l}{blended} & \multicolumn{2}{l}{blended} & \multicolumn{2}{l}{blended} & 2.76 $\pm$ 0.07 & 4.05 $\pm$ 0.09 & \multicolumn{2}{l}{$\textless$ 3$\sigma$} \\
37$_{ 4,33}$-36$_{ 4,32}$ & 337445 & 5.3 $\pm$ 0.1 & 14.5 $\pm$ 0.2 & 2.7 $\pm$ 0.1 & 1.71 $\pm$ 0.06 & \multicolumn{2}{l}{$\textless$ 3$\sigma$} & 2.77 $\pm$ 0.07 & 4.05 $\pm$ 0.09 & \multicolumn{2}{l}{$\textless$ 3$\sigma$} \\
37$_{ 3,34}$-36$_{ 3,33}$ & 338142 & \multicolumn{2}{l}{blended} & \multicolumn{2}{l}{blended} & 1.9 $\pm$ 0.2 & 1.08 $\pm$ 0.0854 & 2.87 $\pm$ 0.06 & 5.49 $\pm$ 0.10 & \multicolumn{2}{l}{$\textless$ 3$\sigma$} \\
\hline
\end{longtable}

\clearpage

\begin{table*}[]
\centering
\caption{Cyanoacetylene}
\begin{tabular}{cccccccccccc}

\hline\hline
 & & \multicolumn{2}{c}{G35.20 A} & \multicolumn{2}{c}{G35.20 B1} & \multicolumn{2}{c}{G35.20 B2} & \multicolumn{2}{c}{G35.20 B3} & \multicolumn{2}{c}{G35.03 A} \\
Transition & Frequency & FWHM & T$_{peak}$ & FWHM & T$_{peak}$ & FWHM & T$_{peak}$ & FWHM & T$_{peak}$ & FWHM & T$_{peak}$ \\
 & (MHz) & (km/s) & (K) & (km/s) & (K) & (km/s) & (K) & (km/s) & (K) & (km/s) & (K) \\
 \hline
\multicolumn{12}{c}{\bf{HC$_3$N $\nu$=0}} \\
\hline
37 - 36 & 336520 & 7.4 $\pm$ 0.1 & 27.6 $\pm$ 0.4 & 4.53 $\pm$ 0.10 & 17.0 $\pm$ 0.3 & $\diamond$1.6 $\pm$ 0.1 & 6.5 $\pm$ 0.4 & 3.32 $\pm$ 0.04 & 47.2 $\pm$ 0.5& \multicolumn{2}{l}{blended} \\
\hline
\multicolumn{12}{c}{\bf{HC$_3$N v6=1}} \\
\hline
37 $_{1}$ - 36 $_{-1 }$& 337070 & \multicolumn{2}{l}{blended} & \multicolumn{2}{l}{$\textless$ 3$\sigma$} & \multicolumn{2}{l}{$\textless$ 3$\sigma$} & 2.4 $\pm$ 0.1 & 2.64 $\pm$ 0.08 & \multicolumn{2}{l}{$\textless$ 3$\sigma$} \\
37 $_{-1}$ - 36 $_{1}$ & 337335 & 5.3 $\pm$ 0.8 & 3.1 $\pm$ 0.3 & \multicolumn{2}{l}{$\textless$ 3$\sigma$} & \multicolumn{2}{l}{$\textless$ 3$\sigma$} & 2.2 $\pm$ 0.1 & 2.1 $\pm$ 0.1 & \multicolumn{2}{l}{$\textless$ 3$\sigma$} \\
\hline
\end{tabular}
\tablefoot{$\diamond$ partially blended with other lines, $\star$ double peaked}
\end{table*}

\begin{table*}[]
\centering
\caption{Thioformaldehyde}
\begin{tabular}{cccccccccccc}

\hline\hline
 & & \multicolumn{2}{c}{G35.20 A} & \multicolumn{2}{c}{G35.20 B1} & \multicolumn{2}{c}{G35.20 B2} & \multicolumn{2}{c}{G35.20 B3} & \multicolumn{2}{c}{G35.03 A} \\
Transition & Frequency & FWHM & T$_{peak}$ & FWHM & T$_{peak}$ & FWHM & T$_{peak}$ & FWHM & T$_{peak}$ & FWHM & T$_{peak}$ \\
 & (MHz) & (km/s) & (K) & (km/s) & (K) & (km/s) & (K) & (km/s) & (K) & (km/s) & (K) \\
 \hline
\multicolumn{12}{c}{\bf{H$_2$CS $\nu$=0}} \\
\hline
10$_{1,10}$ - 9$_{1,9}$ o & 338083 & 5.9 $\pm$ 0.1 & 51.7 $\pm$ 0.9 & 2.7 $\pm$ 0.1 & 27.3 $\pm$ 1.0 & 2.4 $\pm$ 0.1 & 28 $\pm$ 1 & 2.54 $\pm$ 0.06 & 44.3 $\pm$ 0.9 & 6.6 $\pm$ 0.1 & 21.0 $\pm$ 0.3 \\
10$_{1,9}$-9$_{1,8}$ o & 348532 & 5.8 $\pm$ 0.1 & 59.2 $\pm$ 1.3 & 2.6 $\pm$ 0.1 & 31 $\pm$ 1 & 2.3 $\pm$ 0.1 & 31 $\pm$ 1 & 2.5806 $\pm$ 0.0374 & 52.1 $\pm$ 0.7 & 6.33 $\pm$ 0.09 & 21.3 $\pm$ 0.3 \\
\hline
\multicolumn{12}{c}{\bf{H$_2$C$^{33}$S}} \\
\hline
10$_{1,10}$-9$_{1,9}$ & 335160 & 7.5 $\pm$ 0.2 & 3.91 $\pm$ 0.09 & 1.6 $\pm$ 0.5 & 0.4 $\pm$ 0.1 & 1.5 $\pm$ 0.6 & 0.4 $\pm$ 0.2 & 2.5 $\pm$ 0.2 & 1.47 $\pm$ 0.08 & \multicolumn{2}{l}{$\textless$ 3$\sigma$} \\
\hline
\end{tabular}
\tablefoot{$\diamond$ partially blended with other lines}
\end{table*}

\begin{table*}[]
\centering
\caption{Sulfur dioxide}
\begin{tabular}{cccccccccccc}

\hline\hline
 & & \multicolumn{2}{c}{G35.20 A} & \multicolumn{2}{c}{G35.20 B1} & \multicolumn{2}{c}{G35.20 B2} & \multicolumn{2}{c}{G35.20 B3} & \multicolumn{2}{c}{G35.03 A} \\
Transition & Frequency & FWHM & T$_{peak}$ & FWHM & T$_{peak}$ & FWHM & T$_{peak}$ & FWHM & T$_{peak}$ & FWHM & T$_{peak}$ \\
 & (MHz) & (km/s) & (K) & (km/s) & (K) & (km/s) & (K) & (km/s) & (K) & (km/s) & (K) \\
\hline
\multicolumn{12}{c}{\bf{SO$_2$ $\nu$=0}} \\
\hline
23$_{3,21}$-23$_{2,22}$ & 336089 & 7.7 $\pm$ 0.2 & 24.1 $\pm$ 0.5 & 7.8 $\pm$ 0.1 & 19.2 $\pm$ 0.3 & 3.4 $\pm$ 0.2 & 8.0 $\pm$ 0.4 & 4.1 $\pm$ 0.1 & 24.4 $\pm$ 0.7 & 10.0 $\pm$ 0.3 & 21.0 $\pm$ 0.5 \\
16$_{7,9}$-17$_{6,12}$ & 336670 & 8.8 $\pm$ 0.3 & 6.8 $\pm$ 0.2 & 7.3 $\pm$ 0.2 & 6.0 $\pm$ 0.1 & 2.4 $\pm$ 0.1 & 3.2 $\pm$ 0.2 & 3.4 $\pm$ 0.1 & 11.0 $\pm$ 0.3 & 8.4 $\pm$ 0.2 & 10.1 $\pm$ 0.1 \\
18$_{4,14}$-18$_{3,15}$ & 338306 & 8.0 $\pm$ 0.2 & 22.8 $\pm$ 0.5 & 7.6 $\pm$ 0.2 & 25.4 $\pm$ 0.6 & 4.0 $\pm$ 0.3 & 7.5 $\pm$ 0.5 & 3.84 $\pm$ 0.07 & 31.2 $\pm$ 0.5 & 9.8 $\pm$ 0.1 & 25.2 $\pm$ 0.2 \\
20$_{1,19}$-19$_{2,18}$ & 338612 & 5.6 $\pm$ 0.5 & 28.5 $\pm$ 1.6 & 7.5 $\pm$ 0.4 & 27.7 $\pm$ 1.9 & 4.1 $\pm$ 0.6 & 15.3 $\pm$ 0.9 & 3.8 $\pm$ 1.0 & 32 $\pm$ 2 & 10.0 $\pm$ 0.3 & 39.4 $\pm$ 0.9 \\
\hline
\end{tabular}
\end{table*}
\end{landscape}

\clearpage

\section{XCLASS fit errors}
\label{errors}
In this section we give the error values on each of our XCLASS fit results.  They are rarely symmetrical, so they are reported in the format: value, lower error (-), upper error (+), except for column density which is listed as value, lower limit (value-error), upper limit (value+error).  The errors were calculated using the Interval Nested Sampling (INS) error estimator algorithm in XCLASS.

\begin{table*}[!h]
\centering
\caption{Error values for G35.20 A.}
\label{errorsA}
\begin{tabular}{llllllllll}

\hline\hline
Catalog Entry      & Source size & error left & error right & T$_{ex}$ & error left & error right & N$_{col}$ & lower limit & upper limit \\
\hline
CH3OH;v=0          & 0.6   & 0.2   & 0.3   & 164 & 18 & 36  & 3.52E+18 & 6.88E+17 & 7.00E+18 \\
CH3OH;v12=1        & 0.5   & 0.1   & 0.2   & 218 & 24 & 32  & 4.79E+18 & 9.77E+17 & 3.00E+19 \\
CH3OH;v12=2        & 0.39  & 0.09  & 0.11  & 227 & 18 & 73  & 3.78E+18 & 7.93E+17 & 1.00E+19 \\
C-13-H3OH;v=0;     & 1.0   & 0.3   & 0.2   & 121 & 70 & 39  & 1.62E+17 & 4.27E+16 & 3.00E+17 \\
CH3O-18-H;v=0;     & 1.1   & 0.2   & 0.1   & 121 & 62 & 49  & 4.42E+16 & 1.35E+16 & 7.00E+16 \\
CH3CN;v=0;         & 0.45  & 0.10  & 0.15  & 208 & 23 & 32  & 3.91E+16 & 1.16E+16 & 3.00E+17 \\
CH3CN;v8=1;        & 0.6   & 0.1   & 0.2   & 359 & 54 & 41  & 3.22E+16 & 1.10E+16 & 6.00E+16 \\
\hline
\end{tabular}
\end{table*}

\begin{table*}[!h]
\centering
\caption{Error values for G35.20 B1. Star ($\star$) indicates a catalog entry that was not modeled.}
\label{errorsB1}
\begin{tabular}{llllllllll}

\hline\hline
Catalog Entry  & Source size & error left & error right & T$_{ex}$ & error left & error right & N$_{col}$ & lower limit & upper limit \\
\hline
CH3OH;v=0      & 0.3     & 0.2   & 0.2   & 234 & 35  & 63  & 6.99E+17 & 2.16E+17 & 5.00E+18 \\
CH3OH;v12=1    & 0.36    & 0.06  & 0.12  & 153 & 3   & 25  & 1.07E+18 & 3.34E+17 & 5.00E+19 \\
CH3OH;v12=2    & 0.18    & 0.01  & 0.22  & 154 & 1   & 46  & 2.23E+18 & 6.90E+17 & 6.80E+18 \\
C-13-H3OH;v=0; & 0.3     & 0.2   & 0.2   & 113 & 41  & 36  & 3.87E+16 & 1.57E+16 & 1.50E+17 \\
CH3O-18-H;v=0; & 0.3     & 0.1   & 0.2   & 82  & 19  & 35  & 1.45E+16 & 8.20E+15 & 1.20E+17 \\
CH3CN;v=0;     & 0.32    & 0.09  & 0.16  & 129 & 78  & 20  & 7.24E+15 & 5.45E+15 & 1.32E+16 \\
CH3CN;v8=1;    & 0.29    & 0.04  & 0.11  & 283 & 43  & 17  & 3.50E+15 & 2.94E+15 & 7.00E+16 \\
\hline
\end{tabular}
\end{table*}

\begin{table*}[!h]
\centering
\caption{Error values for G35.20 B2. Star ($\star$) indicates a catalog entry that was not modeled.}
\label{errorsB2}
\begin{tabular}{llllllllll}

\hline\hline
Catalog Entry  & Source size & error left & error right & T$_{ex}$ & error left & error right & N$_{col}$ & lower limit & upper limit \\
\hline
CH3OH;v=0      & 0.36    & 0.05 & 0.84 & 136 & 5   & 164 & 5.25E+17 & 1.75E+17 & 5.00E+19 \\
CH3OH;v12=1    & 0.32    & 0.05 & 0.38 & 145 & 0   & 155 & 6.18E+17 & 2.24E+17 & 5.00E+19 \\
CH3OH;v12=2    & 0.15    & 0.03 & 0.24 & 152 & 0   & 148 & 1.06E+18 & 2.77E+17 & 5.00E+19 \\
C-13-H3OH;v=0; & 1.1     & 0.2  & 0.1  & 97  & 34  & 36  & 7.99E+15 & 4.94E+15 & 1.00E+18 \\
CH3O-18-H;v=0; & 0.9     & 0.3  & 0.2  & 75  & 26  & 40  & 1.36E+15 & 1.34E+15 & 2.00E+17 \\
CH3CN;v=0;     & 0.62    & 0.08 & 0.08 & 132 & 7   & 10  & 1.30E+15 & 1.20E+15 & 1.74E+15 \\
CH3CN;v8=1;    & $\star$ &      &      &     &     &     &          &          &          \\
\hline
\end{tabular}
\end{table*}

\begin{table*}[!h]
\centering
\caption{Error values for G35.20 B3}
\label{errorsB3}
\begin{tabular}{llllllllll}

\hline\hline
Catalog Entry  & Source size & error left & error right & T$_{ex}$ & error left & error right & N$_{col}$ & lower limit & upper limit \\
\hline
CH3OH;v=0      & 0.35 & 0.04 & 0.02 & 155 & 6   & 36  & 6.23E+17 & 5.55E+17 & 8.13E+17 \\
CH3OH;v12=1    & 0.36 & 0.02 & 0.07 & 189 & 14  & 14  & 9.81E+17 & 7.62E+17 & 1.12E+18 \\
CH3OH;v12=2    & 0.20 & 0.02 & 0.02 & 172 & 5   & 20  & 1.80E+18 & 1.32E+18 & 2.33E+18 \\
C-13-H3OH;v=0; & 1.20 & 0.06 & 0.03 & 89  & 10  & 9   & 1.63E+16 & 1.20E+16 & 1.93E+16 \\
CH3O-18-H;v=0; & 1.20 & 0.01 & 0.00 & 111 & 43  & 18  & 5.75E+15 & 4.17E+15 & 1.21E+16 \\
CH3CN;v=0;     & 0.35 & 0.06 & 0.04 & 208 & 30  & 30  & 6.46E+15 & 5.21E+15 & 1.06E+16 \\
CH3CN;v8=1;    & 0.8  & 0.2  & 0.2  & 213 & 47  & 27  & 1.30E+15 & 5.69E+14 & 1.00E+18 \\
\hline
\end{tabular}
\end{table*}

\begin{table*}[]
\centering
\caption{Error values for G35.03 A. Star ($\star$) indicates a catalog entry that was not modeled.}
\label{errors3503}
\begin{tabular}{llllllllll}

\hline\hline
Catalog Entry      & Source size & error left & error right & T$_{ex}$ & error left & error right & N$_{col}$ & lower limit & upper limit \\
\hline
CH3OH;v=0          & 0.40    & 0.09 & 0.71 & 142 & 18 & 68  & 1.80E+18 & 3.99E+17 & 1.00E+19 \\
CH3OH;v12=1        & 0.30    & 0.06 & 0.62 & 239 & 40 & 61  & 2.64E+18 & 6.01E+17 & 1.00E+19 \\
CH3OH;v12=2        & 0.20    & 0.05 & 0.52 & 192 & 19 & 108 & 5.06E+18 & 9.95E+17 & 1.00E+19 \\
C-13-H3OH;v=0;     & 0.6     & 0.1  & 0.6  & 120 & 24 & 51  & 5.00E+16 & 1.58E+16 & 1.00E+18 \\
CH3O-18-H;v=0;     & 1.2     & 0.3  & 0.3  & 130 & 21 & 41  & 9.99E+15 & 3.48E+15 & 1.00E+16 \\
CH3CN;v=0;         & 0.30    & 0.06 & 0.40 & 186 & 21 & 114 & 3.04E+16 & 1.04E+16 & 1.00E+19 \\
CH3CN;v8=1;        & 0.30    & 0.05 & 0.20 & 216 & 19 & 54  & 3.57E+16 & 1.35E+16 & 1.29E+19 \\
\hline
\end{tabular}
\end{table*}

\twocolumn
\section{XCLASS analysis details}
\label{xclass}

\subsection{S-bearing}
We detected two S-bearing species with more than one transition in our sources.
\begin{itemize}
  \item Sulfur Dioxide (SO$_2$) - As a species with extended emission, the source size ranges from 1.2$''$ at G35.03 to 2.4$''$ at B2.  The column densities for this species are relatively consistent at 0.6-4.3 x 10$^{16}$ cm$^{-2}$. across G35.20 and 7.5 x 10$^{16}$ cm$^{-2}$. at G35.03. T$_{ex}$ values are from 114~K in B2 to 288~K in B3.  The fits were based on 24 different transitions within the spectral window, five of which were measured in the survey (others were not present or within the noise).  
  \item Thioformaldehyde (H$_2$CS) - This species was observed to have fairly extended emission so the source size ranged from 0.9$''$ for B3 to 2.5$''$ for B1.  The T$_{ex}$ range is on the cooler side at 50~K for B2, less than 100~K for B1, B3, and G35.03, and 165~K at A.  A had the highest column density at 1.4 x 10$^{16}$~cm$^{-2}$., where the others ranged from 1.2 - 6.1 x 10$^{15}$~cm$^{-2}$.
\end{itemize}

\subsection{O-bearing Organics}
\begin{itemize}
  \item Formaldehyde (H$_2$CO) - Only the isotopologues H$_2$C$^{18}$O, HDCO, and D$_2$CO were in the spectral range of these observations.  See subsection D.6.
  \item Formic Acid (HCOOH) - This species was detected at all peaks, but not modeled for B2 because the emission lines were only just above 3$\sigma$.  
  \item Methanol (CH$_3$OH) - the $\nu$=0 state of CH$_3$OH was modeled to a maximum source size of 1$''$, so the source size fits were 0.4-0.6$''$.  The temperature range was from 132-268~K.   The column densities for this species ranged from 0.3-4.8 x 10$^{18}$ cm$^{-2}$.  The vibrationally excited states and isotopologues were modeled separately as their spatial extent is different to the main state and is likely part of different gas - see \S~C.5 and C.6. 
  \item Acetaldehyde (CH$_3$CHO) - the $\nu$=0 state was modeled separately from the vibrationally excited states (see subsection D.5).  This species was observed to have a fairly compact emitting region, so the source size was limited to 1.5$''$.  The best fit source sizes were 0.3-1.2$''$ with relatively high excitation temperatures between 193~K and 300~K.  The column densities ranged from 1.5-17 x 10$^{15}$~cm$^{-2}$.
  \item Methyl Formate (CH$_3$OCHO) - only the $\nu$=0 state was fit as the $\nu$=1 transitions were not in the XCLASS database.  The emission from this species was fairly compact so the model was allowed a maximum source size of 1.2$''$.  The final source sizes were 0.2-1.0$''$ and column densities ranged from 1.9-29 x 10$^{16}$~cm$^{-2}$.  Generally the T$_{ex}$ values were high (182-295~K for B1, B2, and B3), but the excitation temperatures of A and G35.03 were 103~K and 151~K, respectively.
  \item Dimethyl Ether (CH$_3$OCH$_3$) - there was a lot of variation between the sources for this species.  The source size was very compact for A, B1, and B2 (0.3, 0.4, and 0.5$''$, respectively), but more extended for B3 and G35.03 at 0.8$''$.  The temperature differences were also large.  The T$_{ex}$ for B3 was 90~K, 170 and 180~K for B2 and B1, 229~K at A, and 260~K at G35.03.  The column densities were lower for B2 (1.6 x 10$^{16}$~cm$^{-2}$) and higher for the other peaks (8.8 x 10$^{16}$-9.7 x 10$^{17}$~cm$^{-2}$).
  \item Ethanol (C$_2$H$_5$OH) - the trans- and gauche- transitions for ethanol were modeled from a single JPL database entry.  The temperatures varied widely with best fit values of 88~K for B1 and 120~K for B2, and much higher values of 260, 281, and 300~K for B3, A and G35.03, respectively.  The column densities ranged between 0.6 and 7.1 x 10$^{16}$~cm$^{-2}$ range with the lowest at B1 and B2 and the highest at A.  The source sizes were 0.4-0.8$''$.
  \item Ethylene Glycol ((CH$_2$OH)$_2$) - This species was only modeled for G35.20 core A and for G35.03 as it was not detected in any part of core B.  In G35.20 core A and G35.03, the best fit source sizes are 0.6$''$ and 0.2$''$, respectively and the T$_{ex}$ was 172~K for G35.20 and 75~K in G35.03.  The column densities were 3.5 x 10$^{16}$~cm$^{-2}$ in G35.20 core A and 7.8 x 10$^{16}$~cm$^{-2}$ in G35.03.  
\end{itemize}

\subsection{N-bearing Organics}
\begin{itemize}
  \item Cyanoacetylene (HC$_3$N) - the $\nu$=0 state was modeled for all regions and the isotopologue HC$^{13}$CCN $\nu$=0 was coupled with HC$_{3}$N $\nu$=0 to improve the uncertainty (from fitting one transition to fitting two).  The fit for HCC$^{13}$CN $\nu$=0 was also coupled for B3, as this is the only location where this species was detected.  When fits of species are coupled together they share the same temperature and source size, but vary the isotope ratio (and therefore the column density).  See \S~C.5 for excited states.  The column densities for this species ranged from 2.1 x 10$^{14}$ cm$^{-2}$ at B2 to 2.2 x 10$^{15}$ cm$^{-2}$ at G35.03.  The source sizes were fairly consistent, ranging between 0.9 and 1.2$''$ and the T$_{ex}$ range was 132-208~K.
  \item Methyl Cyanide (CH$_3$CN) - the $\nu$=0 state was modeled for all regions, but the isotopologues were not coupled with the main species because their spatial extent was dramatically different.  The modeled source sizes for this species were quite compact: 0.3-0.6$''$ and the temperature range was 124-235~K.  The range of column densities was 1.8-7.2 x 10$^{16}$ cm$^{-2}$.  See subsections D.5 and D.6 for excited states and isotopologues.
  \item Vinyl Cyanide (C$_2$H$_3$CN) - the results for vinyl cyanide are very different for the two regions where it was detected.  G35.20 A has a source size of 0.6$''$, an excitation temperature of 77~K, and a column density of 1.3 x 10$^{16}$ cm$^{-2}$.  On the other end of this source, B3 has a size of 0.8$''$, a T$_{ex}$ of 207~K, and a column density of 7.3 x 10$^{14}$ cm$^{-2}$.  
  \item Ethyl Cyanide (C$_2$H$_5$CN) - this species was only modeled for G35.20 core A, region B3, and G35.03 core A.  The temperatures for this species were all low compared to many of the other species.  At B3, the T$_{ex}$ was 50~K, at A it was 71~K and at G35.03 it was 78~K.  Column densities were 5.0-45 x 10$^{15}$ cm$^{-2}$ with the highest value at B3 and the source sizes were somewhat similar at 0.3$''$ for B3, 0.5$''$ for G35.03, and 0.6$''$ for A.
\end{itemize}

\subsection{H, N, and O-bearing Organics}
\begin{itemize}
  \item Isocyanic Acid (HNCO) - the fit was based on a single strong transition so some assumptions were made.  The source size was fixed based on the 3$\sigma$ level emission and the column density was modeled at temperatures of 50~K and 100~K.  These temperatures are reasonable for a more extended emitting region, as the bulk of the emission is not likely to come from very near the heating source.  The resulting column densities are between 2.2 x 10$^{15}$ cm$^{-2}$ and 2.0 x 10$^{16}$ cm$^{-2}$ at 50~K and range from 9.4 x 10$^{14}$ cm$^{-2}$ to 7.5 x 10$^{16}$ cm$^{-2}$ at 100~K.
  \item Formamide (NH$_2$CHO) - the $\nu$=0 transitions were fit with the NH$_2$$^{13}$CHO transitions coupled with the same parameters except abundance.  The temperatures for this species were comparatively low, ranging from 43-100~K.  The best fit source sizes were between 0.5$''$ (at B1) and 0.8$''$ at G35.03.  The range of column densities is 0.3-7.6 x 10$^{15}$ cm$^{-2}$.
\end{itemize}

\subsection{Vibrationally excited transitions}
High energy transitions were modeled uncoupled to the main state as those in our sources are observed to emit from a much smaller area than lower energy transitions (see Figures~\ref{maps3520} and \ref{maps3503}) and many are not observed in B1 or B2.  Vibrationally excited emission from HC$_{3}$N is only found toward cores A and B3 and weakly toward B1. See Figure~\ref{HC3Nvibe}.

\begin{itemize}
  \item Methanol (CH$_3$OH) - the v$_{12}$=1 and v$_{12}$=2 excited states were modeled separately on the assumption that the different excited states are emitted in different gas, since the spatial extent of each excited state grows more compact with increased energy (see section 2.3 and \citet{Alvaro2014} Figure 6).  The source sizes for the v$_{12}$=2 are more compact than the v$_{12}$=1 states and generally have a higher temperature.  The range of column densities between the two states are similar at 0.3-4.8 x 10$^{18}$ cm$^{-2}$ for v$_{12}$=1 and 1.9-5.1 x 10$^{18}$ cm$^{-2}$ for v$_{12}$=2.
  \item Cyanoacetylene (HC$_3$N) - each of the vibrational states ($\nu_6$=1, $\nu_7$=1, $\nu_7$=2) were modeled separately and the source size was more compact with higher excitation.  No vibrationally excited states were modeled for B2, as they were not detected in the observations, and only the $\nu_7$=1 state was modeled for B1 and G35.03.  The $\nu_6$=1 state was modeled for A and B3 and was coupled with HCC$^{13}$CN with the $^{12}$C/$^{13}$C isotope ratio fixed at 50.  The resulting temperatures based on these 2-3 transitions was 200~K at A and 365~K at B3 with source sizes of 0.6 and 0.4$''$.  Both peaks had similar column densities at 3.1 and 2.7 x 10$^{15}$ cm$^{-2}$, only slightly more than those of the $\nu$=0 state.  
  The $\nu_7$=1 state was also modeled coupled with the three different $^{13}$C isotopologues for A and B3 with the isotope ratio fixed at 50.  The $\nu_7$=1 source size is more compact for all modeled sources, but still somewhat extended at 0.8-1.0$''$.  The excitation temperatures are 25-80~K higher than the $\nu$=0 state, ranging from 194~K at G35.03 to 283~K for both A and B3.  The abundances of the $\nu_7$=1 state are similar to that of the $\nu$=0 state, but for A, it is about 0.45, for B1 and B3 the abundances are almost equal, and for G35.03, the $\nu_7$=1 abundance is about twice the abundance of the $\nu$=0 state.  
  The $\nu_7$=2 state was only modeled for A and B3 where the emission becomes very compact, 0.1 and 0.3$''$, the temperature is very hot, 420-450~K, and the abundances are 1.3 and 0.7 times higher than the $\nu$=0 state.
  \item Methyl Cyanide - CH$_3$CN v$_8$=1 was modeled separately since its spatial distribution was significantly different from that of the $\nu$=0 transitions.  The v$_8$=1 emission was not modeled for region B2 since it were not detected to a significant degree.  Temperatures for this species are generally high, ranging from 215-360~K with compact source sizes of 0.3-0.6$''$.  The column densities for this excited species are 0.4-3.6 x 10$^{16}$~cm$^{-2}$ which are similar to those of the $\nu$=0 state.
  \item Acetaldehyde (CH$_3$CHO) - the $\nu_{15}$=1 and $\nu_{15}$=2 excited states were modeled as more compact emission sources than the $\nu$=0 state.  Fits were made for the $\nu_{15}$=1 state for all sources except B2, but only for A and B3 for the $\nu_{15}$=2 state.  In all sources the temperature increases with increasing excitation and the source size decreases.  The column densities for both excited states range from 3.5-9.8 x 10$^{15}$~cm$^{-2}$ though the column density for the $\nu_{15}$=2 state at B3 is 5.2 x 10$^{16}$~cm$^{-2}$.
\end{itemize}

\subsection{Isotopologues and Deuteration}
\begin{itemize}
  \item Formaldehyde (H$_2$CO) - Only the isotopologues HDCO, D$_2$CO and H$_2$C$^{18}$O were in the spectral range of these observations.  These were modeled separately where the H$_2$C$^{18}$O fit was based on 6 transitions and the HDCO fit was based on a single transition.  For H$_2$C$^{18}$O the size ranges from 0.24-0.75$''$ and the temperatures are 26~K for B2 to 275~K for A.  The range of column densities is 0.4-4.6 x 10$^{15}$~cm$^{-2}$.  The abundances for HDCO are also in this order 0.15-1.6 x 10$^{15}$ cm$^{-2}$ at 50~K and 0.3-16~x~10$^{14}$~cm$^{-2}$ at 100~K with the source size fixed at 1.5$''$.  D$_2$CO was not fitted as only single weak lines were detected.
  \item Methanol - CH$_3$$^{18}$OH and $^{13}$CH$_3$OH were modeled uncoupled to the main isotopologue because their spatial extent differs somewhat from the main isotopologue.  The isotope ratios reached were 20-80 for CH$_3$OH/$^{13}$CH$_3$OH and 180-320 for CH$_3$OH/CH$_3$$^{18}$OH at peaks B2, B3, and G35.03.  The isotope ratio is less ideal for G35.30N A and B1 at 40 and 80, respectively, but the source size is also significantly different. Deuterated methanol (CH$_2$DOH) was not modeled because it was not in the XCLASS database, so analysis for this species was completed using Cassis.
  \item Methyl Cyanide - CH$_3$$^{13}$CN, and CH$_2$DCN were modeled separately from the $\nu$=0 emission because of their differing spatial distribution.  The CH$_2$DCN was not modeled for region B2 since it were not detected to a significant degree.
  \item Ethyl Cyanide - CH$_3$CHDCN was only detected at G35.20 A and was modeled for that location.
\end{itemize}

\section{Line ID xclass fits}
\label{xclassfits}
Presented below are the spectra of each peak with the original data, the full XCLASS fit, and fits of selected species alone in different colors.\\
\onecolumn
\begin{landscape}
\begin{figure}
   \resizebox{\hsize}{!}
            {\includegraphics{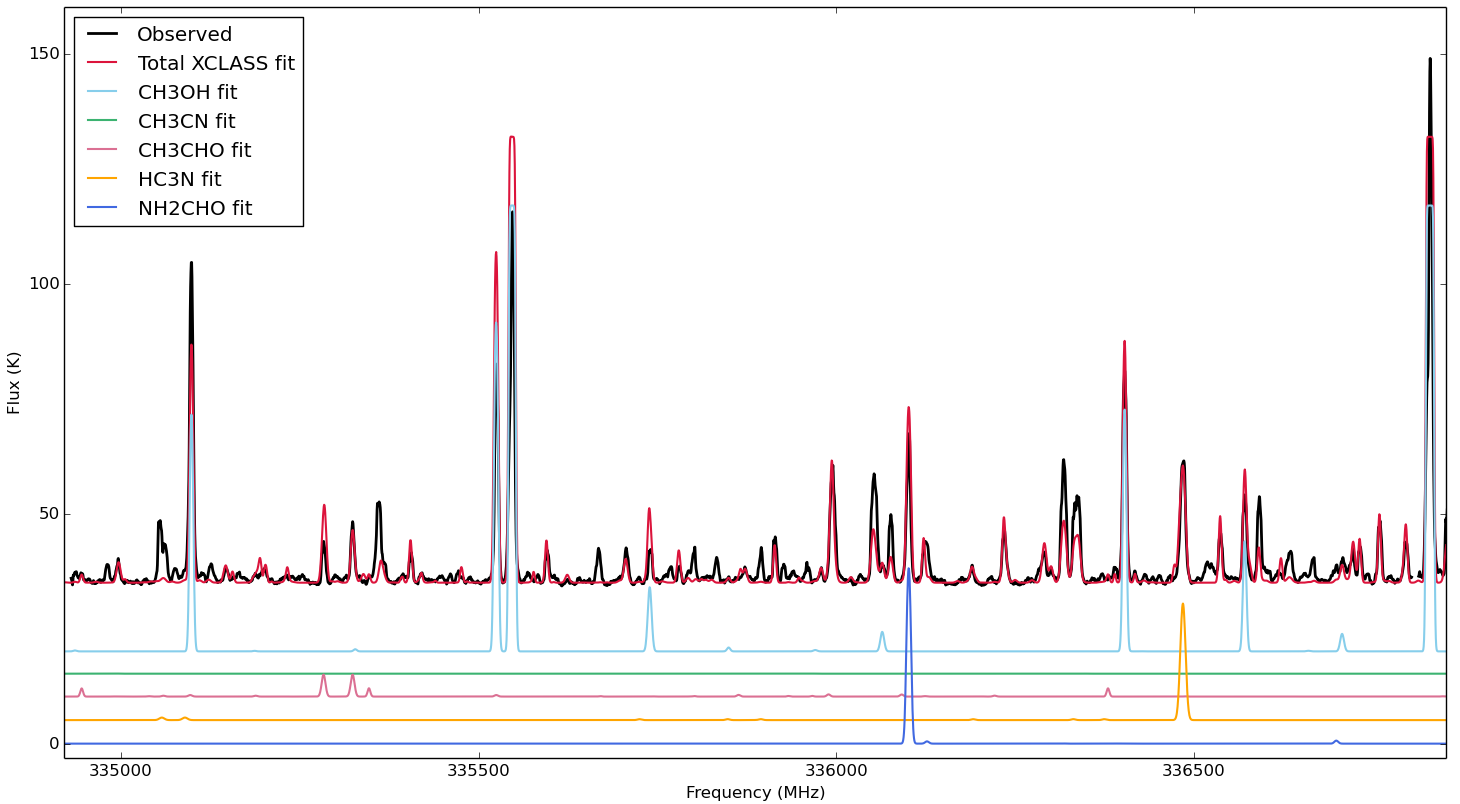}}
   \caption{G35.20 peak A spectral window 1 (334.9-336.8 GHz), XCLASS total fit, plus selected species.}
\end{figure}

\begin{figure}
   \resizebox{\hsize}{!}
            {\includegraphics{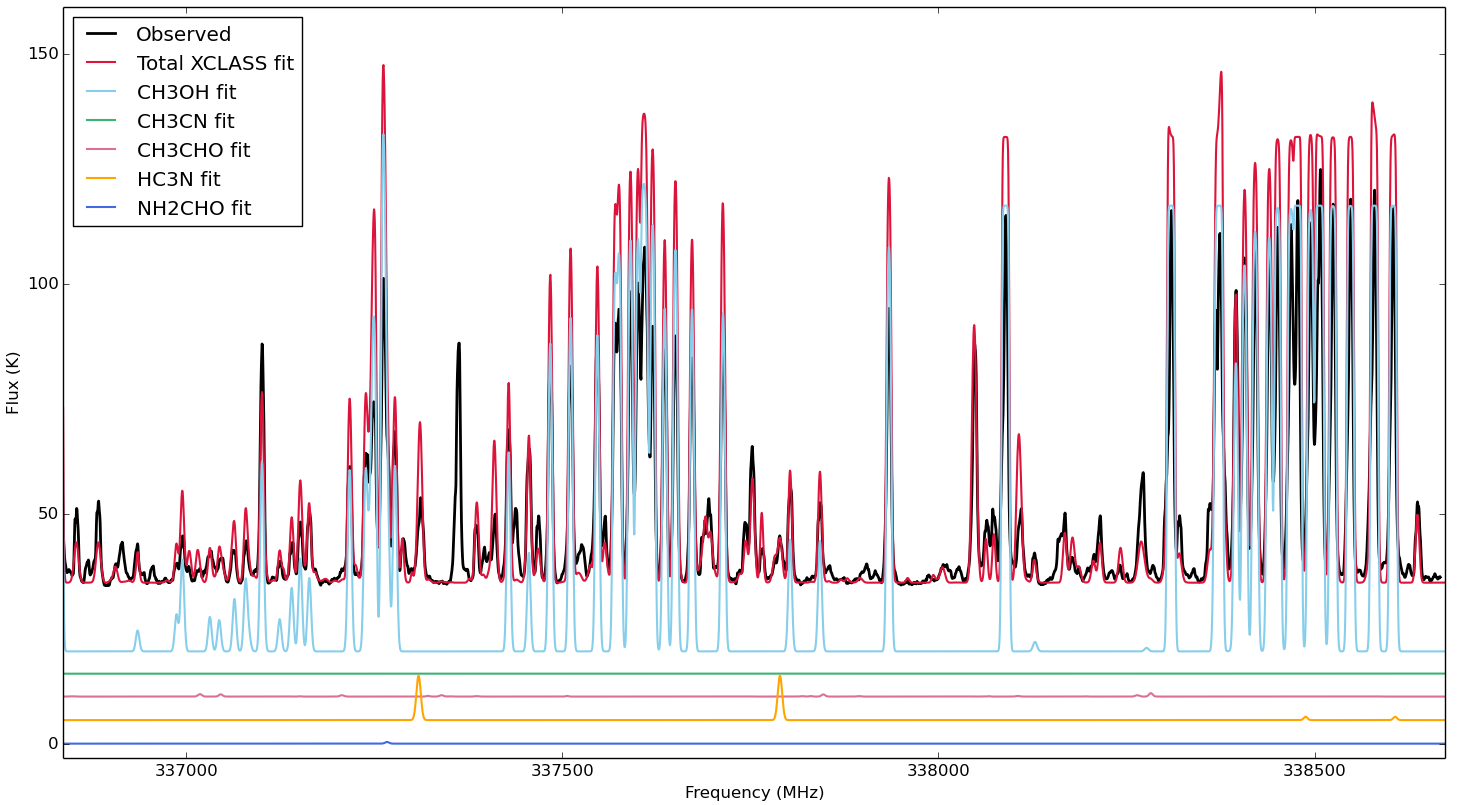}}
   \caption{G35.20 peak A spectral window 0 (336.8-338.7 GHz), XCLASS total fit, plus selected species.}
\end{figure}

\begin{figure}
   \resizebox{\hsize}{!}
            {\includegraphics{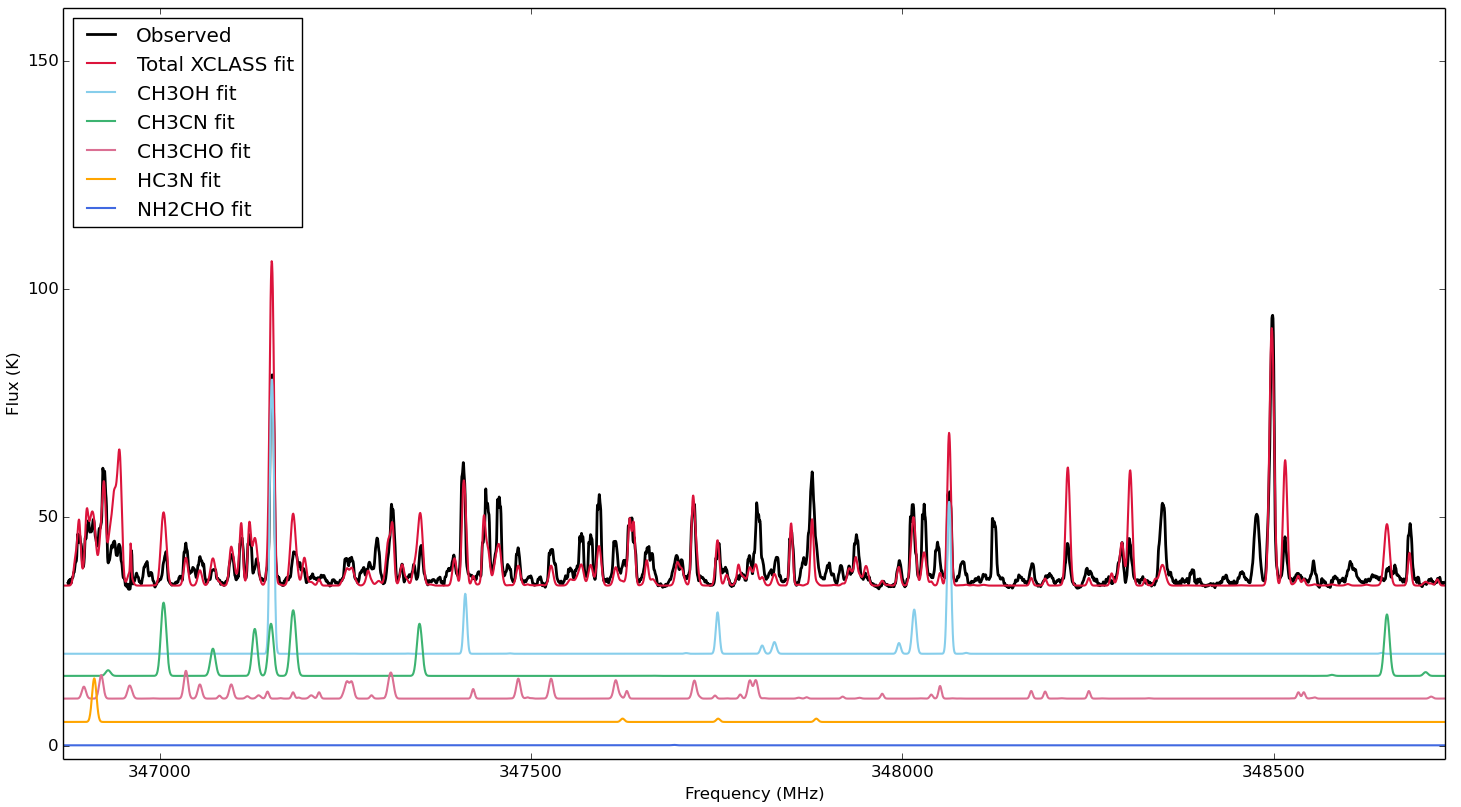}}
   \caption{G35.20 peak A spectral window 3 (346.9-348.7 GHz), XCLASS total fit, plus selected species.}
\end{figure}

\begin{figure}
   \resizebox{\hsize}{!}
            {\includegraphics{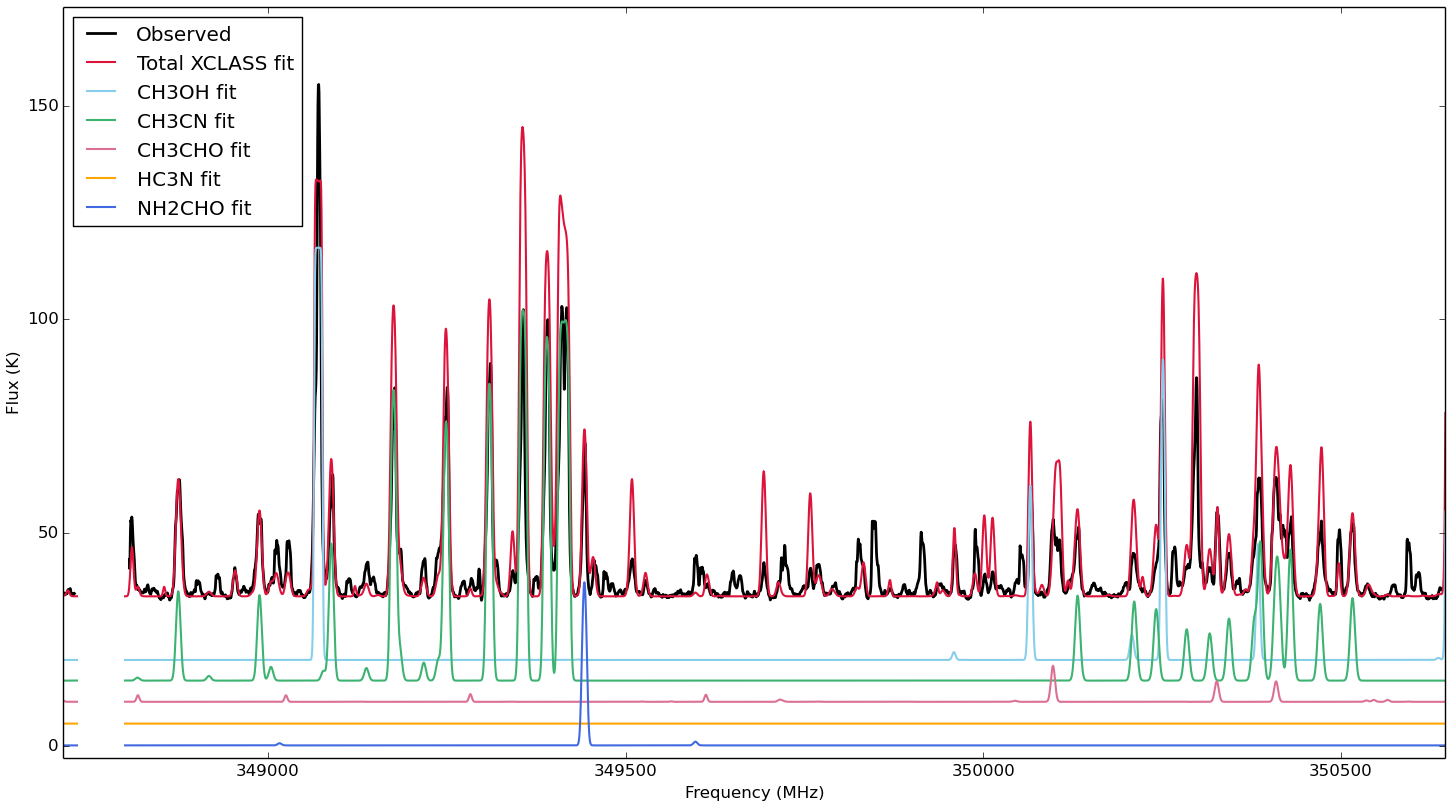}}
   \caption{G35.20 peak A spectral window 2 (348.8-350.7 GHz), XCLASS total fit, plus selected species.}
\end{figure}

\clearpage

\begin{figure}
   \resizebox{\hsize}{!}
            {\includegraphics{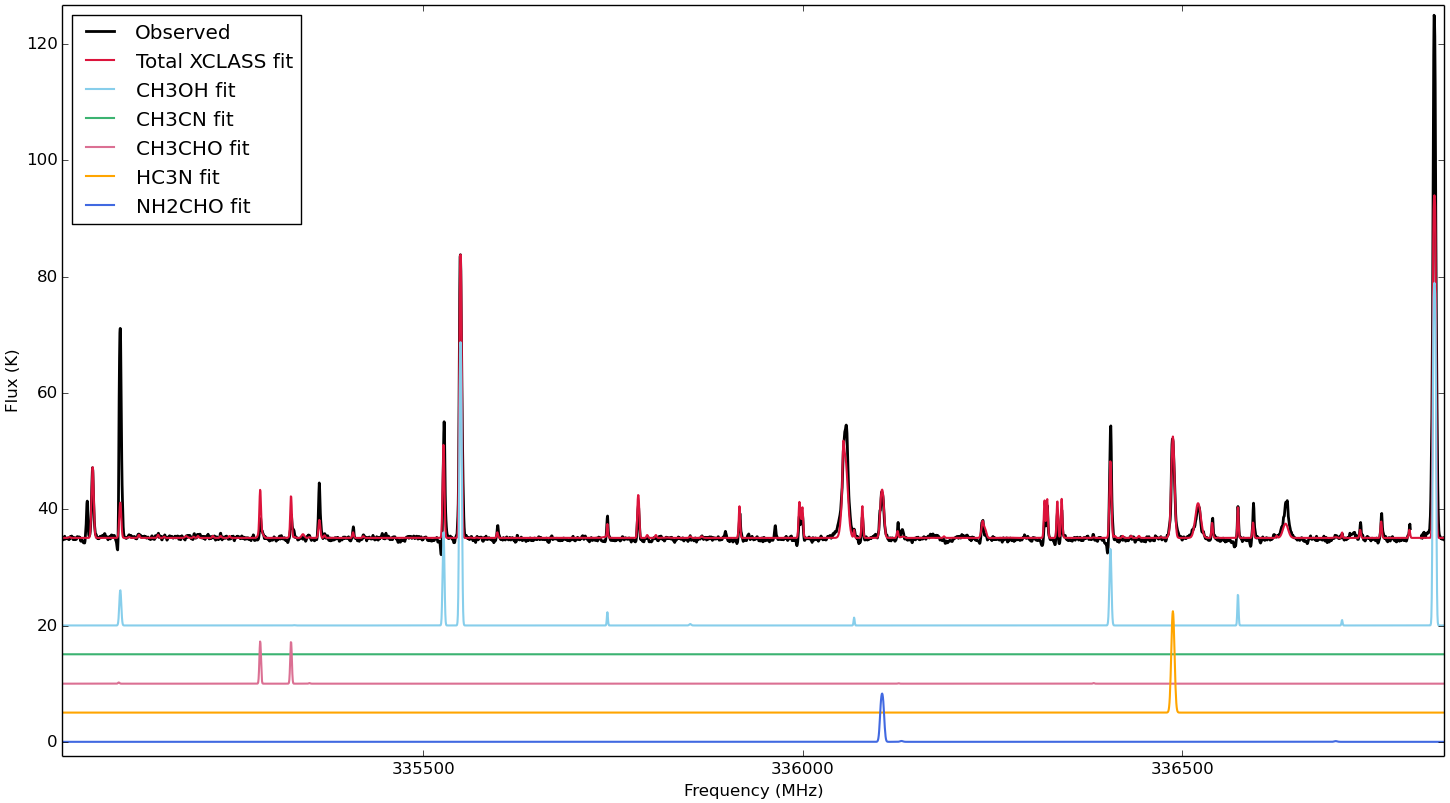}}
   \caption{G35.20 peak B1 spectral window 1 (334.9-336.8 GHz), XCLASS total fit, plus selected species.}
\end{figure}

\begin{figure}
   \resizebox{\hsize}{!}
            {\includegraphics{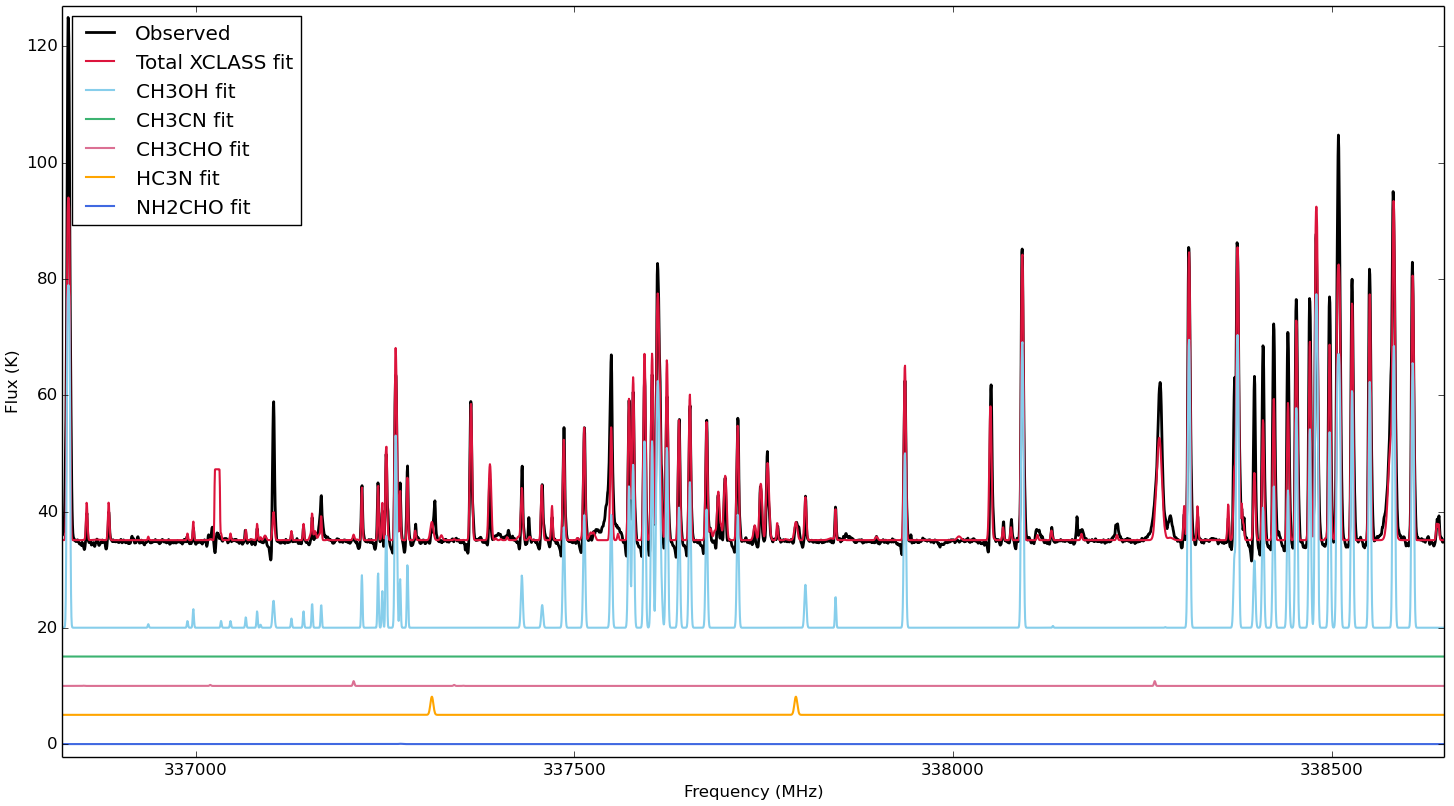}}
   \caption{G35.20 peak B1 spectral window 0 (336.8-338.7 GHz), XCLASS total fit, plus selected species.}
\end{figure}

\begin{figure}
   \resizebox{\hsize}{!}
            {\includegraphics{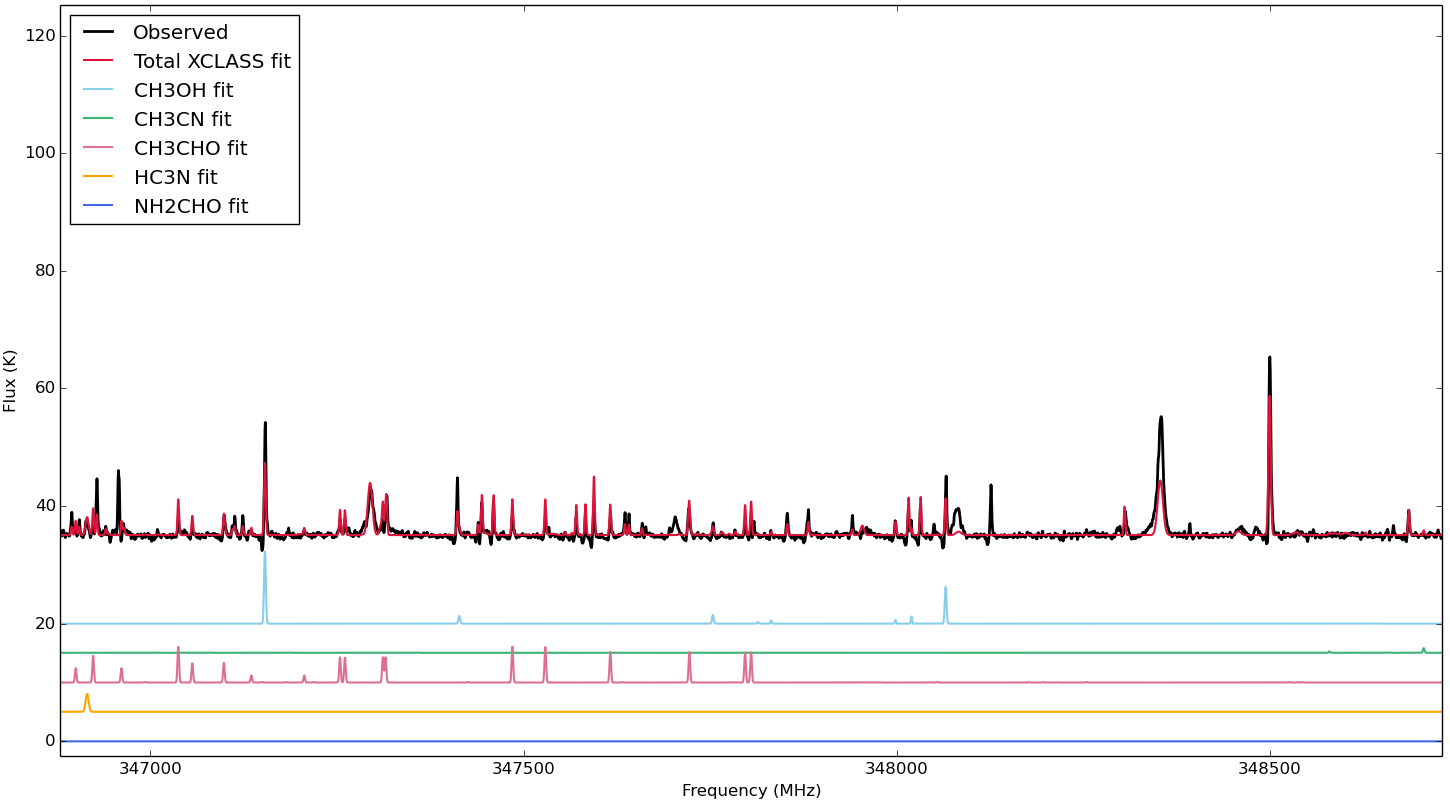}}
   \caption{G35.20 peak B1 spectral window 3 (346.9-348.7 GHz), XCLASS total fit, plus selected species.}
\end{figure}

\begin{figure}
   \resizebox{\hsize}{!}
            {\includegraphics{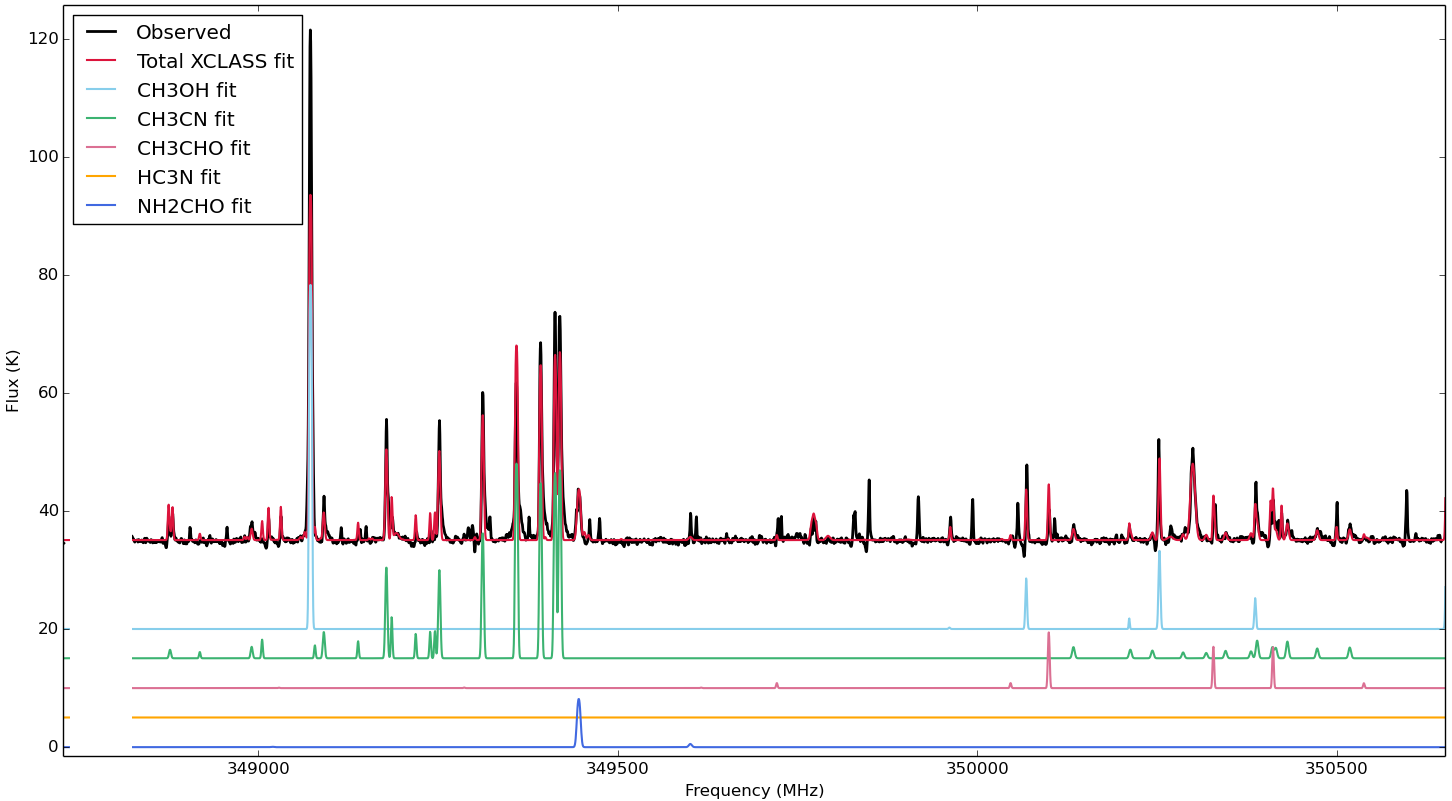}}
   \caption{G35.20 peak B1 spectral window 2 (348.8-350.7 GHz), XCLASS total fit, plus selected species.}
\end{figure}

\clearpage

\begin{figure}
   \resizebox{\hsize}{!}
            {\includegraphics{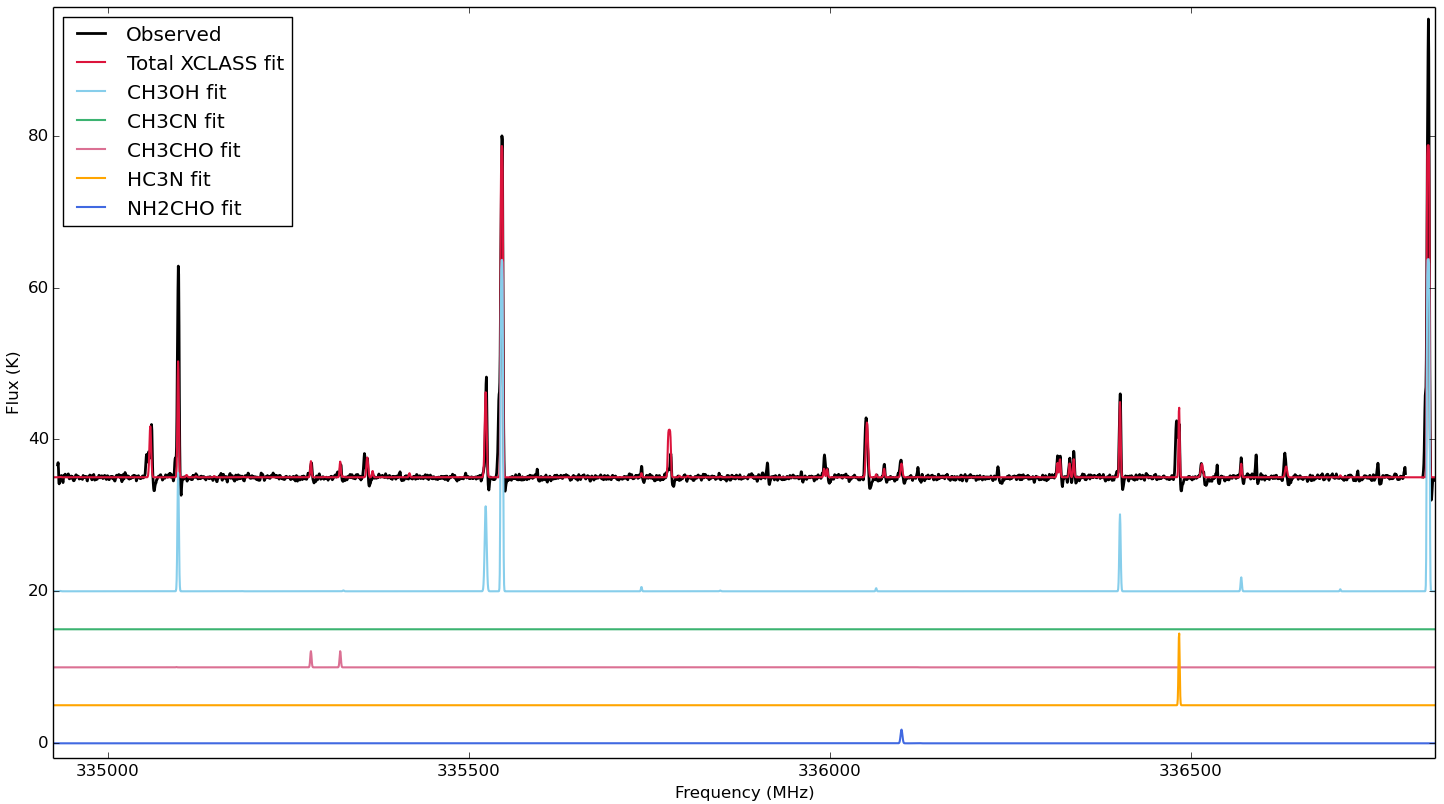}}
   \caption{G35.20 peak B2 spectral window 1 (334.9-336.8 GHz), XCLASS total fit, plus selected species.}
\end{figure}

\begin{figure}
   \resizebox{\hsize}{!}
            {\includegraphics{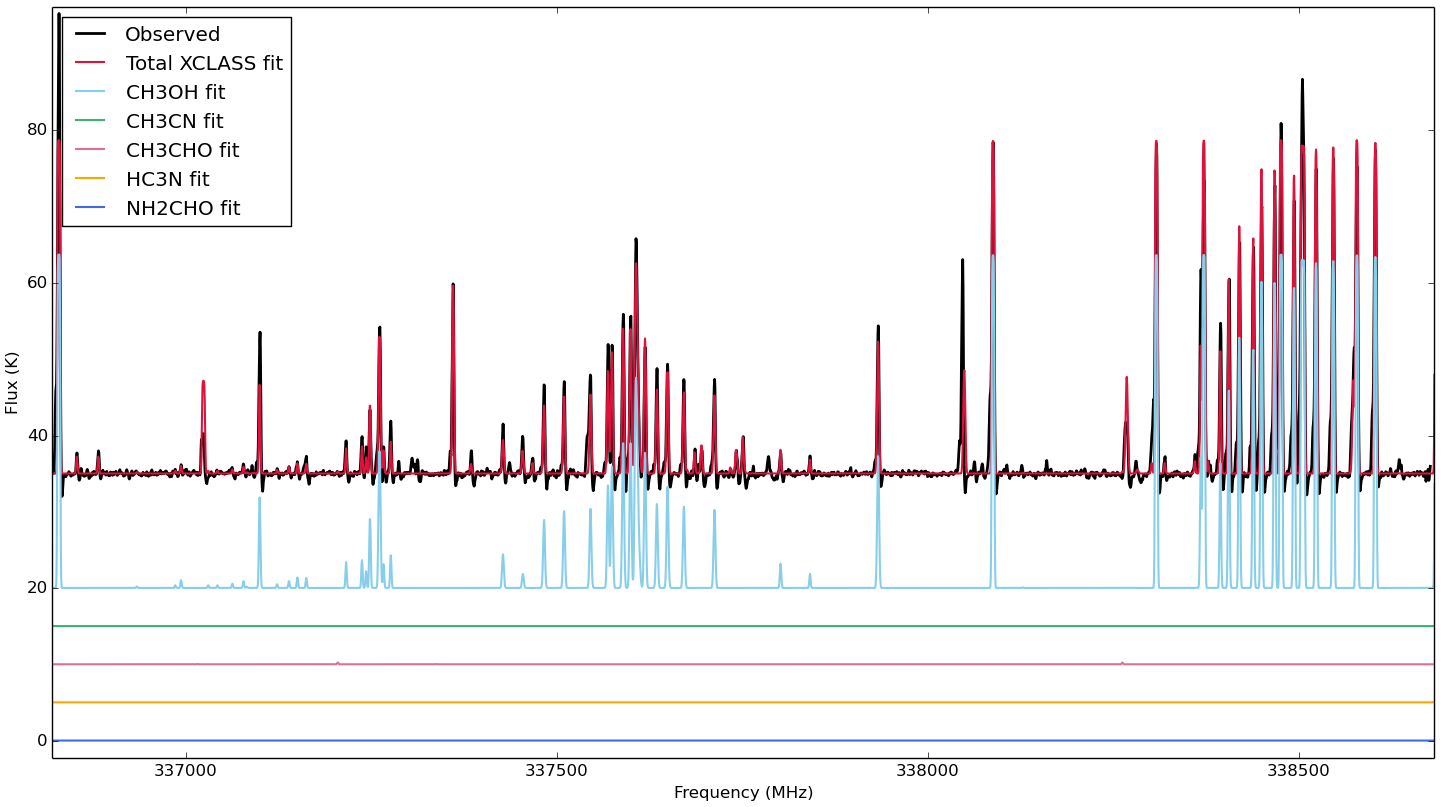}}
   \caption{G35.20 peak B2 spectral window 0 (336.8-338.7 GHz), XCLASS total fit, plus selected species.}
\end{figure}

\begin{figure}
   \resizebox{\hsize}{!}
            {\includegraphics{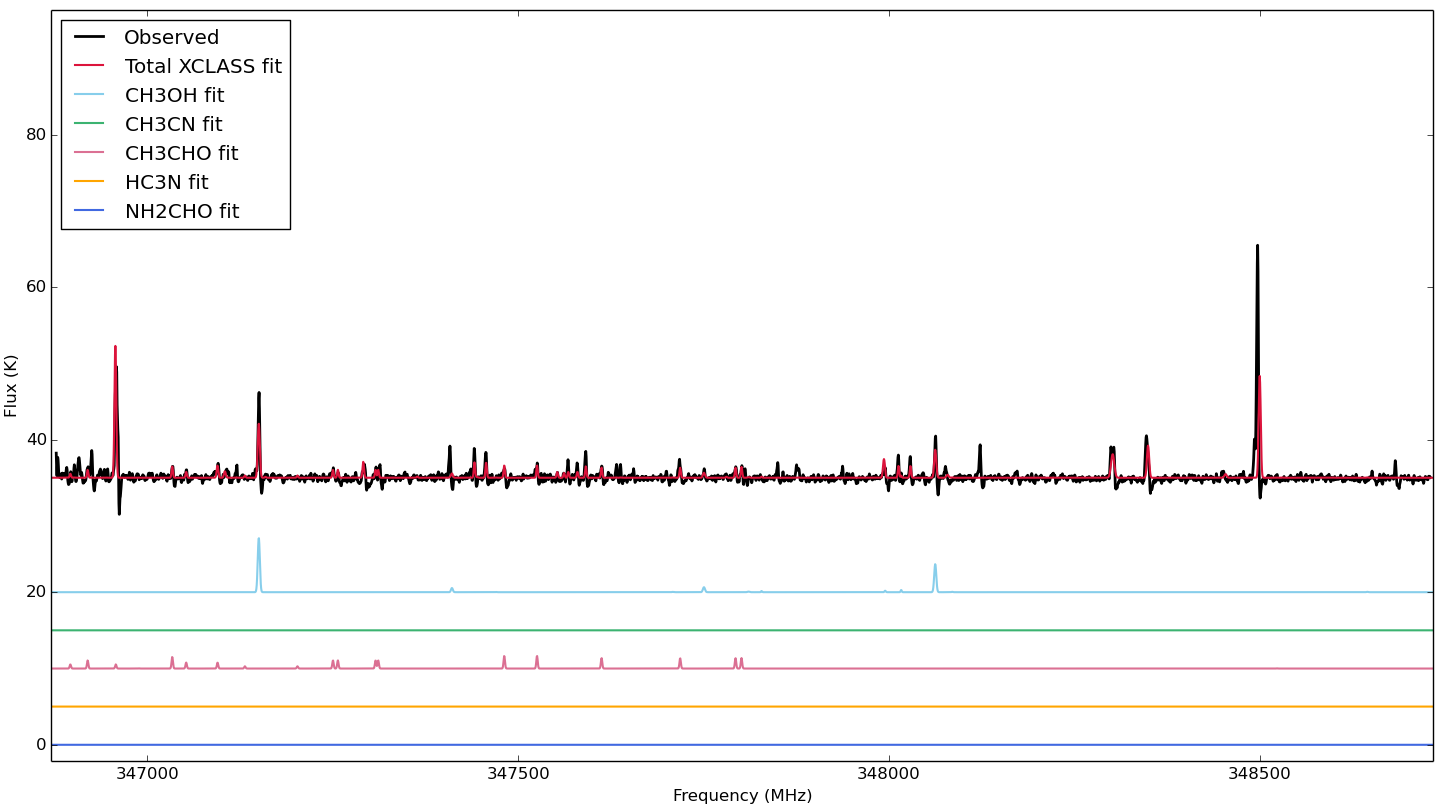}}
   \caption{G35.20 peak B2 spectral window 3 (346.9-348.7 GHz), XCLASS total fit, plus selected species.}
\end{figure}

\begin{figure}
   \resizebox{\hsize}{!}
            {\includegraphics{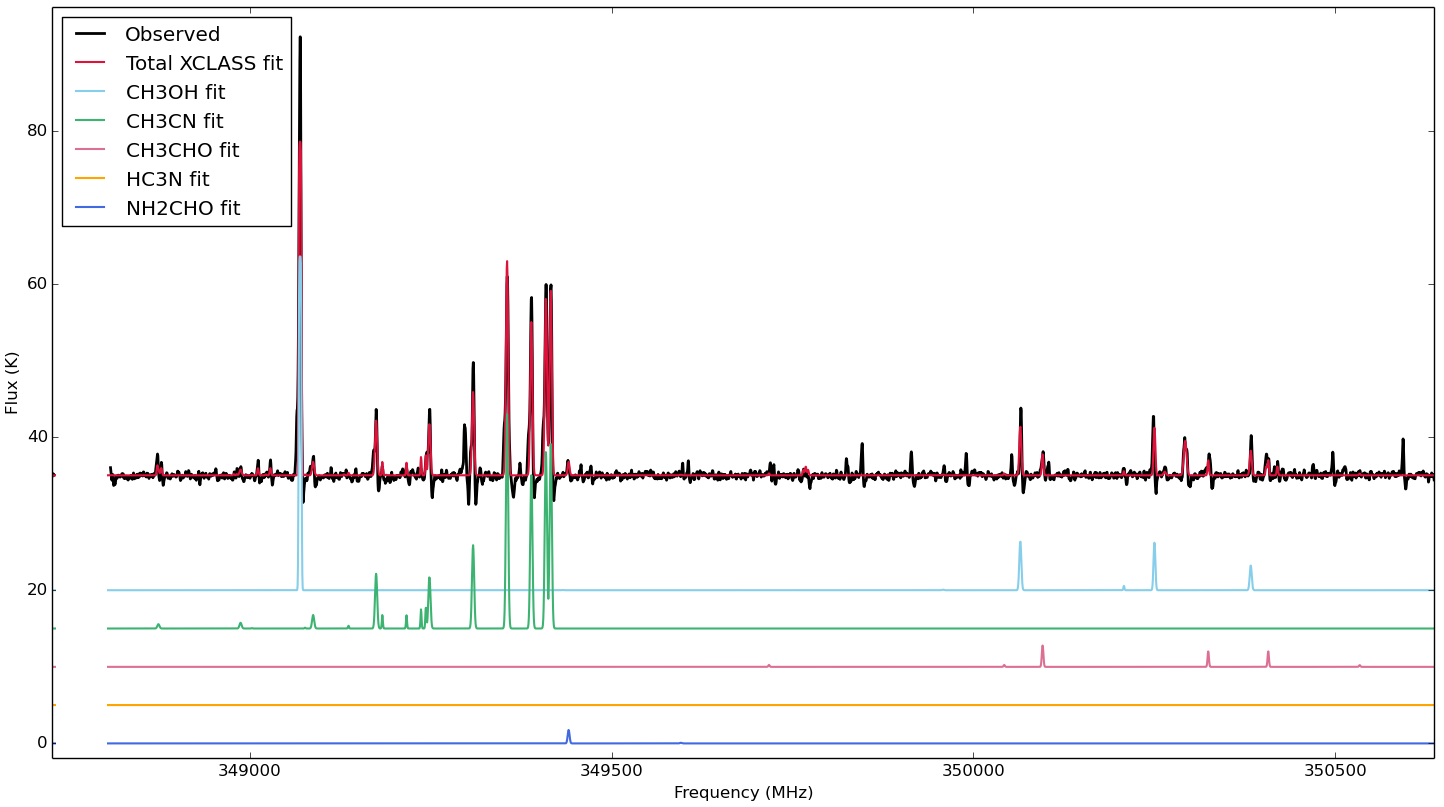}}
   \caption{G35.20 peak B2 spectral window 2 (348.8-350.7 GHz), XCLASS total fit, plus selected species.}
\end{figure}

\clearpage

\begin{figure}
   \resizebox{\hsize}{!}
            {\includegraphics{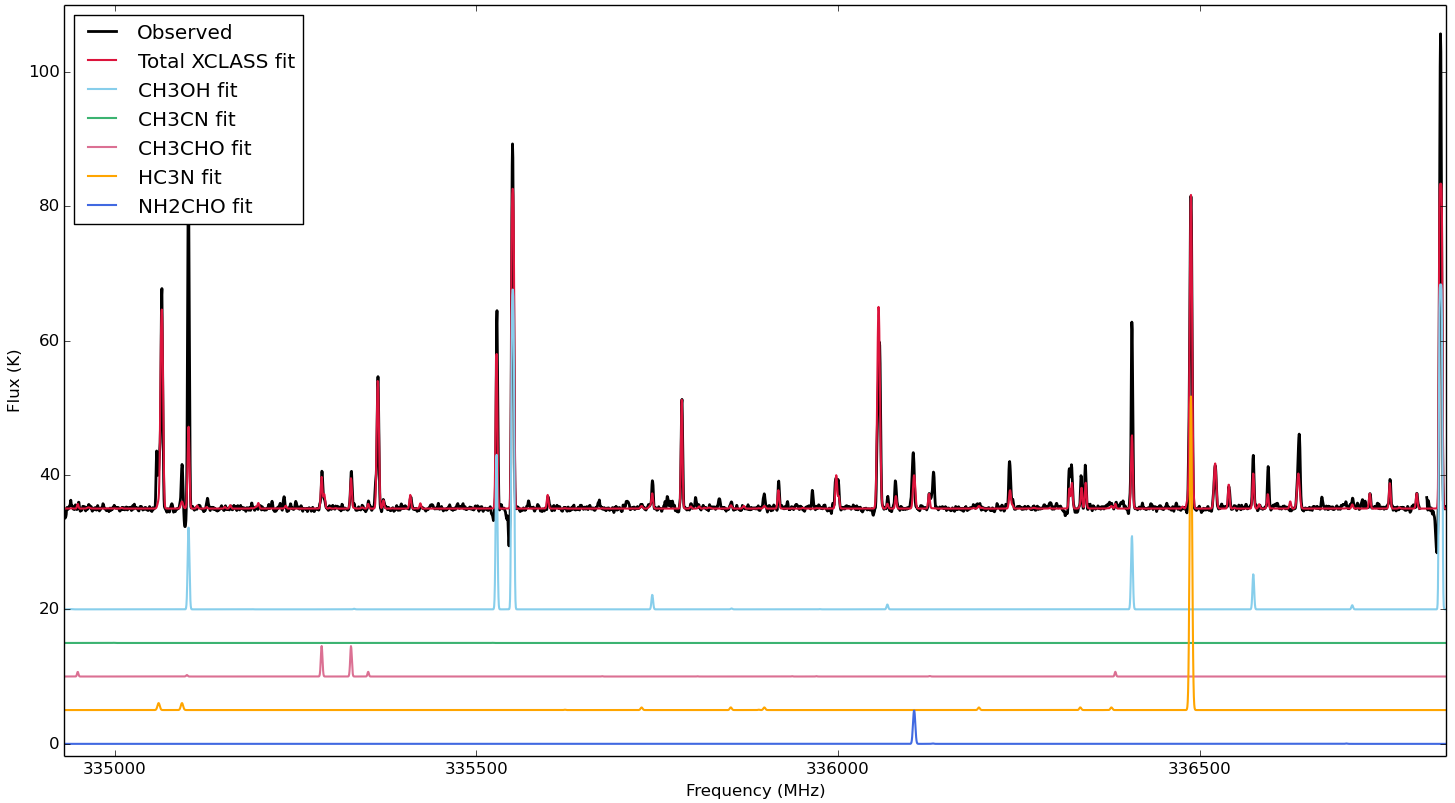}}
   \caption{G35.20 peak B3 spectral window 1 (334.9-336.8 GHz), XCLASS total fit, plus selected species.}
\end{figure}

\begin{figure}
   \resizebox{\hsize}{!}
            {\includegraphics{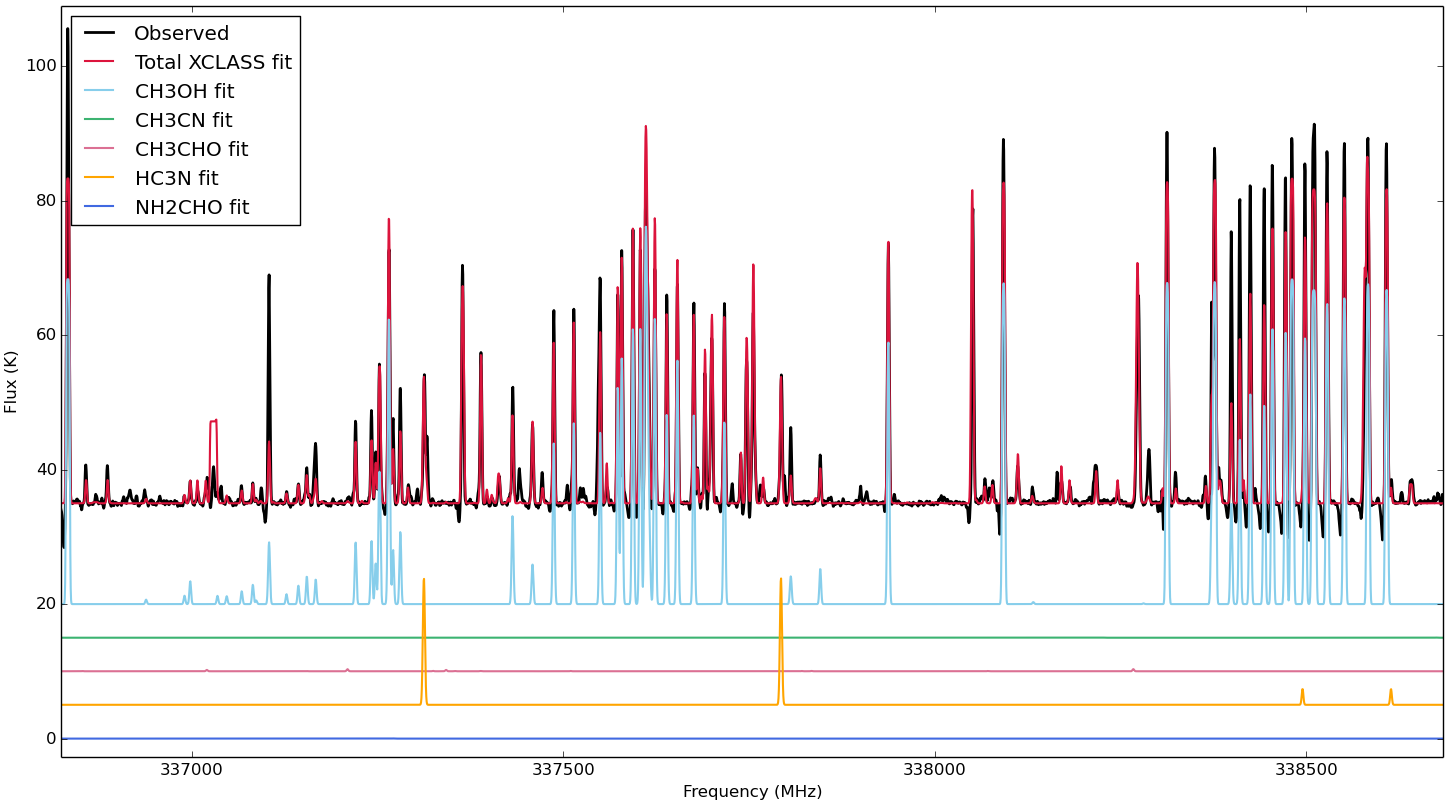}}
   \caption{G35.20 peak B3 spectral window 0 (336.8-338.7 GHz), XCLASS total fit, plus selected species.}
\end{figure}

\begin{figure}
   \resizebox{\hsize}{!}
            {\includegraphics{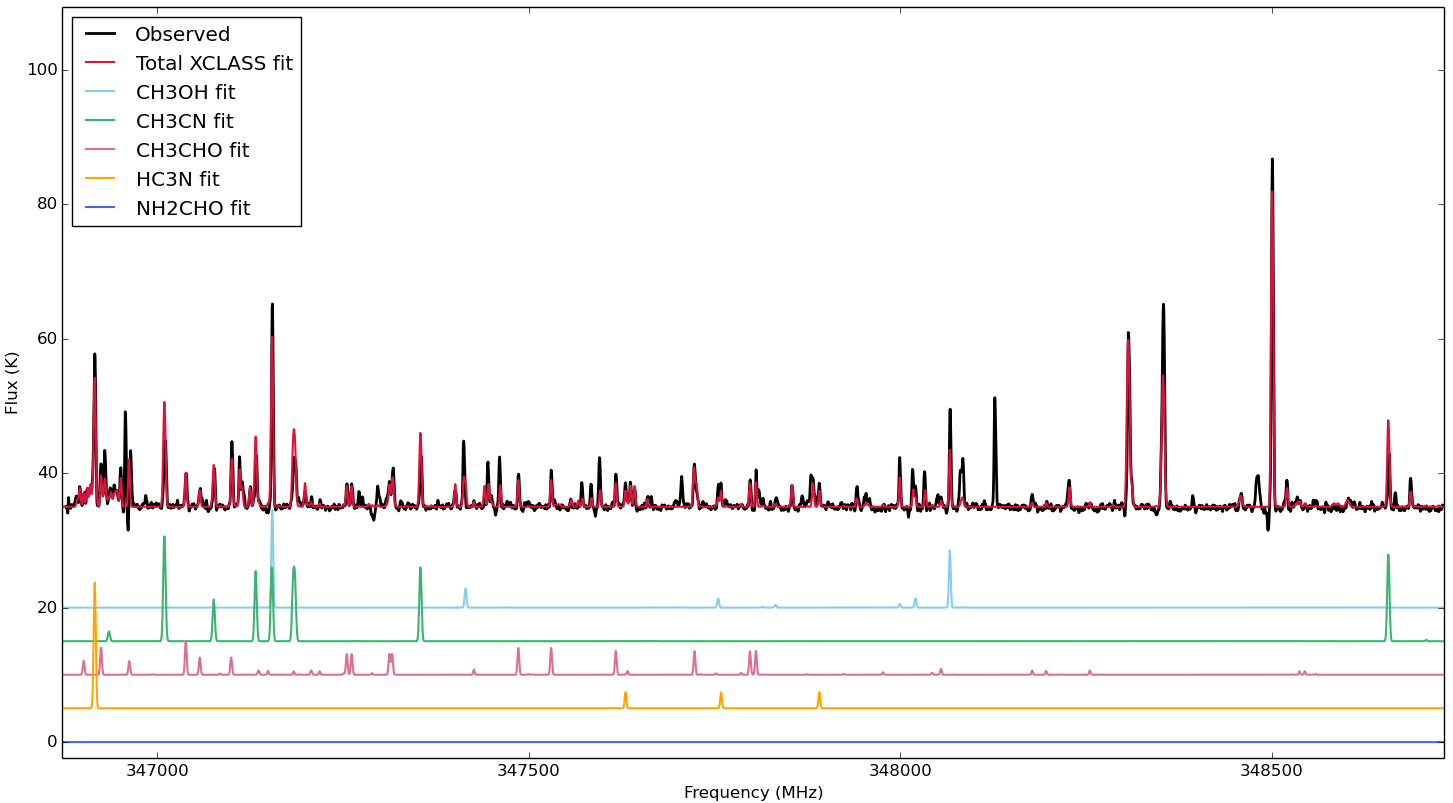}}
   \caption{G35.20 peak B3 spectral window 3 (346.9-348.7 GHz), XCLASS total fit, plus selected species.}
\end{figure}

\begin{figure}
   \resizebox{\hsize}{!}
            {\includegraphics{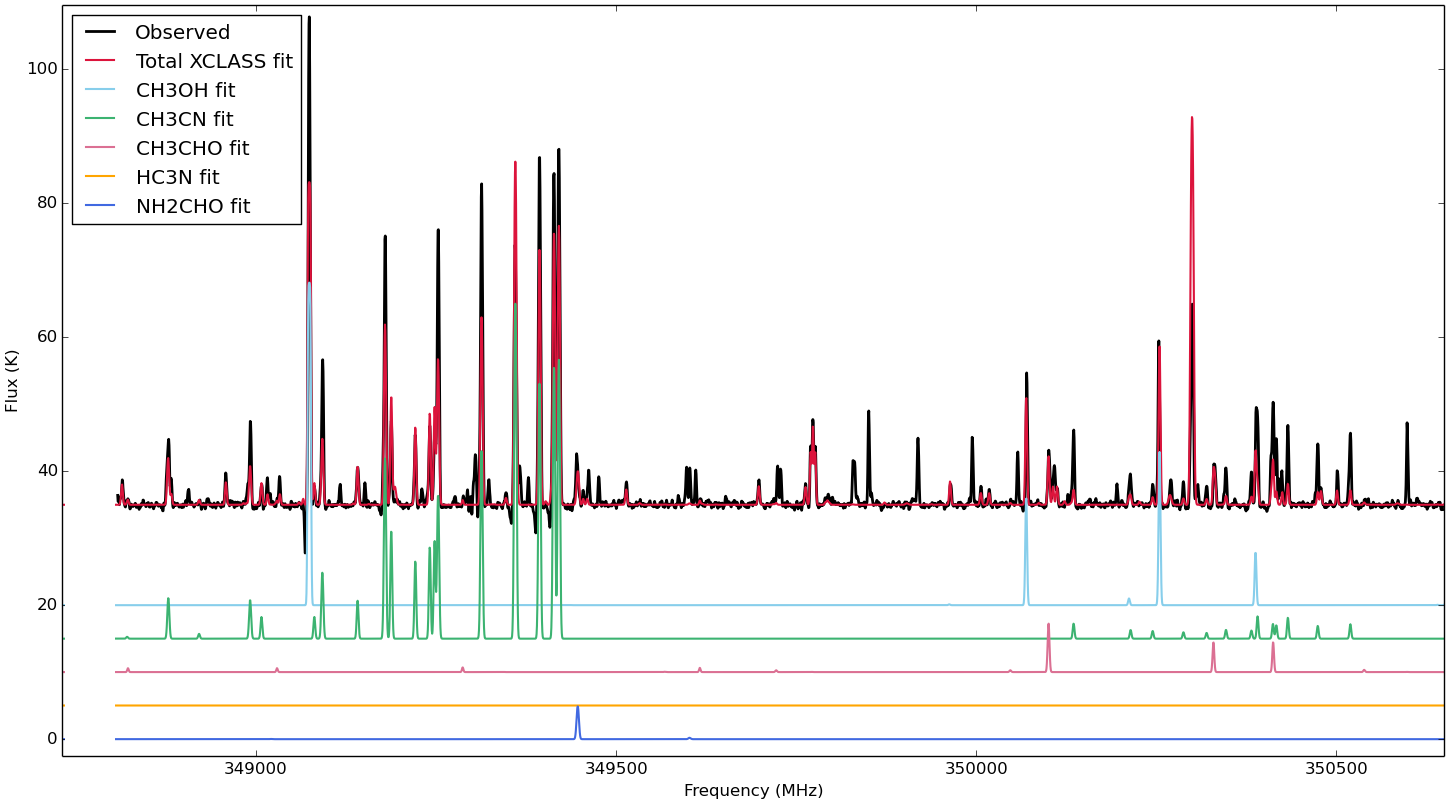}}
   \caption{G35.20 peak B3 spectral window 2 (348.8-350.7 GHz), XCLASS total fit, plus selected species.}
\end{figure}

\clearpage

\begin{figure}
   \resizebox{\hsize}{!}
            {\includegraphics{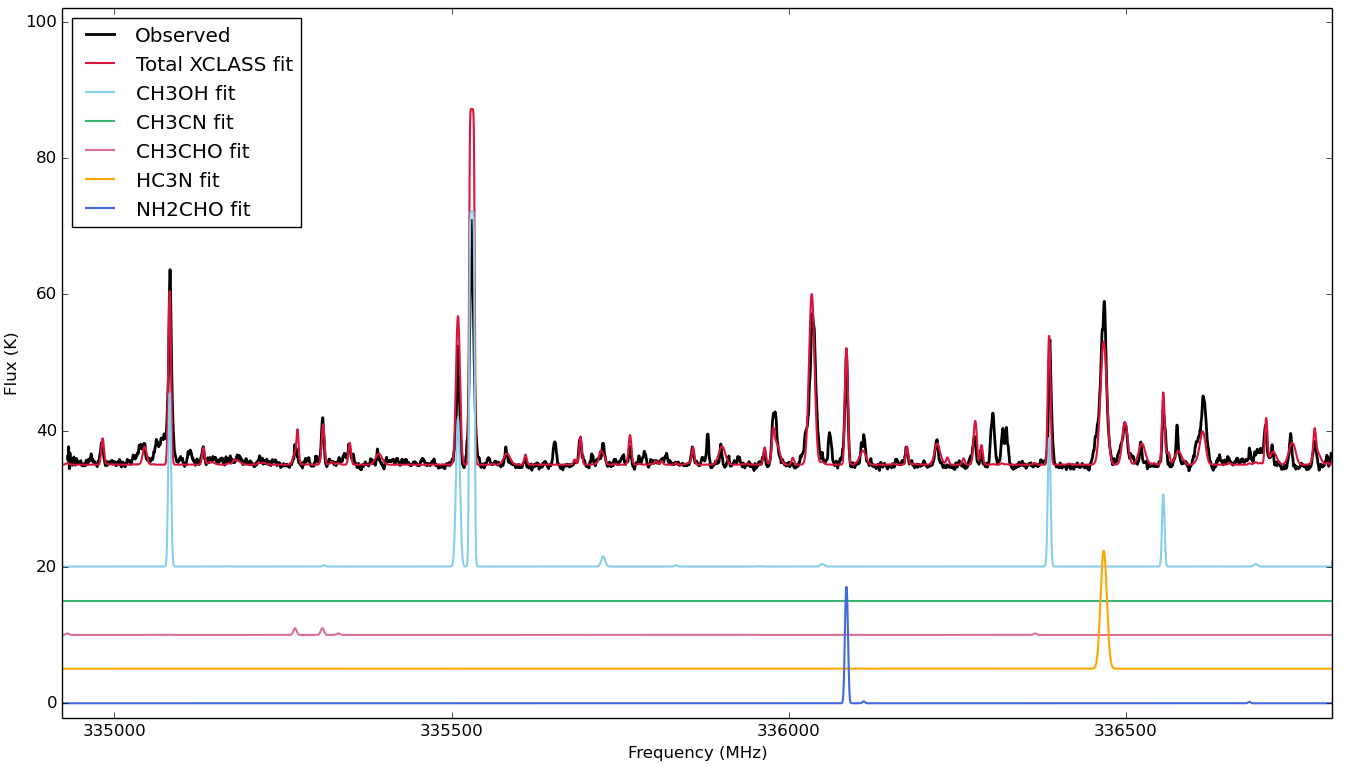}}
   \caption{G3503 spectral window 1 (334.9-336.8 GHz), XCLASS total fit, plus selected species.}
\end{figure}

\begin{figure}
   \resizebox{\hsize}{!}
            {\includegraphics{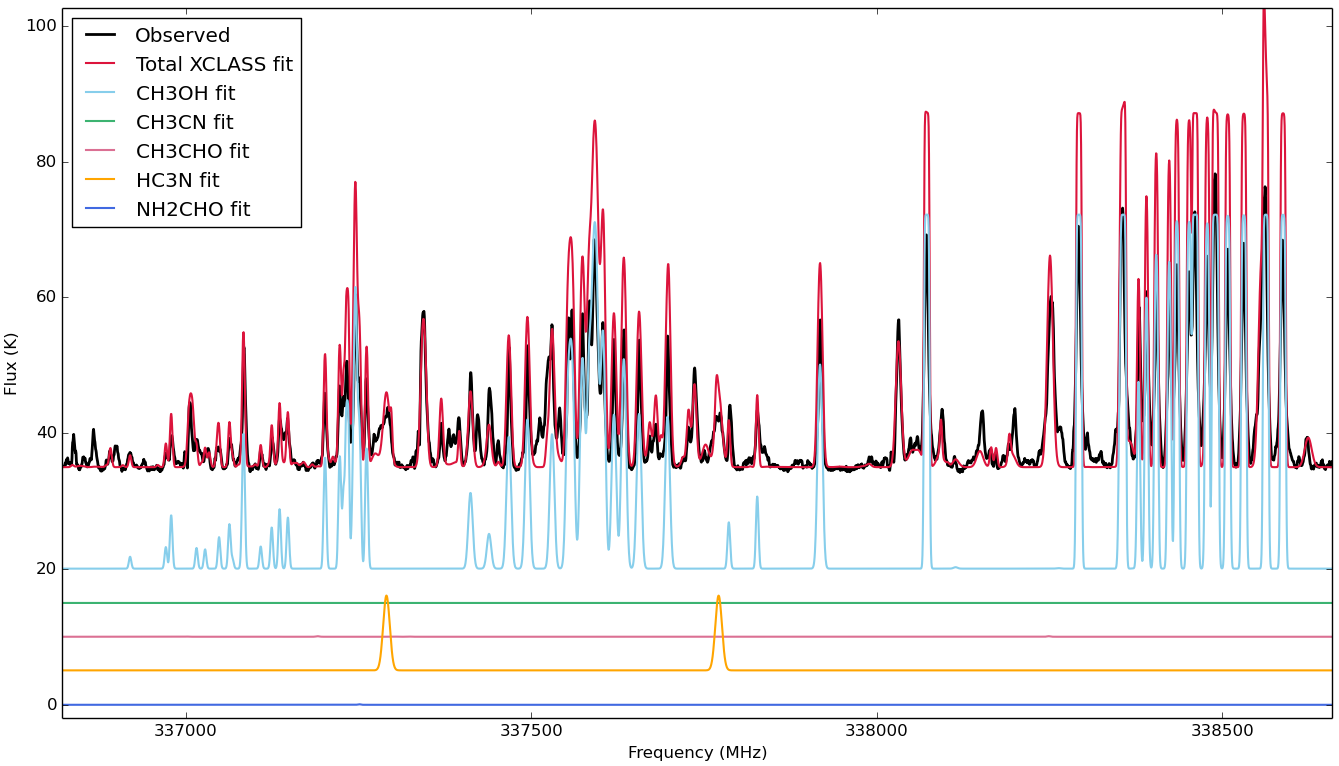}}
   \caption{G3503 spectral window 0 (336.8-338.7 GHz), XCLASS total fit, plus selected species.}
\end{figure}

\begin{figure}
   \resizebox{\hsize}{!}
            {\includegraphics{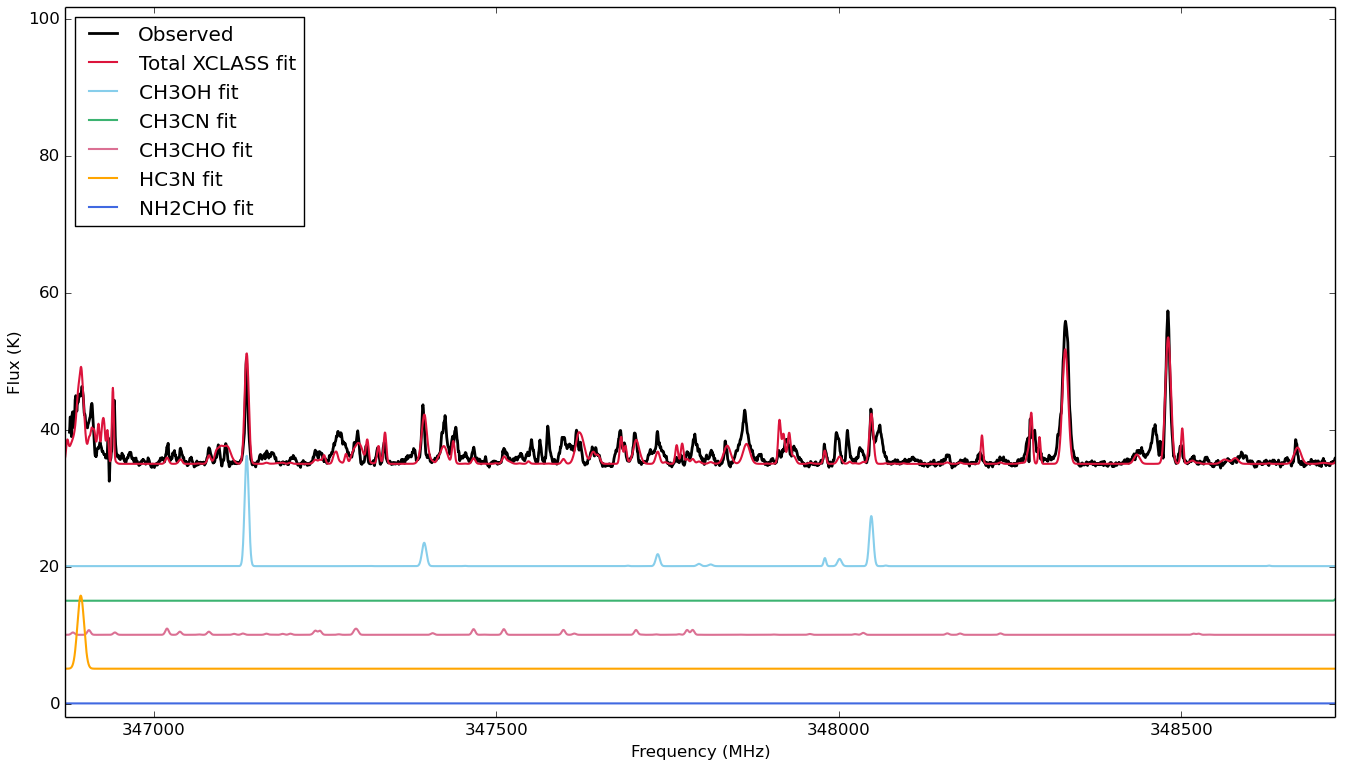}}
   \caption{G3503 spectral window 0 (336.8-338.7 GHz), XCLASS total fit, plus selected species.}
\end{figure}

\begin{figure}
   \resizebox{\hsize}{!}
            {\includegraphics{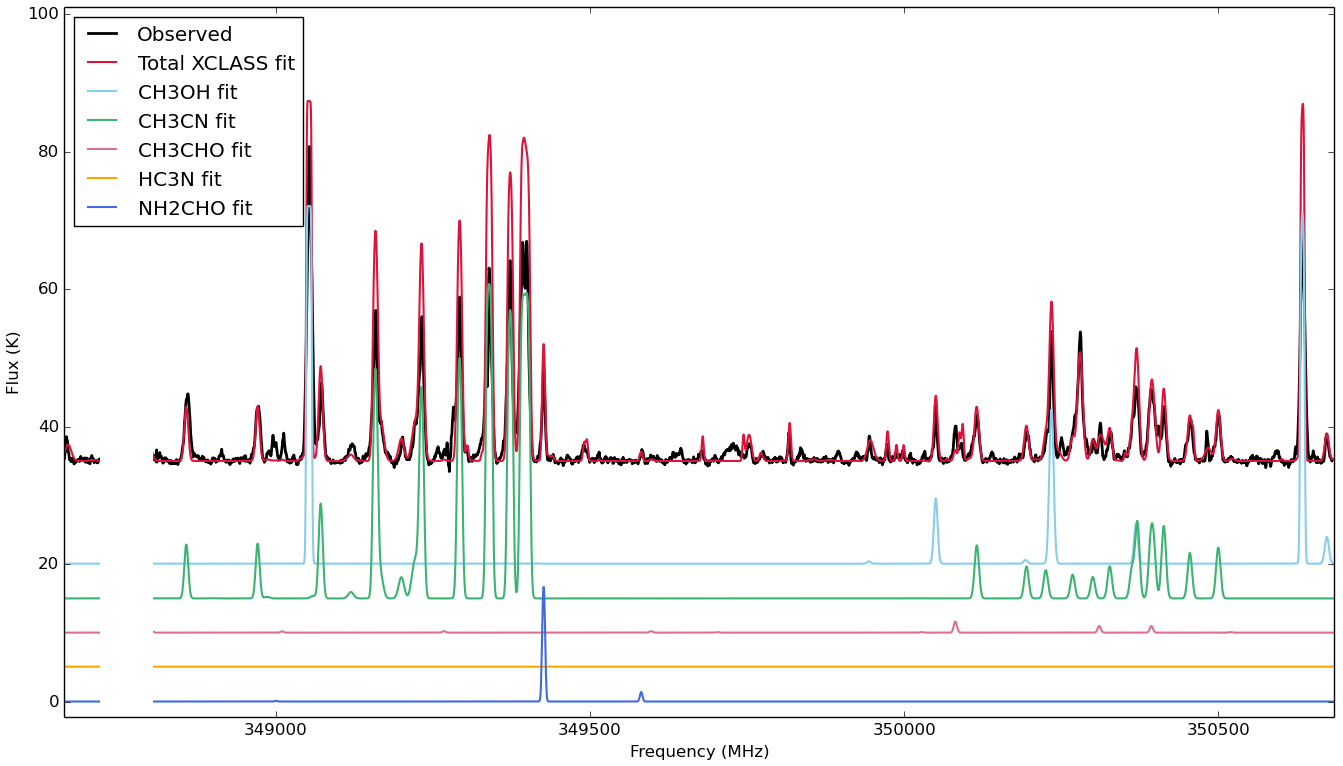}}
   \caption{G35.03 spectral window 2 (348.8-350.7 GHz), XCLASS total fit, plus selected species.}
\end{figure}

\clearpage

\end{landscape}
\twocolumn

\clearpage

\end{appendix}

\end{document}